# A National Study of Postdoctoral Research Fellows in South Africa

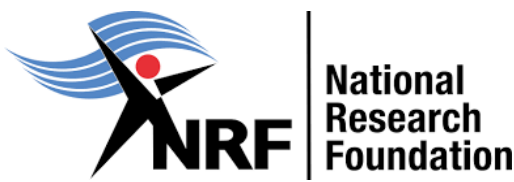
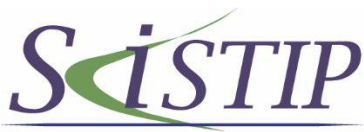

# Table of contents

# List of tables

# List of figures

# Abbreviations

| | |
|---|---|
| APC | article-processing charge |
| ASSAf | Academy of Science of South Africa |
| BEE | black economic empowerment |
| COVID-19 | coronavirus disease of 2019 |
| CPUT | Cape Peninsula University of Technology |
| CUT | Central University of Technology |
| CeSTII | Centre for Science, Technology and Innovation Indicators |
| CEO | Chief Executive Officer |
| DHET | Department of Higher Education and Training |
| DSI | Department of Science and Innovation |
| DST | Department of Science and Technology |
| DUT | Durban University of Technology |
| HE | higher education |
| HCD | Human Capital Development |
| MUT | Mangosuthu University of Technology |
| nGAP | New Generation of Academics Programme |
| NMU | Nelson Mandela University |
| NRF | National Research Foundation |
| NWU | North-West University |
| MoA | memorandum of agreement |
| MS | Microsoft |
| OBP | Onderstepoort Biological Products |
| PI | principal investigator |
| R&D | research and development |
| REC: SBER | Research Ethics Committee for Social, Behavioural and Educational Research |
| RU | Rhodes University |
| SAIAB | South African Institute for Aquatic Biodiversity |
| SAMRC | South African Medical Research Council |
| SciSTIP | DSI-NRF Centre of Excellence in Scientometrics and Science, Technology and Innovation Policy |
| SMU | Sefako Makgatho Health Sciences University |
| SPU | Sol Plaatje University |
| SU | Stellenbosch University |
| TUT | Tshwane University of Technology |
| UCT | University of Cape Town |
| UFH | University of Fort Hare |
| UFS | University of the Free State |
| UJ | University of Johannesburg |
| UL | University of Limpopo |
| UNISA | University of South Africa |
| UKZN | University of KwaZulu-Natal |
| UP | University of Pretoria |
| UWC | University of the Western Cape |
| UNIVEN | University of Venda |
| UNIZULU | University of Zululand |
| VUT | Vaal University of Technology |
| Wits | University of the Witwatersrand |
| WSU | Walter Sisulu University |

# Acknowledgements

The National Research Foundation and the Oppenheimer Memorial Trust are acknowledged as main funders of this study.

The team wishes to thank all colleagues at the universities who responded positively to our requests for information on the postdoctoral research fellows at their institutions. To those postdoctoral research fellows who took the time and effort to complete our questionnaire, we extend our sincerest appreciation.

Additionally, the team wishes to thank the following members of staff at CREST for their contributions to the project:

- Rein Treptow
- Annemarie Visagie
- Marthie van Niekerk

29 February 2024

# Abstract

This report provides the first comprehensive analysis of postdoctoral research fellows ("postdocs") in South African public universities. It combines an analysis of existing data with the analysis of primary data collected in the form of a survey of institutions on the postdocs they host, a bibliometric study of the research output of postdocs, and an individual survey of postdocs. Main findings are that the number of postdocs has been increasing steadily from 2016 to 2022. The number of postdocs varies across universities, with larger research-intensive universities hosting more postdocs. In terms of demographics, the proportion of black African postdocs has increased; the proportion of female postdocs has remained lower than that of males; there is an increasing proportion of older postdocs; and more than 60% of postdocs are foreign-born. The bibliometric analysis of the postdocs' publication output shows that it increased substantially from 2005 to 2022. Some main results of the individual survey are that a postdoc position is taken primarily to enhance future employment prospects, especially a permanent academic position. However, securing such positions is reported as challenging, which is supported by the result that one in every four postdocs has held multiple consecutive postdoc positions, and postdocs in general, but especially non-South Africans, perceive the job market as poor. Postdocs plan to leave South Africa primarily to seek better job opportunities, but also due to immigration rules or visa issues, which constitute major challenges for non-South Africans. On a positive note, postdocs are mostly satisfied with the opportunities they are given to work on interesting projects and to interact with high-quality researchers. However, most postdocs desire to contribute to teaching and supervision, but often lack the opportunity. Dissatisfaction stems mostly from low levels of remuneration, and difficulties created by the precarious nature of their positions – especially lack of employment benefits and access to financial and other services in the broader society. A lack of support for training and career development in their hosting institutions was also highlighted. Factors contributing to a successful postdoc experience include quality research, supportive colleagues, mentoring, and communication.

# Chapter 1: Introduction and background

## 1.1 Background and rationale

This national study of postdoctoral research fellows ("postdocs") in South Africa was conducted by the DSI-NRF Centre of Excellence in Scientometrics and Science, Technology and Innovation Policy (SciSTIP), on commission of the National Research Foundation (NRF). It is the first comprehensive national study to establish a number of facts about postdocs at South African research institutions (RIs). These institutions are primarily universities, but also include science councils and other public research organisations, such as national research facilities. Besides gathering information about such matters as the actual number of postdocs at public universities, their fields of research, sources of funding, remuneration levels, their publication output and key demographics, the study also established more qualitative aspects of their career trajectory and experiences as postdocs in the higher education system.

The study is of national importance and was urgently needed to answer several questions where there are significant gaps in our current understanding of the postdocs in the country, as well as to address important policy and research-funding imperatives. Over the past number of years, the South African higher education sector has invested increasingly in postdocs as a significant human resource capability. However, it is also the case that, until now, we had very little information on postdocs at a national level. We did not have such basic information as the numbers of fellows by scientific field, their research specialisations, their country of origin or source of funding. In addition, we had no demographic information (gender, race, nationality, age) of these fellows that would tell us more about how to design interventions and funding support in critical areas.

It is also the case that the status of postdocs has been a contested issue for some time now. Anecdotal evidence suggests that there are many questions around their job descriptions, conditions of service, and the like that require attention, as there do not seem to be good-practice guidelines across sectors, and within the higher education sector, in this regard. Most of the information that we needed was either (1) only available at the institutional level; or (2) not readily available and needed to be gathered through this sector-wide, national study. The study provides the required up-to-date and accurate evidence that all postdoc-hosting RIs and national agencies require to fully realise the potential of postdocs as an important group of emerging scholars.

In the remainder of this section, we provide a brief literature review of (1) the relevant policy and strategic documents that informed our study; and (2) the results of previous research on the state and dynamics of postdocs in the South African science system. We conclude this section with a summary of the knowledge gaps that emerged from our review.

### 1.1.1 Policy and strategic documents that informed the study

The 1997 Education white paper 3: programme for the transformation of higher education (Department of Higher Education and Training [DHET], 1997:33), under the heading "Research", drew attention to the importance of increased access of female students to postdoctoral positions as a means of increasing the pool of researchers and improving the demographic representation of staff in higher education. Four years later, the 2001 National plan for higher education (Department of Education [DoE], 2001) asserted that there is a competitive edge to be derived from intensive postdoctoral training. As such, South Africa needs to improve on the quality and output of its postdocs to support national research, assist with equity goals and ensure that universities and the nation can meet global challenges (Vranas & Hendry 2013).

The 2013 White paper for post-school education and training (DHET, 2013:35) refers to postdocs as an important component of building the research capacity of the country's universities. The first dedicated reference in South African policy documents to advancing the postdoc appear in the 2016 Human capital development (HCD) strategy for research, innovation and scholarship (Department of Science and Technology[1] [DST], 2016). In this document it is suggested that the increased PhD output should be complemented with a focus on the demand-side of human capital, which includes an increase in the availability of, amongst others[2], postdoctoral fellowships. The DST committed itself to ensuring that support for "a significant scaling up of South Africa's postdoctoral cadre" receives high priority through the NRF. Higher education institutions (HEIs) and science councils were to be requested to consider institutional interventions that they could make to complement DST/NRF efforts to increase the domestic and international attractiveness of postdoc contracts in South Africa (DST, 2016). The DST itself suggested a focused programme to secure large numbers of postdoc candidates from abroad. It was committed to engage key bilateral partners with large science systems in pursuit of cooperation agreements that would see meaningful numbers of recent doctoral graduates moving to South Africa for two- to three-year terms, ideally with a sharing of costs. Sandwich programmes were to be strengthened to ensure the retention of the internationally trained doctoral graduates. Three years later, the DST (2019b) proposed easing immigration rules to attract postdoctoral researchers to South Africa. The Academy of Science of South Africa (ASSAf, 2018) also called for the relaxing of immigration requirements to encourage international PhD graduates to stay on in postdoc positions to increase the internationalisation of engineering education in South Africa.

At the same time, the DST (2019b) expressed the intention of government to "support and retain an appropriate percentage of each annual graduating PhD cohort" (which has shown sustained increases) in the "Postdoctoral Fellowship Programme[3]" for "eventual absorption into the academic and research system". To improve demographic representation among established researchers, a significant number of those who will be "retained" as such, will be "black and women doctoral graduates, particularly South Africans". The policy further expresses a concern that the "contribution" of postdocs hosted at universities and science councils (the number of which have increased) "has not been optimised because their status has not been defined". Thus, the policy expresses the intent of the DST and the DHET to "formalise a set of guidelines on how to optimise the contribution of postdoctoral fellows" (DST, 2019b).

The reasons for these policy's foci on postdocs originate from two main concerns. The first is that postdoctoral fellowships enable doctoral graduates, particularly those from designated groups, to obtain further experience in research and innovation in preparation for research and development (R&D) positions and future leadership roles within the South African national system of innovation (NSI) (NRF, 2018). With specific reference to the academic sector, HEIs invest in postdocs as one avenue to produce highly skilled and qualified academics (Holley *et al*., 2018; Simmonds & Bitzer, 2018), and the former DST (2016) expressed its intention in its 2016 HCD strategy for research, innovation and scholarship, to create "career awards" for recent postdoctoral researchers who wish to pursue academic careers. This was to be undertaken in conjunction with universities and public research institutions, to retain and nurture this category of emerging researchers in those institutions (DST, 2016). However, no evidence has been found of the implementation of such career awards, thereby supporting Kerr's (2021) suspicion that postdocs "might actually not be part of the academic pipeline, despite these fellowships being framed as further research training for PhDs, and valuable preparation for an academic career".

---

[1] The Department of Science and Technology (DST) was renamed the Department of Science and Innovation (DSI) on 26 June 2019.

[2] Entry-level positions, development programmes, and internship programmes.

[3] As far as could be determined, such a formal programme does not (yet) exist.

The recent policy focus seems to be more on postdocs' role in alleviating supervisory or "internal" bottlenecks, "augmenting" the supervisory and research capacity of universities, and boosting the country's PhD productivity (DST, 2016, 2019b). The 2019 White paper on science, technology and innovation (DST, 2019b) identifies a "supervisory bottleneck" created by the much higher increase, from 1996 to 2014 in the number of enrolled students (350%) than in the number of PhD-qualified staff (65%). As Breier and Herman (2017) explain, South African universities face a conundrum: they need more academics with PhDs, from historically disadvantaged population groups in particular. In order to have more staff with PhDs, they need to produce more PhD graduates. But in order to produce more PhD graduates, they need more staff with PhDs to supervise. According to the DST (2019b), this conundrum has been partly addressed through the increased number of postdocs who "make an invaluable contribution to the research system by mentoring postgraduate students". Similarly, increased use of postdocs has been recommended by ASSAf (2018) as a means to increase doctoral completion rates in engineering in South Africa, by such postdocs "strengthening support".

It is clear from this brief overview of the policy context that inform discussions around the postdoc, that there are many unanswered questions that needed to be addressed by our national study. In particular, the exact position and value of the postdoc in the academic pipeline – from post-graduate students through to emerging scholars and finally established scholars – required further attention.

### 1.1.2   Results of previous research on postdocs in the South African science system

In 2020, the first author of this report conducted a scoping review of scholarship on postdocs across the world (Prozesky, 2021). It emerged from that review that our current knowledge and understanding of the South African postdoc are based on some (limited) statistical knowledge gathered through the annual, national surveys of R&D (from here on referred to as the R&D surveys) undertaken by the Centre for Science, Technology and Innovation Indicators (CeSTII) at South Africa's Human Sciences Research Council (HSRC). In addition, a few surveys have been conducted, the most recent being the National Tracer Study of Doctoral Graduates in South Africa, conducted by SciSTIP in 2021 (Mouton *et al*., 2022; from here on referred to as the "PhD Tracer Study"). Smaller (case) studies have also been undertaken. For the purpose of this proposal, we summarise the salient results and findings of these studies below.

1. The most recent R&D survey shows that in 2019, approximately 2 800 postdocs were 'employed' in the South African science system (HSRC–CeSTII, 2021). This represents a significant increase from 357 in 2003 (HSRC–CeSTII, 2005).
2. In the PhD Tracer Study, one-fifth of all doctoral graduates who had studied full-time indicated that they had accepted a postdoc upon completion of their studies. Although the proportion of PhD graduates who accepted a postdoc increased from 16% in the earliest (2000–2004) graduation window to the 2010–2014 window, it seems to have stabilised at 22% (Mouton *et al*., 2022).
3. The PhD Tracer Study also confirmed international trends where individuals accept a second or even third term as a postdoc, because of a lack of permanent academic and scientific positions in the national science system. In response to a specific question, nearly 30% of all doctoral graduates who had accepted a postdoc on completion of their studies indicated that they had no choice but to do so, as no relevant vacancies were available. This percentage has increased in the recent past, which suggests that the absorptive capacity of the South African higher-education and science system to employ highly skilled doctoral graduates and postdocs has already reached a point of saturation (Mouton *et al.*, 2022).
4. The main rationale for postdocs' appointments is that they conduct research and thereby contribute to knowledge production. To underscore this point, our analysis of R&D survey results from 2007 to 2019 (HSRC–CeSTII, 2014, 2019 & 2021) shows that South African postdocs spend on average 92% of their time performing research, which is much more than the percentage of time that PhD students (55%) and researchers (24%) in the same sector spend on research. Recent studies have shown that postdocs are also increasingly required not only to do research and contribute to knowledge production (the main rationale for their appointment) but increasingly to assist permanent academics with other duties, *inter alia,* teaching, supervision, community engagement and administrative duties. It is, therefore, not surprising that many universities currently interpret the role of the postdoc as providing additional support for their own research agendas, teaching responsibilities and related tasks.
5. Various sources show that the NRF has supported over time on average about one-third of all postdocs in the system. However, the available data also shows that the NRF's relative support for postdocs may have declined in recent years. Our own calculations show that, from 2012 to 2019, the NRF funded on average 734 postdoctoral fellows per annum, which amounts to an average of 34% of the postdocs in the country's higher education sector over the same period. But in the most recent year for which we have data (2019), the NRF funded only 27% (780) of the 2 867 postdoctoral fellows in the country's higher education sector. In 2018, 395 of the 799 postdoctoral fellows that the NRF funded were South Africans, 73 were permanent residents and 331 (41%) were non-South Africans (DST, 2019a).

The existing statistical and quantitative information (mostly sourced from the R&D surveys and SciSTIP's PhD Tracer Study) remains incomplete in many respects. SciSTIP is in the position to conduct bibliometric analyses around the productivity, research collaborations, networks and citation visibility of the postdocs that were identified through its PhD Tracer Study. This is not a straightforward exercise, as it will require some matching and integration of diverse data sources. However, the sample of postdocs from its PhD Tracer Study also does not cover the entire (or the majority) population of current postdocs in the system. In the next section, we list the key research questions that remained unanswered prior to our national study of postdocs in South Africa, and which out study therefore addressed.

## 1.2   Key research questions

Our literature review allowed us to formulate the following key research questions that our study addressed.
1. What is the quantum and distribution of postdocs in the public university system, by scientific discipline and by key demographics (chronological age, career age; gender, country of birth, race, citizenship/ residency status in South Africa, and country of PhD training)?
2. Which local and international agencies and institutions, aside from the NRF, are providing financial support to postdocs at South African public universities?
3. How productive are South African postdocs in terms of research publications (journal articles, books, book chapters and conference proceedings), benchmarked against the average productivity of permanent academics?

4. How does that productivity correlate with key demographics of the postdocs, and how does it differ across universities and scientific disciplines?
5. In 2022, what were postdocs' work situation, job satisfaction, challenges, as well as career development and career intentions?

## 1.3    Structure of the report

This report is divided into four main sections:

1. The first section – Chapter 2 – provides detailed information on the design of the study.
2. The second section – Chapter 3 and 4 – provides information based on the data submitted by the universities where Chapter 3 focuses on descriptive data on the number and composition of postdocs from 2016 to 2022, while Chapter 4 focuses on the publications produced by postdocs over the same period.
3. The third section – Chapter 5 – provides the results of the survey of postdocs hosted by South African public universities.

We include in this report a detailed table of contents to assist the reader in navigating the various sub-sections of each of the sections described above.

# Chapter 2: Study design and methodology

## 2.1    Introduction

In terms of research design, our study consisted of three components: two primary-data-collection processes (Components 1 and 3), and an analysis of existing bibliometric data (Component 2). As most universities routinely collect and store data on their postdocs, in Component 1 we requested universities to provide us with a selection of these data. We therefore refer to this component as "institutional data gathering". It provided us with a quantitative overview of the scale and nature of postdocs in South Africa (Research Questions 1 and 2), without having to ask postdocs themselves to provide these data.

Component 2, the bibliometric study, was aimed at addressing Research Questions 3 to 4. It depended on Component 1 producing up-to-date and comprehensive basic (biographical) data on postdocs, as these data had to be linked to SciSTIP's existing bibliometric databases to allow for bibliometric analysis.

The second primary-data-collection process (Component 3) involved collecting additional (anonymous) data directly from postdocs, through a web-based survey. This survey provided us with the more granular and in-depth data required to comprehensively address Research Question 5.

The main challenges that we foresaw were that too few universities would provide us with usable data; and/or that a low survey response rate could limit the external validity of our results and the extent to which we can analyse the data with sophisticated statistical methods. Also, if the institutional data gathering did not produce the necessary data on postdocs, we would have been prevented from conducting the bibliometric analysis required to address some of the research questions. We have attempted to address these issues through extensive engagement with universities (and postdocs). The nature and extent of this engagement is discussed in the section below.

## 2.2    Prior engagement with universities and postdocs

To achieve the goal of establishing a holistic picture of the state of play of postdocs in the NSI, it is crucial that as many of the South African research institutions (RIs) that host postdocs, cooperate in our study. Given that the majority (we estimate around 94%) of these are universities, our focus was on them.

Since July 2022, we have been engaging with South African universities to (1) build rapport with them; (2) negotiate their cooperation and access to their data on the postdocs they host; (3) plan our research more effectively to avoid setbacks; and (4) ensure that they as well as the postdocs they host, share in the benefits of the research as far as possible. This engagement has taken various forms, which we discuss in more detail in the remainder of this section.

In early July, SciSTIP and most of the project team members were involved in convening the first (since 2015) Annual National Postdoctoral Research Forum. All RIs in the country were invited to send representatives to participate in the two days of discussions that were held in Stellenbosch. The Forum provided an opportunity for the project team to brief the representatives of the postdoc offices and related entity structures about the planned study, and how their participation could be of benefit to them. The Forum also provided a space for a discussion with postdocs who attended, to refine our data-collection instrument for Component 3.

Later in July 2022, we emailed the deputy vice-chancellors with a research portfolio, or research directors, of most South African universities to alert them to the study and asking them whether they would be willing to assist and cooperate with us in conducting the study. We also asked them the following: (1) whether there is a central location at each university where a database/list of all postdocs is gathered and maintained; and (2) what categories/types of information about the postdocs are included in such a database. The responses, obtained from the majority (19) of the universities, were extremely positive. Most of the universities indicated that they have a central database that contains at least some information on the postdocs they host, and in most cases, we were also provided with the relevant contact person at each university.

Third, a letter was drafted by the project team, which reiterated the background to the study, as well as its rationale and importance. It confirms that the study involves a national survey that has the official support and endorsement of the NRF. The letter concludes with the statement, "I trust that your university will therefore agree to participate in the completion of this study that is of national importance". In the last week of July 2022, the letter was provided to the top management of the funder (the NRF), together with the email addresses of the relevant contact person at each university (as determined earlier in July) to whom they should send the letter.

After agreeing with the content of the letter, it was countersigned by the Chief Executive Officer of the NRF (Prof Fulufhelo Nelwamondo), and Deputy CEO: RISA (Dr Eugene Lottering). Dr Eugene Lottering's office (Research Innovation Support and Advancement [RISA]) at the NRF then distributed the letter to the email addresses we provided.

We obtained ethical clearance for the study from Stellenbosch University's Research Ethics Committee for Social, Behavioural and Educational Research (REC:SBER). Another eleven universities indicated that we needed to follow additional processes to apply for permission and/or ethics clearance from their institutions before we could proceed with data collection. To comply with these requirements, we had to stagger our data collection process, and run some of the study components in parallel, to accommodate the universities' requirements.

## 2.3   Component 1: Institutional data gathering

After the necessary ethical clearance and/or institutional permission had been granted by the universities, the deputy vice-chancellors or research directors, and the contact person they nominated, were emailed with a more detailed request for the data, including a data collection template. Universities were given the option to complete the template or, if more convenient to them, provide us with the data in another, existing format. The data we requested are listed in Table .

**Table 2.1. Data on postdocs that were requested from universities**

| Surname | |
|---|---|
| First names | |
| ORCID number | |
| Email address (*2022 postdocs only*) | |
| Demographics | Year of birth |
| | Gender (*M/F/Other*) |
| | Country of birth |
| | Race (*only if country of birth is SA*) |
| | Citizenship / residency status in SA (*Citizen / Permanent resident / Temporary resident*) |

| PhD | Year awarded |
| | Name of awarding university/ies |
| | Title of thesis |
| Name of host academic department | |
| 1st year of appointment | |
| Last year of appointment | |
| Names of sources of funding of appointment | Primary source |
| | Secondary source |
| | Other source(s) |
| Total amount of funding (R) | |

As Table indicates, some personal information was collected, and much of the information could be considered confidential. First names, surnames and/or ORCID numbers were collected for the sole purpose of identifying the postdocs' publications in SciSTIP's existing databases, to allow for bibliometric analysis (Component 2). After they served that purpose, they were deleted from our institutional data set. The email addresses of the 2022 postdocs were only used to administer our web-based survey (see Component 3 below). The other data were required to answer the key research questions listed in Chapter 1, as agreed upon with the funder. In particular, our analysis of the demographic data will be useful to the NRF in their design of interventions and funding support in critical areas. The NRF also collects data on the students, postdocs and researchers that it funds (estimated at a third of all postdocs in South Africa), and our data collection template and the categories of the variables have been aligned to those the NRF collects.

The identity of the postdocs in our institutional data set are protected by the fact that all our analyses have been conducted at an aggregated level (i.e., no personal information of the postdocs have been made public), and we avoided publicly communicating results in a way that may damage any institution's reputation. As a token of gratitude for their cooperation, and to assist them in their research planning, we plan to provide to those universities with at least 100 postdocs with a customised fact sheet that benchmarks the aggregated results on their own postdocs, with the national results on the higher education sector as a whole.

## 2.4    Component 2: Bibliometric study

Component 1 provided us with sufficient data on postdocs to allow for bibliometric analysis. The institutional data were linked to two of SciSTIP's existing bibliometric databases. The first of these, the Clarivate Analytics™ Web of Science (WoS), represents the total output in WoS-indexed journals. The Centre for Research on Evaluation, Science and Technology (CREST), which hosts SciSTIP, receives the individual records under a licensing agreement with Clarivate Analytics, providing us with access to more than 80 million records covering the period since 1980. CREST is the only centre on the African continent that has access to the full micro-data of the WoS database. The second source of existing data for the bibliometric component of our study is CREST's proprietary database, SA Knowledgebase. It includes the publication output of South African academics, as published in all the journals recognised for subsidy by the DHET [WoS, the Proquest IBSS list, the Norwegian list, Scopus (since 2016), and local South African journals on the DHET list].

## 2.5 Component 3: Web-based survey

To establish more granular aspects of the status and experiences of postdocs, we conducted a survey of all 2022 postdocs in South Africa.

### 2.5.1 Developing the questionnaire

The questionnaire (see [Annexure C](#)) was constructed using the literature identified in a scoping review (Prozesky, 2021) as a starting point, thereby allowing us to compare our results with those found internationally. To this aim, questionnaires used in previous multi-national and other national surveys on postdocs were consulted. These questionnaires were either freely available (e.g., van Benthem *et al.*, 2020; Nature & Shift Learning, 2020; Ribeiro *et al.*, 2017; McConnell *et al.*, 2018), or their items could be deduced from the publications that were produced from the results (e.g., Richards, 2012; Powell, 2012; Davis, 2005). The items were then compiled in a database of approximately 300 items, from which the project team members selected (and adapted, where necessary) what they considered to be the most relevant ones for the South African context. The selection and adaptation (refinement) of the items were informed by (1) a discussion with postdocs at the 2022 Postdoctoral Research Forum; (2) two research team members' own experiential knowledge (Dr Mothapo, as the Head: Postdoctoral Research Support at SU, and Dr van Schalkwyk, as a current postdoc, also at SU); and (3) previous small-scale surveys of, or qualitative research on, postdocs in South Africa (Vranas & Hendry; Simmonds & Bitzer, 2018).

We are quite aware of the demands made on individuals to complete questionnaires of this nature. For this reason, we kept the time to complete the questionnaire as short as possible. We estimated (using [Versta Research's "point" system](#) that accurately predicts questionnaire length) that it would take, on average, less than 20 minutes to complete. No fields were set as "required", thereby allowing respondents to "skip" any questions that they did not want to answer. While completing the questionnaire, respondents could withdraw from participating in the survey, by simply closing their browser. However, once they had submitted a completed questionnaire online, they were no longer able to withdraw their responses as the survey was anonymous and therefore their responses cannot be linked back to them.

### 2.5.2 Administering the questionnaire: recruitment and response rates

After the necessary ethical clearance and/or institutional permission had been granted by the relevant universities, the deputy vice-chancellors or research directors, and the contact person they nominated, were emailed with a request for the first names, surnames, and email addresses of the 2022 postdocs. Although this personal, confidential information was collected to administer the questionnaire, the survey was strictly anonymous, and no personal, biographical information (e.g., gender, race, age, or nationality), IP (internet protocol) addresses or email addresses of respondents were collected through the questionnaire.

The recruitment email that was sent to potential respondents (see [Annexure A](#)) included a link to a respondent information leaflet (see [Annexure B](#)). Together, the content of the email and leaflet, adapted from REC: SBER's template for online surveys, provided respondents with all the relevant information they needed to grant their informed consent to participate in the survey. To allow for comparison with survey results on postdocs in other countries, the questionnaire touches on some sensitive subjects, such as income bracket and experiences of discrimination. Potential respondents were alerted to this fact in the information leaflet. We obtained consent from the potential respondents by means of an electronic consent process (i.e., they clicked on the link to the questionnaire to indicate their consent).

Five universities (the University of KwaZulu-Natal [UKZN]; North-West University [NWU]; Sefako Makgatho Health Sciences University [SMU]; Sol Plaatje University [SPU]; and University of Mpumalanga) elected not to provide the names and contact details of the postdocs they hosted in 2022, and therefore those postdocs were not recruited. However, some of the respondents (see "Other" in Table .2 below) indicated a host institution that was different from the one indicated by the universities that did provide names and contact details. Among these institutions are four of the universities that did not formally participate in the survey, as well as the South African Medical Research Council (SAMRC), South African Institute for Aquatic Biodiversity (SAIAB) and Onderstepoort Biological Products (OBP). We assume that the respondents who indicated these institutions as their hosts (one in each case) were incorrectly listed as 2022 postdocs of other institutions that did participate in the survey.

Table 2.2. Recruitment dates and numbers, and number of responses and response rates, by institution (20 March 2023)

| University | Date of recruitment | No. of recruitments | No. of responses | Response rate |
|---|---|---|---|---|
| Walter Sisulu University (WSU) | 02-Feb 2023 | 20 | 8 | **40%** |
| University of Cape Town (UCT) | 20-Jan 2023 | 172 | 62 | **36%** |
| University of Pretoria (UP) | Jan & Feb 2023 | 274 | 96 | **35%** |
| University of South Africa (UNISA) | 20-Jan 2023 | 101 | 26 | **26%** |
| Central University of Technology (CUT) | 18-Nov 2022 | 14 | 3 | **21%** |
| University of the Free State (UFS) | 31-Oct 2022 | 164 | 32 | **20%** |
| Rhodes University (RU) | 01-Dec 2022 | 129 | 26 | **20%** |
| Vaal University of Technology (VUT) | 14-Nov 2022 | 17 | 3 | **18%** |
| Mangosuthu University of Technology (MUT) | 20-Jan 2023 | 17 | 3 | **18%** |
| Cape Peninsula University of Technology (CPUT) | 20-Jan 2023 | 57 | 10 | **18%** |
| Tshwane University of Technology (TUT) | 23 & 24-Jan 2023 | 67 | 12 | **18%** |
| University of Johannesburg (UJ) | 03-Feb 2023 | 378 | 68 | **18%** |
| University of Zululand (UNIZULU) | 18-Nov 2022 | 29 | 5 | **17%** |
| Stellenbosch University (SU) | 24-Nov 2022 | 293 | 49 | **17%** |
| University of the Witwatersrand (WITS) | 28-Nov 2022 | 152 | 24 | **16%** |
| Nelson Mandela University (NMU) | 08-Mar 2023 | 130 | 21 | **16%** |
| University of Venda (UNIVEN) | 18 & 19-Nov 2022 | 22 | 3 | **14%** |
| Durban University of Technology (DUT) | 08-Dec 2022 | 52 | 6 | **12%** |
| University of the Western Cape (UWC) | 26-Jan 2023 | 128 | 14 | **11%** |
| University of Limpopo (UL) | 18-Nov 2022 | 10 | 1 | **10%** |
| University of Fort Hare (UFH) | 15-Feb 2023 | 34 | 1 | **3%** |
| **Sub-total** |  | **2260** | **473** | **21%** |
| Other[4] |  |  | 7 |  |
| Missing |  |  | 27 |  |
| **TOTAL** |  |  | **507** |  |

Four respondents' data were deleted because they (1) answered no questions; or (2) did not meet the selection criteria for the survey (e.g., in 2022 they were affiliated to a non-South African host institution, or to no host institution). The total number of cases that were analysed after data processing (including cleaning) is therefore 503. However, as mentioned earlier (and according to good ethics practice), respondents were allowed to "skip" questionnaire items that they did not want to answer. In the case of sensitive topics, such as income or discrimination, an explicit option, "prefer not to answer" was also included. In addition, some questionnaire items were only applicable to a sub-set of respondents. These three categories of "missing" responses were excluded from our analyses, and the number of valid responses to questionnaire items therefore varies but is indicated for each result (in the "Total" in tables, or as a footnote below figures).

---

[4] One respondent each reported the following seven institutions from whom respondents were not formally recruited: NWU, UKZN, SPU, SMU, SAMRC, SAIAB, and OBP.

Universities were given the option to administer their own survey, using our questionnaire. UP was the only institution that made use of this option. They used a different platform (Qualtrics, while we used SurveyMonkey), and in the process of transferring the Microsoft (MS) Word version of the questionnaire to that platform, there were some accidental omissions of questions or options, and slight adaptations were made to the questionnaire. Also, in 2023, we adapted the wording of the recruitment email and information leaflet to the fact that we were no longer asking about the respondents' "current" postdoc position, but the position they held the previous year. In February 2023, it was decided to also adapt the wording of our original questionnaire to reflect this fact.

Thus, there were in fact three versions of the questionnaire. The version provided in [Annexure C](Annexure C) is the primary one that was completed by three-quarters (76%) of the respondents; one-fifth (19%) completed the UP version; and 4% completed the most recent version, adapted for 2023. It should be noted that UP send out reminders to their 2022 postdocs to complete the questionnaire, while this approach was not followed for the other universities. Interestingly, though, the response rate for UP, although relatively high, is only the third highest among the universities. This may be the result of the university not offering an inducement to encourage participation to potential respondents.

We considered an inducement necessary to maximise the response rate (and therefore the external validity and possibilities for statistical analysis of the data) amongst a relatively small population whose time is relatively constrained. It is also aligned with the practice of surveys among postdocs conducted elsewhere (e.g., Nature & Shift Learning, 2020; McConnell et al. 2018). If respondents were interested in entering into a prize draw for a R2 500 Takealot voucher, they were redirected (upon completion of the questionnaire) to a separate site where they could enter their email address into a drawing for a chance to win. However, this information was not linked to their responses, and was treated as confidential. We also took care to ensure that the inducement would not unduly influence an informed choice about participation, nor undermine a potential participant's assessment of risk of harm. Respondents who wanted us to share with them a report detailing the results of the survey, were asked to email a request to the first author of this report, thereby again preventing any linking of their responses with their name and email address.

### 2.5.3  Data processing, analysis, and management

The data were processed and analysed using a combination of IBM® SPSS® Statistics and MS Excel. The questionnaire was comprised mainly of closed-ended, pre-coded items that did not require much data processing. The few open-ended items, as well as specified "other" options, were quantified and coded and, in the case of "other" options, added to the existing response categories. In some cases, the respondents' specified "other" options were directly quoted and integrated, as illustration, with the results of the analysis of the quantitative data. The final item in the questionnaire requested respondents to share any additional comments or remarks. A total of 241 respondents provided such comments, ranging from a few words to lengthy paragraphs. These were analysed thematically, language edited, and integrated – as verbatim quotes

– with the results of the analysis of the quantitative data. The processed data other than those collected through the closed-ended, pre-coded items are collectively referred to in the results as the "qualitative data".

We completed SU's Division for Information Governance's privacy impact assessment for this project, which indicated that our survey is medium risk, because we needed to collect "identified personal information". The names and email addresses of the survey respondents that were recruited have been deleted from storage.

We did not collect any paper data, and the electronic survey data are stored in a password-protected MS Teams folder, which allows sharing of data among team members while adequately protecting the data from unauthorised access, loss and/or corruption. The anonymous survey data will be stored for future use after completion of this survey, as they could serve as a baseline if we decide to repeat the survey in future to determine changes over time.

### 2.5.4 A profile of respondents: institutional affiliation and scientific domain in 2022

As Table 2.2 below shows that the majority of respondents in 2022 were hosted by UP (20%), followed by UJ (14%), UCT (13%) and SU (10%). The remaining 43% of the respondents were hosted primarily by UFS (7%), RU, UNISA and WITS (at 5% each).

**Table 2.2. Host institutions of respondents in 2022**

|  | N | % |
|---|---|---|
| UP | 94 | 20% |
| UJ | 68 | 14% |
| UCT | 62 | 13% |
| SU | 49 | 10% |
| UFS | 32 | 7% |
| RU | 26 | 5% |
| UNISA | 26 | 5% |
| WITS | 24 | 5% |
| NMU | 21 | 4% |
| UWC | 14 | 3% |
| TUT | 12 | 3% |
| CPUT | 11 | 2% |
| WSU | 8 | 2% |
| DUT | 6 | 1% |
| UNIZULU | 5 | 1% |
| MUT | 3 | 1% |
| UNIVEN | 3 | 1% |
| VUT | 3 | 1% |
| CUT | 2 | 0% |
| Other public university[5] | 6 | 1% |
| Other institution[6] | 3 | 1% |
| **Total** | **478** | **100%** |

---

[5] One respondent each reported NWU, UFH, UKZN, UL, SPU, or SMU.
[6] One respondent each reported SAMRC, SAIAB, or OBP.

If one classifies the public universities by type, we find that almost two-thirds of the respondents (64%) were hosted by one of twelve traditional "research" universities (especially UP and UCT – see Figure 2.1). Trailing far behind, at 28%, are six comprehensive universities (mostly UJ and UNISA); while the seven universities of technology (but in particular TUT and CPUT) hosted 8% of the respondents in 2022.

**Figure 2.1. Types of public university that hosted respondents in 2022**

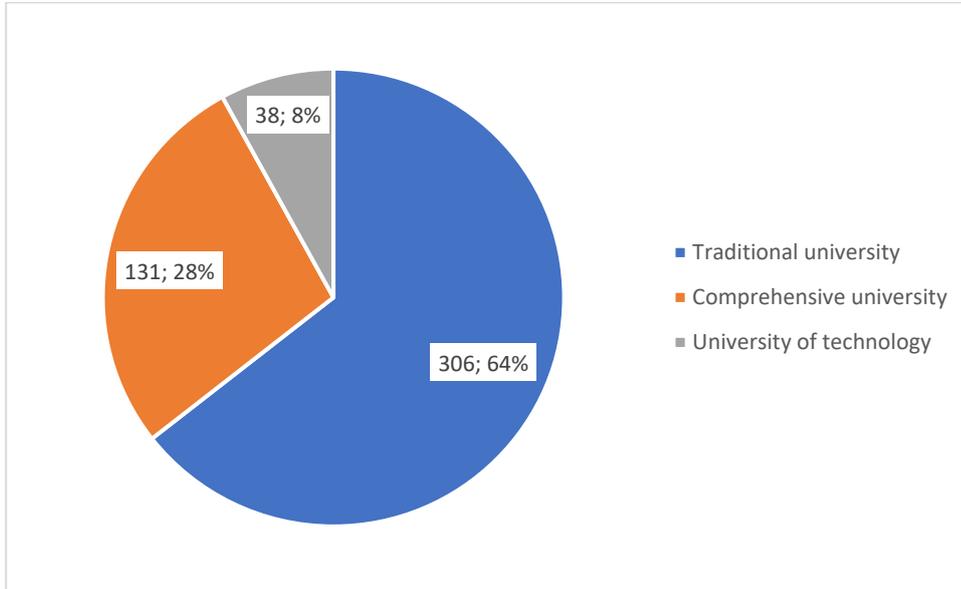

n=475

Respondents were asked in an open-ended question to describe the scientific field or discipline that best matches the work that they primarily conducted in their 2022 postdoc position. Given that postdocs have historically traditionally been appointed in the fields of science, technology, engineering and mathematics, it is not surprising that (as Figure shows) the larger percentages of the respondents are found in the natural sciences, health sciences and agriculture. Slightly more surprising is their relatively high level of representation in the social sciences as well as arts and humanities.

**Figure 2.2. Main science domains within which respondents worked in 2022**

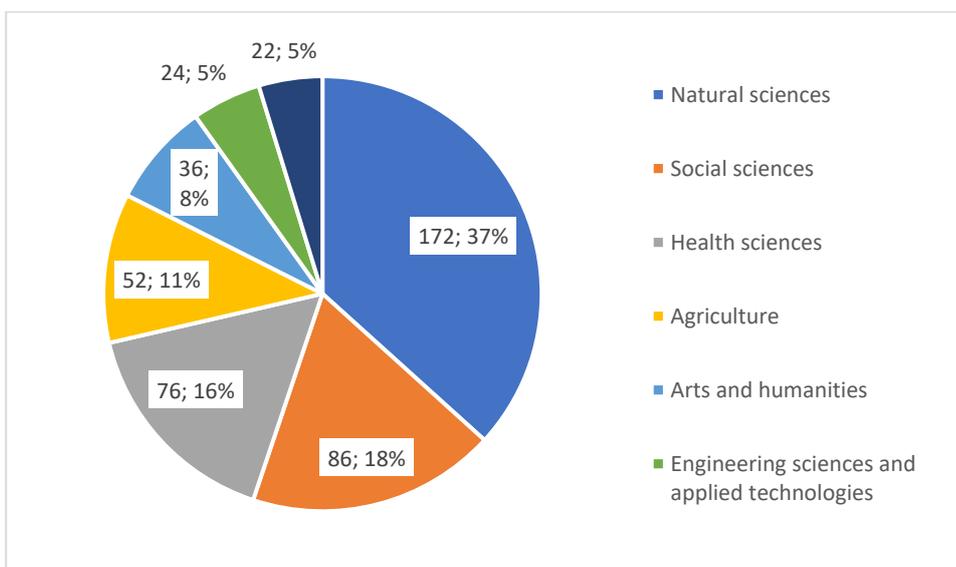

n=468

## 2.6 Bibliometric study

To analyse demographic information such as gender, age at time of publication, nationality, and population group, we linked the postdocs with their publications in the South African Knowledgebase (SAK) database on distinct fields including names, institution, and date of birth (where available). The SAK contains South African publications that are submitted to the DHET for subsidy and covers all submissions made between 2005 and 2022 at the individual author level. Unlike most bibliometric databases, SAK contains information about authors such as gender, age, region, and race that can facilitate answering research questions that are difficult or impossible to answer using other bibliometric databases currently available (e.g., Clarivate Web of Science, Scopus or Dimension). In the cases where we could not link the postdocs with certainty to SAK authorships, typically due to the absence of author and/or postdoc first names or date of birth, we used the South African Thesis Database (SATD) as a source to recover the missing postdoc data. This enabled us to complete the coupling with SAK. The CREST South African Thesis Database contains a collection of the MSc and PhD theses of the South African universities. This database currently covers the theses of all Masters' and doctoral graduates between 2005 and 2021.

## 2.7 Ethical issues concerning data management and analysis

We completed SU's Division for Information Governance's privacy impact assessment for this project, which indicated that our study is medium risk, because we needed to collect identified personal information and de-identified special personal information. We ensured, and continue to ensure, that we implement the appropriate organisational and technological security measures to protect this information.

We did not collect any paper data, and the electronic data are stored in a password-protected Microsoft Team, which allows sharing of data among team members while adequately protecting the data from unauthorised access, loss and/or corruption. The anonymised data will be stored for future use after completion of this study, as the data could serve as a baseline if we decide to repeat the study in future to determine changes over time.

# Chapter 3: Results from an analysis of the institutional data

## 3.1 Introduction

Over the past number of years, the South African higher education sector has invested increasingly in postdocs as a significant human resource capability. However, it is also the case that, until now, we had very little information on postdocs at a national level. We did not have such basic information as the numbers of fellows by scientific field, their research specialisations, their country of origin or source of funding. In addition, we had no demographic information (gender, race, nationality, age) of these fellows that would tell us more about how to design interventions and funding support in critical areas.

## 3.2 Summary of key findings

### 3.2.1 Total number of postdocs in the SA public university system

The 2013 White Paper for Post-School Education and Training (DHET, 2013:35) refers to postdocs as important for building the research capacity of the country's universities. The 2016 Human Capital Development (HCD) Strategy for Research, Innovation and Scholarship (Department of Science and Technology[7] [DST], 2016) suggests that an increase in PhD graduates should be complemented by an increase in the availability of postdoctoral fellowships. The DST committed itself to "a significant scaling up of South Africa's postdoctoral cadre" through the NRF.

Figure 3.1 shows the number of postdocs for all 26 public universities for the period 2016 to 2022. It shows that as per the strategic intention of government, there has been a steady increase in the number of postdocs from 1,788 in 2016 to 2,432 in 2022. The figure also shows a marginal decline in the number of postdocs in 2020 compared with 2019.

**Figure 3.1. Total number of postdocs in South African public universities, 2016-2022**

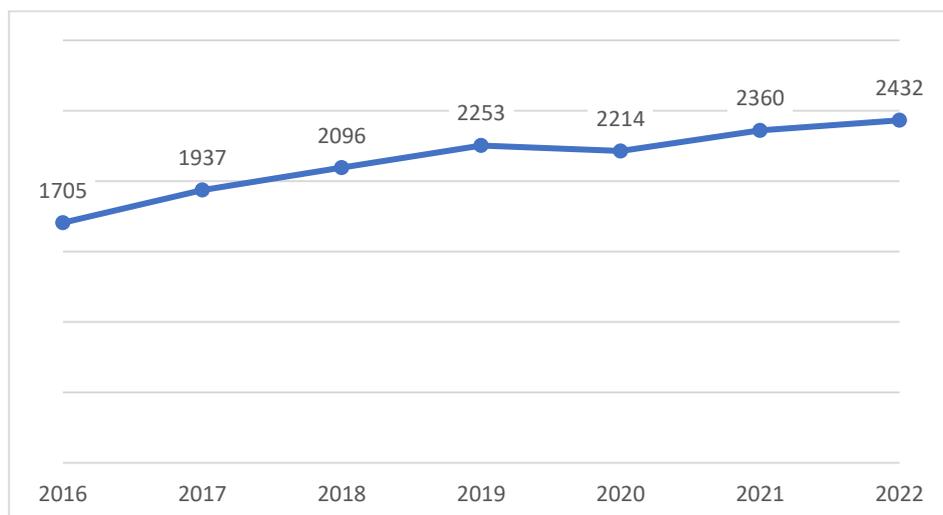

---
[7] The Department of Science and Technology (DST) was renamed the Department of Science and Innovation (DSI) on 26 June 2019.

Table 3.1 shows comparative data for the number of postdocs from the institutional data and the most recent data for the HSRC's CeSTII R&D survey. The data show consistently lower numbers of postdocs in the case of the institutional data collected by this study when compared with the R&D survey. This is to be expected given that the institutional data excluded science councils and other national research organisations, including those such as the MRC that host relatively high numbers of postdocs, while the R&D survey makes no distinction between different types of research organisation.

Table 3.1. Number of postdocs: Institutional data vs R&D Survey (2016–2020*)

| Year | Institutional data total | R&D survey total | Difference |
|---|---|---|---|
| 2016 | 1705 | 2 471 | +766 |
| 2017 | 1937 | 2 741 | +804 |
| 2018 | 2096 | 2 727 | +631 |
| 2019 | 2253 | 2 867 | +614 |
| 2020 | 2214 | 2 978 | +764 |

*Latest available data from the CeSTII R&D Survey at the time of writing

### 3.2.2 Number of postdocs by race and gender

The DST (2019b) expressed its intention to "support and retain an appropriate percentage of each annual graduating PhD cohort" (which has shown sustained increases) in the "Postdoctoral Fellowship Programme[8]" for "eventual absorption into the academic and research system". According to the DSI, to improve demographic representation among established researchers, a significant number of doctoral graduates will be "black and women doctoral graduates, particularly South Africans".

Table 3.2 shows that the proportion of Black African postdocs has increased from 17,73% in 2016 to 32,30% in 2022, resulting in Black African postdocs making up the second-largest racial group amongst South African postdocs after white postdocs (53,1% in 2022). The proportion of white postdocs has declined while in the case of the two least represented race groups, the proportion of Coloured postdocs has also declined (marginally) while the number of Indian postdocs has remained largely unchanged.

Table 3.2. Proportion of postdocs by race, 2016–2022

| Race | 2016 | 2017 | 2018 | 2019 | 2020 | 2021 | 2022 |
|---|---|---|---|---|---|---|---|
| Black African | 17,73% | 19,90% | 21,78% | 23,31% | 24,64% | 27,71% | 32,30% |
| Coloured | 10,20% | 9,39% | 8,63% | 7,63% | 7,12% | 7,14% | 8,85% |
| Indian | 5,52% | 6,21% | 6,08% | 5,61% | 5,70% | 5,43% | 5,34% |
| White | 66,56% | 64,49% | 63,51% | 63,45% | 62,54% | 59,71% | 53,51% |
| **Total** | **100,00%** | **100,00%** | **100,00%** | **100.00%** | **100,00%** | **100,00%** | **100,00%** |

Table 3.3 shows that in terms of gender, the proportion of female postdocs between 2016 and 2022 has remained lower than that of males, and within a range of 41-44% of postdocs being female.

Table 3.3. Proportion of postdocs by gender, 2016-2022

| Year | Female | Male |
|---|---|---|
| 2016 | 44% | 56% |

---

[8] As far as could be determined, no such formal programme exists.

| 2017 | 43% | 57% |
| 2018 | 43% | 57% |
| 2019 | 41% | 59% |
| 2020 | 42% | 58% |
| 2021 | 41% | 59% |
| 2022 | 41% | 59% |

### 3.2.3 Distribution of postdocs by age

Figure 3.2 shows the proportion of postdocs by age-bands of 10 years for the period 2016 to 2022. The data shows that the proportion of older postdocs is increasing, most notably in the age band 41-50 where the number of postdocs increased from 188 in 2016 to 394 in 2022.[9] The youngest reported postdoc was 22 (in 2016) while the oldest was 78 (in 2021).

**Figure 3.2. Age distribution of postdocs, 2016–2022**

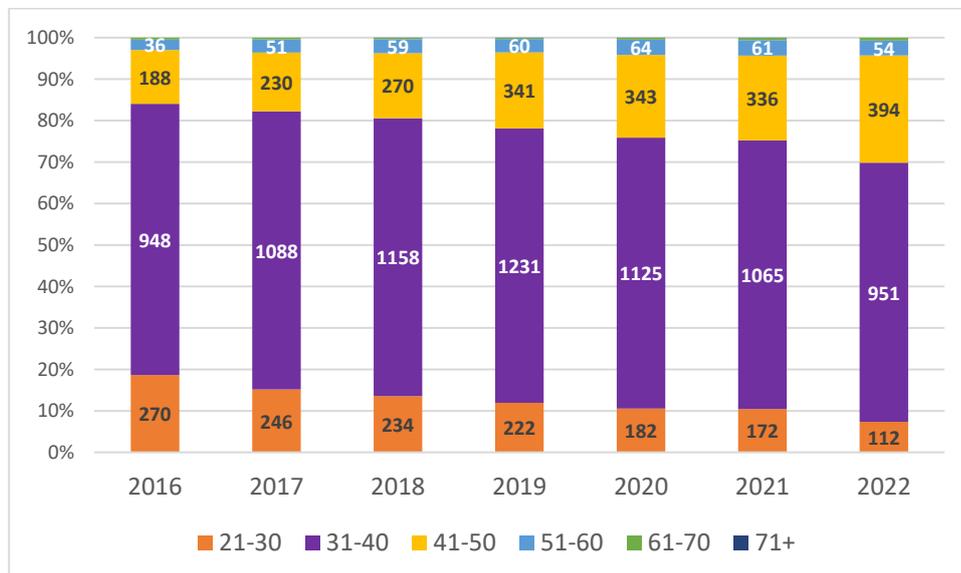

### 3.2.4 Number of postdocs by nationality

The DST committed itself to ensuring that support for "a significant scaling up of South Africa's postdoctoral cadre" receives high priority through the NRF. Higher education institutions (HEIs) and science councils were to be requested to consider institutional interventions that they could make to complement DST/NRF efforts to increase the domestic and international attractiveness of postdoc contracts in South Africa (DST, 2016). The DST itself suggested a focused programme to secure large numbers of postdoc candidates from abroad. It was committed to engage key bilateral partners with large science systems in pursuit of cooperation agreements that would see meaningful numbers of recent doctoral graduates moving to South Africa for two- to three-year terms, ideally with a sharing of costs. Sandwich programmes were to be strengthened to ensure the retention of the internationally trained doctoral graduates. Three years later, the DST (2019b) proposed easing immigration rules to attract postdoctoral researchers to South Africa. The Academy of Science of South Africa (ASSAf, 2018) also called for the relaxing of immigration requirements to encourage international PhD graduates to stay on in postdoc positions to increase the internationalisation of engineering education in South Africa.

---

[9] It should be noted that age data for a relatively large number of postdocs was not specified in the available data: 24% in 2016; 23% in 2017; 21% in 2018 and 2019; 23% in 2020; 32% in 2021; and 37% in 2022.

More than 60% of postdocs are foreign-born. However, the DSI/NRF have not been successful in increasing the proportion of foreign-born postdocs. The percentage of foreign-born postdocs has remained constant: 2016: 61%; 2017: 64%; 2018: 62%; 2019: 64%; 2020: 62%; 2021: 62%; 2022: 61%. Figure 3.3 shows the 110 countries of birth of postdocs hosted by South African universities for the period 2016 to 2022.

**Figure 3.3. Nationalities of postdocs hosted by South African universities, 2016–2022**

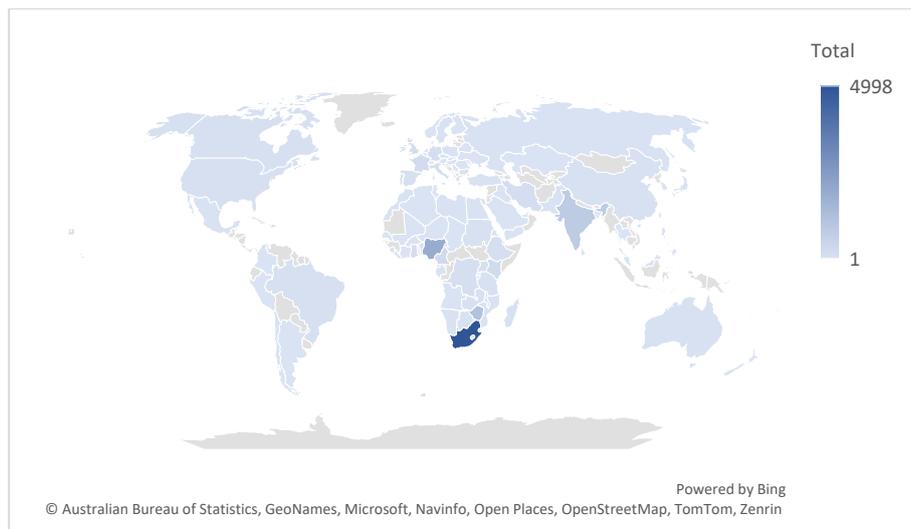

Table 3.4 shows that most postdocs were born in either South Africa, Nigeria or Zimbabwe, followed by a relatively steep decline in representation for other countries of birth. The data also show marked decreases between 2016 and 2022 in the number of postdocs born in North America, Europe and India. Growth is mainly driven by Nigeria and Zimbabwe, with a few other African countries making marginal contributions.

**Table 3.4. Top 30 'nationalities' of postdocs hosted by South African public universities, 2016–2022**

| Country of birth | 2016 | 2017 | 2018 | 2019 | 2020 | 2021 | 2022 | Total |
|---|---|---|---|---|---|---|---|---|
| South Africa | 604 | 640 | 724 | 731 | 754 | 770 | 775 | **4998** |
| Nigeria | 130 | 191 | 237 | 308 | 334 | 354 | 367 | **1921** |
| Zimbabwe | 96 | 136 | 135 | 162 | 176 | 216 | 257 | **1178** |
| India | 132 | 133 | 146 | 133 | 112 | 108 | 86 | **850** |
| Cameroon | 38 | 44 | 49 | 53 | 50 | 50 | 46 | **330** |
| Kenya | 24 | 39 | 45 | 53 | 50 | 52 | 33 | **296** |
| United Kingdom | 52 | 49 | 38 | 44 | 33 | 28 | 17 | **261** |
| France | 44 | 50 | 48 | 38 | 26 | 23 | 17 | **246** |
| Ghana | 15 | 17 | 23 | 30 | 35 | 35 | 31 | **186** |
| Iran | 16 | 20 | 28 | 25 | 31 | 32 | 19 | **171** |
| Germany | 33 | 32 | 31 | 21 | 17 | 14 | 15 | **163** |
| Italy | 32 | 33 | 28 | 21 | 13 | 12 | 9 | **148** |
| Ethiopia | 16 | 22 | 23 | 24 | 18 | 17 | 23 | **143** |
| DRC | 14 | 22 | 22 | 21 | 21 | 21 | 17 | **138** |
| Uganda | 14 | 16 | 17 | 24 | 22 | 21 | 19 | **133** |
| Spain | 21 | 24 | 21 | 20 | 12 | 11 | 4 | 113 |
| USA | 33 | 21 | 19 | 10 | 8 | 12 | 8 | 111 |
| Canada | 19 | 15 | 14 | 17 | 13 | 11 | 9 | 98 |
| Lesotho | 7 | 11 | 17 | 17 | 16 | 12 | 12 | 92 |
| Brazil | 6 | 11 | 12 | 20 | 11 | 16 | 13 | 89 |
| Netherlands | 11 | 11 | 15 | 17 | 15 | 9 | 9 | 87 |
| Malawi | 9 | 11 | 11 | 14 | 10 | 13 | 17 | 85 |
| China | 13 | 9 | 8 | 14 | 13 | 11 | 7 | 75 |
| Sudan | 10 | 7 | 6 | 9 | 8 | 16 | 18 | 74 |
| Tanzania | 7 | 8 | 11 | 10 | 8 | 11 | 11 | 66 |
| Madagascar | 6 | 11 | 12 | 13 | 9 | 9 | 5 | 65 |
| Zambia | | 5 | 8 | 9 | 11 | 12 | 16 | 61 |
| Australia | 13 | 10 | 11 | 11 | 5 | 5 | 3 | 58 |
| Botswana | 4 | 5 | 5 | 6 | 5 | 9 | 13 | 47 |
| Namibia | 4 | 5 | 7 | 8 | 7 | 5 | 8 | 44 |

## 3.3   Number of postdocs by university

The number of postdocs by university for the period 2016 to 2022 is shown in Table 3.5. As is to be expected, there are more postdocs at the larger, research-intensive universities such as Stellenbosch University, University of Cape Town, University of Pretoria and Wits. Notably, the two Cape-based universities have witnessed a decline in their number of postdocs between 2016 and 2022, while the Gauteng-based universities have maintained or increased their number of postdocs over the same period. Another Gauteng-based university, the University of Johannesburg, has more than doubled its number of postdocs from a base in 2016 virtually equivalent to that of the University of Pretoria.

**Table 3.5. Number of postdocs by university, 2016-2022**

| Institution | 2016 | 2017 | 2018 | 2019 | 2020 | 2021 | 2022 |
|---|---|---|---|---|---|---|---|
| Cape Peninsula University of Technology | 2 | 1 | 3 | 27 | 64 | 76 | 81 |
| Central University of Technology | 3 | 5 | 10 | 18 | 17 | 13 | 16 |
| Durban University of Technology | 5 | 8 | 11 | 15 | 17 | 29 | 31 |
| Mangosuthu University of Technology | 3 | 5 | 5 | 10 | 13 | 15 | 18 |
| Nelson Mandela University | 12 | 23 | 52 | 44 | 65 | 109 | 140 |
| Rhodes University | 83 | 61 | 87 | 98 | 104 | 101 | 112 |
| Stellenbosch University | 317 | 323 | 334 | 330 | 347 | 336 | 295 |
| Tshwane University of Technology | 2 | 7 | 15 | 58 | 59 | 51 | 49 |
| University of Cape Town | 359 | 373 | 362 | 373 | 348 | 334 | 297 |
| University of Fort Hare | 33 | 91 | 83 | 44 | 8 | 27 | 30 |
| University of Johannesburg | 204 | 291 | 298 | 392 | 414 | 444 | 457 |
| University of Limpopo | 7 | 6 | 4 | 4 | 5 | 11 | 20 |
| University of Pretoria | 209 | 226 | 261 | 264 | 300 | 299 | 303 |
| University of South Africa | 18 | 18 | 32 | 35 | 41 | 44 | 127 |
| University of the Free State | 106 | 131 | 142 | 171 | 111 | 182 | 146 |
| University of the Western Cape | 115 | 127 | 136 | 100 | 53 | 36 | 129 |
| University of the Witwatersrand | 220 | 237 | 251 | 252 | 222 | 241 | 150 |
| University of Venda | 7 | 12 | 13 | 18 | 17 | 15 | 11 |
| University of Zululand | 0 | 3 | 11 | 25 | 28 | 19 | 26 |

| Vaal University of Technology | 13 | 13 | 18 | 14 | 18 | 14 | 17 |
| Walter Sisulu University | 6 | 5 | 4 | 5 | 5 | 12 | 28 |

Data for University of Mpumalanga, University of KwaZulu-Natal, North-West University, Sefako Makgatho Health Science University and Sol Plaatjie University were not available.

Table 3.6 takes into account the number of academic staff employed by universities in 2021 to normalise the number of postdocs in relation to the size of a university. The normalised data show that for the traditional research universities, including Rhodes University, there are 4 or 5 per members of the academic staff for every postdoc hosted. At the University of Johannesburg there are 3 academic staff members for every one postdoc, while the ratio is also relatively low for universities such as Free State and Nelson Mandela when compared to their peers.

**Table 3.6. Number of postdocs to one permanent academic staff member, 2021**

| University | Permanent academic staff 2021 | Postdocs 2021 | No. of staff per postdoc |
| --- | --- | --- | --- |
| University of Johannesburg | 1363 | 444 | 3 |
| University of Cape Town | 1182 | 334 | 4 |
| Stellenbosch University | 1402 | 336 | 4 |
| Rhodes University | 371 | 101 | 4 |
| University of Pretoria | 1318 | 299 | 4 |
| University of Free State | 864 | 182 | 5 |
| University of Witwatersrand | 1218 | 241 | 5 |
| Nelson Mandela University | 727 | 109 | 7 |
| Cape Peninsula University of Technology | 780 | 76 | 10 |
| University of Fort Hare | 333 | 27 | 12 |
| Mangosuthu University of Technology | 228 | 15 | 15 |
| Tshwane University of Technology | 922 | 51 | 18 |
| University of Zululand | 346 | 19 | 18 |
| University of Western Cape | 691 | 36 | 19 |
| Central University of Technology | 316 | 13 | 24 |
| Durban Institute of Technology | 710 | 29 | 24 |
| Vaal University of Technology | 365 | 14 | 26 |
| University of Venda | 433 | 15 | 29 |
| University of South Africa | 1861 | 44 | 42 |
| University of Limpopo | 676 | 11 | 61 |
| Walter Sisulu University | 908 | 12 | 76 |

## 3.4  Postdoc funding

Of the data available on the sources of funding for postdoc fellowships in 2022 (n=1,657), 27% (n=441) were funded by the NRF. The other most common source of funding for postdoc is from institutional funds, although it is difficult to quantify the proportion accurately because of how the data were provided by the universities. An approximation based on certain assumptions (including that if a university is listed as the funder, then the funding is assumed to be from institutional sources) is that 46% (n=756) of postdocs were funded from institutional funds in 2022.

# Chapter 4: Results from an analysis of the bibliometric data

## 4.1 Introduction

The request to public universities for data on their postdocs (Component 1 of this project) provided us with sufficient data on postdocs to allow for bibliometric analysis. The institutional data were linked to two of SciSTIP's existing bibliometric databases: (1) the Clarivate Analytics™ Web of Science (WoS), which represents the total output in WoS-indexed journals, and (2) existing CREST's proprietary database, SA Knowledgebase, which includes the publication output of South African academics, as published in all the journals recognised for subsidy by the DHET [WoS, the Proquest IBSS list, the Norwegian list, Scopus (since 2016), and local South African journals on the DHET list].

## 4.2 Summary of key findings

Figure 4.1 shows that the number of articles published by postdocs in journals accredited by the DHET increased from 30 in 2005 to 5,455 in 2022. Figure 4.2 shows that postdocs at the traditional universities (UCT, UKZN, SU, UP and Wits) published more articles relative to other universities in the system, with the exception of UJ whose postdocs published more articles during the same period than postdocs hosted by Wits.

**Figure 4.1. Total number of articles published by postdocs, 2005–2022**

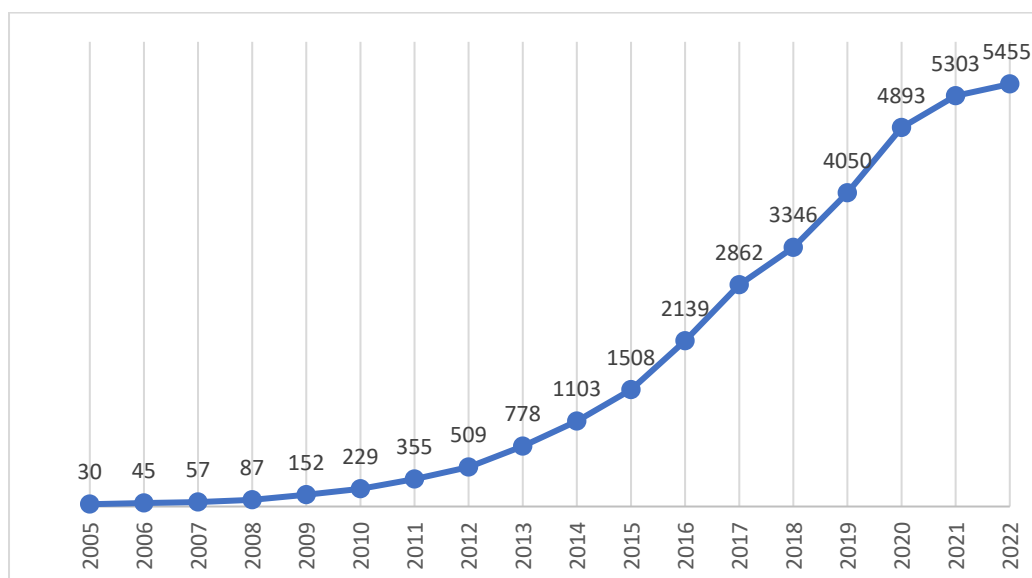

Figure 4.2. Total number of publications by postdocs per university, 2005–2022

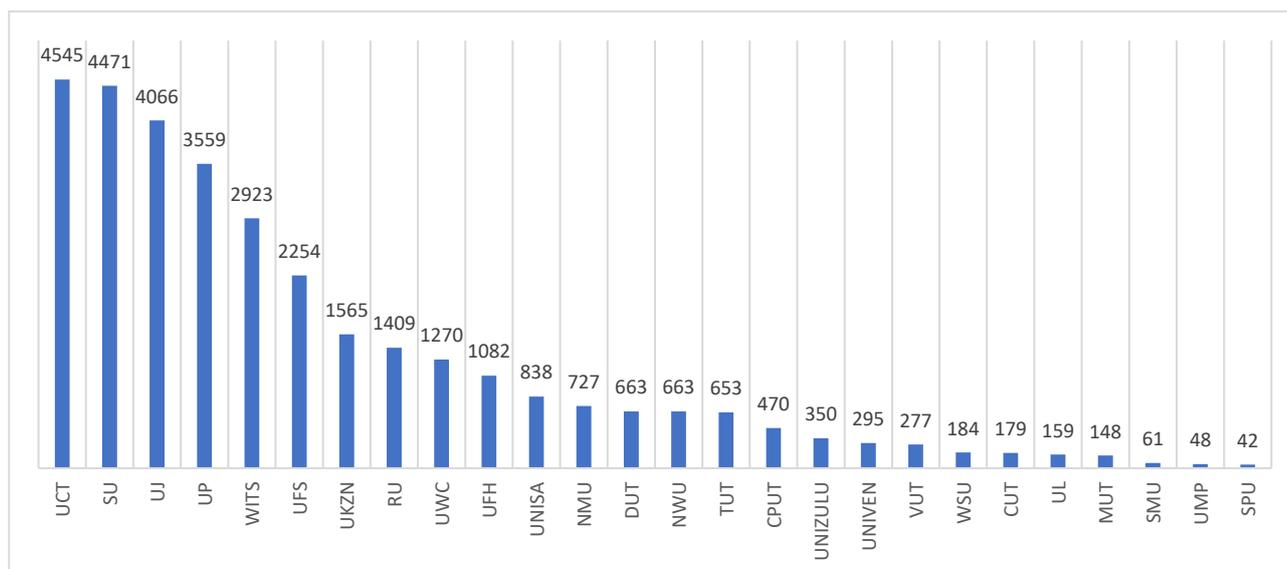

Table 4.1 shows the number of publications produced by postdocs for the period 2005 to 2022. With very few exceptions, the data shows the increase in the number of publications that postdocs contribute to the publications output of the public university system. Table 4.2 normalises the publication output by taking into account the number of postdocs at each university. While there are few unusually high outputs for the number of postdocs at one or two institutions in some years, in general the publication output is within the expected range given that many universities require their postdocs to produce two publications per year.

Table 4.1. Number of publications by postdocs per university, 2005–2022

|  | 2005 | 2006 | 2007 | 2008 | 2009 | 2010 | 2011 | 2012 | 2013 | 2014 | 2015 | 2016 | 2017 | 2018 | 2019 | 2020 | 2021 | 2022 | Total |
|---|---|---|---|---|---|---|---|---|---|---|---|---|---|---|---|---|---|---|---|
| CPUT | 0 | 0 | 0 | 0 | 3 | 4 | 6 | 12 | 15 | 12 | 11 | 13 | 17 | 14 | 34 | 77 | 122 | 130 | **470** |
| CUT | 0 | 0 | 0 | 0 | 0 | 0 | 0 | 0 | 0 | 1 | 4 | 9 | 11 | 18 | 19 | 35 | 43 | 39 | **179** |
| DUT | 0 | 0 | 0 | 0 | 0 | 0 | 3 | 1 | 3 | 7 | 12 | 14 | 55 | 52 | 82 | 102 | 162 | 170 | **663** |
| MUT | 0 | 0 | 0 | 0 | 0 | 0 | 0 | 0 | 0 | 1 | 2 | 10 | 6 | 6 | 22 | 29 | 72 | | **148** |
| NMU | 0 | 0 | 1 | 2 | 2 | 4 | 7 | 5 | 11 | 29 | 32 | 32 | 45 | 71 | 98 | 136 | 118 | 134 | **727** |
| NWU | 0 | 0 | 0 | 1 | 2 | 6 | 4 | 5 | 5 | 13 | 13 | 31 | 44 | 91 | 91 | 108 | 133 | 116 | **663** |
| RU | 2 | 4 | 1 | 1 | 8 | 7 | 19 | 20 | 43 | 40 | 65 | 103 | 162 | 175 | 186 | 215 | 209 | 149 | **1409** |
| SMU | 0 | 0 | 0 | 0 | 0 | 0 | 0 | 0 | 0 | 1 | 3 | 3 | 3 | 3 | 10 | 12 | 26 | | **61** |
| SPU | 0 | 0 | 0 | 0 | 0 | 0 | 0 | 0 | 0 | 0 | 0 | 0 | 0 | 0 | 3 | 11 | 17 | 11 | **42** |
| SU | 11 | 6 | 13 | 22 | 20 | 36 | 50 | 91 | 122 | 197 | 223 | 344 | 411 | 492 | 533 | 606 | 653 | 641 | **4471** |
| TUT | 1 | 0 | 0 | 2 | 0 | 1 | 8 | 13 | 18 | 18 | 33 | 44 | 44 | 64 | 95 | 112 | 91 | 109 | **653** |
| UCT | 1 | 5 | 14 | 7 | 35 | 43 | 65 | 89 | 147 | 193 | 269 | 383 | 423 | 512 | 517 | 619 | 610 | 613 | **4545** |
| UFH | 0 | 4 | 1 | 5 | 0 | 22 | 36 | 37 | 45 | 76 | 103 | 120 | 168 | 79 | 124 | 133 | 76 | 53 | **1082** |
| UFS | 1 | 3 | 4 | 6 | 8 | 10 | 12 | 19 | 25 | 36 | 54 | 102 | 142 | 224 | 269 | 402 | 441 | 496 | **2254** |
| UJ | 0 | 0 | 0 | 7 | 7 | 9 | 23 | 32 | 51 | 104 | 145 | 279 | 313 | 339 | 547 | 623 | 715 | 872 | **4066** |
| UKZN | 0 | 9 | 4 | 15 | 8 | 18 | 25 | 30 | 54 | 47 | 54 | 70 | 110 | 173 | 225 | 222 | 268 | 233 | **1565** |
| UL | 1 | 0 | 0 | 0 | 0 | 0 | 0 | 2 | 0 | 2 | 4 | 1 | 14 | 10 | 11 | 28 | 51 | 35 | **159** |
| UMP | 0 | 0 | 0 | 0 | 0 | 0 | 0 | 0 | 0 | 0 | 0 | 0 | 0 | 1 | 5 | 5 | 14 | 23 | **48** |
| UNISA | 0 | 0 | 2 | 3 | 2 | 3 | 1 | 2 | 7 | 15 | 23 | 33 | 54 | 74 | 109 | 136 | 151 | 223 | **838** |
| UNIVEN | 0 | 0 | 0 | 0 | 0 | 0 | 4 | 4 | 3 | 6 | 9 | 15 | 26 | 59 | 62 | 39 | 32 | 36 | **295** |
| UNIZULU | 1 | 0 | 0 | 0 | 1 | 0 | 0 | 0 | 1 | 1 | 15 | 12 | 27 | 43 | 57 | 84 | 61 | 47 | **350** |
| UP | 9 | 9 | 11 | 8 | 39 | 39 | 44 | 67 | 97 | 130 | 184 | 283 | 323 | 366 | 411 | 486 | 575 | 478 | **3559** |
| UWC | 0 | 4 | 3 | 7 | 10 | 12 | 22 | 24 | 52 | 81 | 91 | 133 | 88 | 122 | 150 | 198 | 145 | 128 | **1270** |
| VUT | 0 | 0 | 0 | 0 | 0 | 0 | 0 | 0 | 1 | 5 | 7 | 6 | 41 | 51 | 52 | 44 | 45 | 25 | **277** |
| WITS | 3 | 1 | 3 | 1 | 7 | 15 | 26 | 56 | 78 | 87 | 154 | 96 | 330 | 301 | 342 | 434 | 502 | 487 | **2923** |
| WSU | 0 | 0 | 0 | 0 | 0 | 0 | 0 | 0 | 0 | 3 | 1 | 11 | 1 | 6 | 19 | 6 | 28 | 109 | **184** |

Table 4.2. Number of publications per postdoc per university, 2020–2022

| | *2020* | | | *2021* | | | *2022* | | |
|---|---|---|---|---|---|---|---|---|---|
| | Pub.'s | Postdocs | Pub.'s/postdoc | Pub.'s | Postdocs | Pub.'s/postdoc | Pub.'s | Postdocs | Pub.'s/postdoc |
| CPUT | 77 | 64 | 1,20 | 122 | 76 | 1,61 | 130 | 81 | 1,60 |
| CUT | 35 | 17 | 2,06 | 43 | 13 | 3,31 | 39 | 16 | 2,44 |
| DUT | 102 | 17 | 6,00 | 162 | 29 | 5,59 | 170 | 31 | 5,48 |
| MUT | 22 | 13 | 1,69 | 29 | 15 | 1,93 | 72 | 18 | 4,00 |
| NMU | 136 | 65 | 2,09 | 118 | 109 | 1,08 | 134 | 140 | 0,96 |
| RU | 215 | 104 | 2,07 | 209 | 101 | 2,07 | 149 | 112 | 1,33 |
| SU | 606 | 347 | 1,75 | 653 | 336 | 1,94 | 641 | 295 | 2,17 |
| TUT | 112 | 59 | 1,90 | 91 | 51 | 1,78 | 109 | 49 | 2,22 |
| UCT | 619 | 348 | 1,78 | 610 | 334 | 1,83 | 613 | 297 | 2,06 |
| UFH | 133 | 8 | 16,6 | 76 | 27 | 2,81 | 53 | 30 | 1,77 |
| UFS | 402 | 111 | 3,62 | 441 | 182 | 2,42 | 496 | 146 | 3,40 |
| UJ | 623 | 414 | 1,50 | 715 | 444 | 1,61 | 872 | 457 | 1,91 |
| UL | 28 | 5 | 5,60 | 51 | 11 | 4,64 | 35 | 20 | 1,75 |
| UNISA | 136 | 41 | 3,32 | 151 | 44 | 3,43 | 223 | 127 | 1,76 |
| UNIVEN | 39 | 17 | 2,29 | 32 | 15 | 2,13 | 36 | 11 | 3,27 |
| UNIZULU | 84 | 28 | 3,00 | 61 | 19 | 3,21 | 47 | 26 | 1,81 |
| UP | 486 | 300 | 1,62 | 575 | 299 | 1,92 | 478 | 303 | 1,58 |
| UWC | 198 | 53 | 3,74 | 145 | 36 | 4,03 | 128 | 129 | 0,99 |
| VUT | 44 | 18 | 2,44 | 45 | 14 | 3,21 | 25 | 17 | 1,47 |
| WITS | 434 | 222 | 1,95 | 502 | 241 | 2,08 | 487 | 150 | 3,25 |
| WSU | 6 | 5 | 1,20 | 28 | 12 | 2,33 | 109 | 28 | 3,89 |

Figure 4.3 shows the broad fields of study in which articles published by postdocs for the period 2006 to 2022. Most articles are published in the natural sciences (34%), followed by the health sciences (19%) and the social sciences (19%), while the agricultural sciences, the humanities and arts, and multidisciplinary fields are the least common fields of study in which postdocs publish. Table 4.3 confirms the data in Figure 4.3 to the extent that the top 30 journals in which postdoc publish are either in the natural sciences or health science, or are more generalist but typically publish empirically-based research (e.g. *PLOS One* and *South African Journal of Science*).

**Figure 4.3. Percentage of articles by broad field of study, 2006–2022**

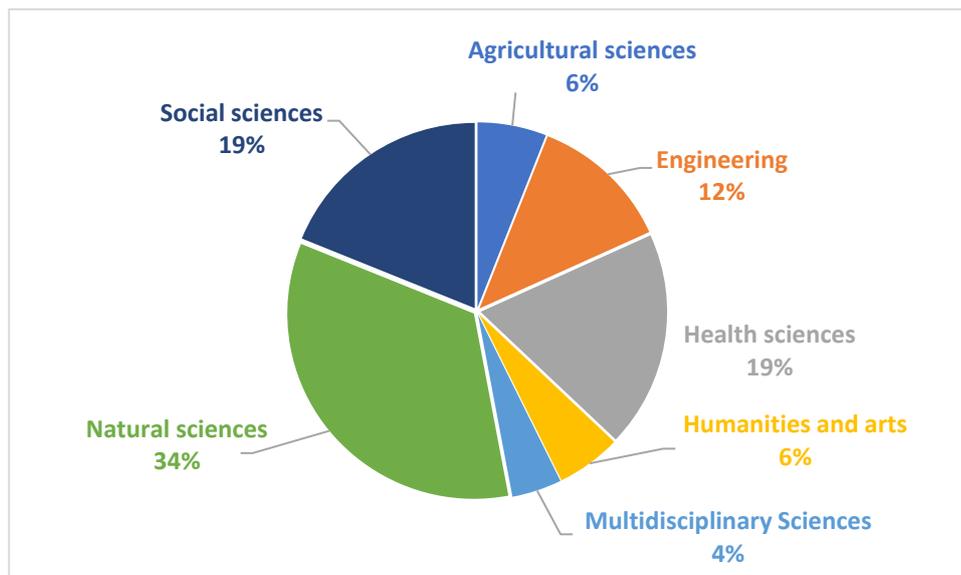

**Table 4.3. Top 30 journals in which postdocs published articles, 2005–2022**

| Journal title | No. of articles |
|---|---|
| PLoS ONE | 385 |
| Monthly Notices of The Royal Astronomical Society | 315 |
| Scientific Reports | 264 |
| Molecules | 220 |
| South African Journal Of Science | 148 |
| South African Journal of Botany / Suid Afrikaanse Tydskrif vir Plantkunde | 143 |
| International Journal of Environmental Research and Public Health | 138 |
| Sustainability | 130 |
| Journal of Molecular Liquids | 108 |
| Heliyon | 104 |
| Journal of Human Ecology | 102 |
| Astronomy and Astrophysics | 99 |
| Journal Of Alloys And Compounds | 93 |
| HTS Teologiese Studies-Theological Studies | 92 |
| Journal of Molecular Structure | 87 |
| Science of The Total Environment | 87 |
| International Journal of Molecular Sciences | 85 |
| African Journal of Marine Science | 82 |
| Journal of Ethnopharmacology | 81 |
| Gender and Behaviour | 77 |
| International Journal of Advanced Manufacturing Technology | 76 |
| Materials Today: Proceedings | 75 |
| BMJ Open | 74 |
| South African Medical Journal (SAMJ) | 74 |
| Frontiers in Microbiology | 73 |
| IEEE Access | 73 |
| South African Journal Of Botany | 73 |

| Acta Horticulturae | 72 |
|---|---|
| Frontiers in Immunology | 71 |
| Acta Crystallographica Section E: Structure Reports Online | 65 |

# Chapter 5: Results from an analysis of the survey data

This chapter reports on the results of a mainly descriptive analysis of the data generated by respondents completing a mostly structured survey questionnaire. The chapter is comprised of eight sections, the first of which describes the career trajectories of the survey respondents prior to holding their postdoc position in 2022. The second section describes the first of two sets of characteristics of the postdoc positions held by survey respondents in 2022, namely conditions of service (remuneration and benefits, work roles and requirements, and other contractual issues). The second set of characteristics of the postdoc positions held by survey respondents in 2022 (the third section) includes their reasons for taking that postdoc position, and satisfaction with both the position and their supervisor.

In the fourth section, the focus then broadens beyond the position held by the respondents in 2022, to a description of respondents' experiences of discrimination and harassment until the time of the survey. Similarly, the fifth section considers the factors respondents view as contributing, in general, to a successful postdoc experience, based on their experience until the time of the survey. The sixth section is more future-oriented, with its description of respondents' career expectations and plans, but also the challenges they experience in terms of career progression, and how well prepared they consider themselves for their future careers. The last of the results sections (the seventh section) considers migration-related issues. Some of it describes respondents' plans to migrate from South Africa, but most of the section details the issues postdocs have experienced during their migration to South Africa. The report concludes with the eights section that summarises the key findings and provides recommendations. Annexure D also provides results, namely of an analysis of the respondents' assessment of the survey and the questionnaire.

## 5.1 Postdoc career trajectories

In this section, we report on the respondents' career trajectories up until the time of the survey, with a focus on the phenomenon of "serial" postdocs. Approximately three-quarters (74%) of the respondents indicated that they moved directly after their doctoral graduation to a postdoc position. This explains why (as shown in Table below) three-quarters (75%) had not held another postdoc position right before they assumed their 2022 fellowship position. The table further shows that, of the 127 (or 25%) of respondents that had held another postdoc position (from here on referred to as "serial postdocs"), half (n=63) had done so in South Africa and at the same institution; a third (n=42) in South Africa but at a different institution. A small minority (n=22) had held another postdoc position in another country.

Table 5.1. Whether and where respondents held another postdoc position right before starting the position held in 2022

|  |  |  | N | % |
|---|---|---|---|---|
| No |  |  | 376 | 75% |
| Yes | In South Africa | At a different institution | 63 | 13% |
|  |  | At the same institution | 42 | 8% |
|  | In another country |  | 22 | 4% |
| **Total** |  |  | **503** | **100%** |

It should be noted that some of the respondents who had not held another postdoc position right before they started their 2022 position, may have held other positions, as one respondent explained in the qualitative data:

> I am not at the onset of an academic career. I have taught and supervised a large number of students at the same university for the past 12 years in the position of part-time temporary lecturer. The postdoctoral grant came to me in the sense that the head of my department suggested I apply for the fellowship. Through the five years that followed the granting of my PhD, I applied for the same postdoctoral fellowships a number of times. Each of these applications were unsuccessful, so it came as a big and welcome surprise when the department invited me to apply.

Another respondent who had not held another postdoc position before, mentioned[10] that, "because I am a 59-year-old white man, I have no expectation of securing an academic post. Nevertheless, I would appreciate an element of job security, having taught, and supervised in two academic departments for the past 12 years". A third respondent remarked (in relation to the insufficient level of postdocs' remuneration) that "sometimes, we are coming from formal or permanent employment. The current stipend implies that we all have completed our PhD straight after undergraduate or are under the age of 24 years or so. And that is hardly the case". It is clear from such comments that not all postdoc positions are early-researcher fellowships for recent doctoral graduates. To complicate matters further, at least one respondent referred to holding a "part-time" postdoctoral fellowship.

The 127 serial postdocs, identified as such in Table above, were asked three further questions concerning their postdoc career trajectory. First, they were requested to report the total number of postdoc positions that they had held since completing their doctoral degree (including their 2022 position, thus the minimum number for them would be two). During data processing, the 376 respondents who had not held another postdoc position before (see Table above), were coded on this variable as having held only one postdoc position since completing their doctoral degree. As already reported, they constitute by far the majority (75%) of the respondents, followed by 17% who had held two postdoc positions since completing their doctoral degree, and 7% who had held three, as Figure below shows. Four respondents had held four postdoc positions each since completing their doctoral degree, and one respondent reported having held five.

---

[10] As a specification of the academic work role in which they would most prefer to work, other than the ones listed in the questionnaire, when thinking of their future, long-term career plans.

**Figure 5.1. Total number of postdoc positions respondents had held since completing their doctoral degree**

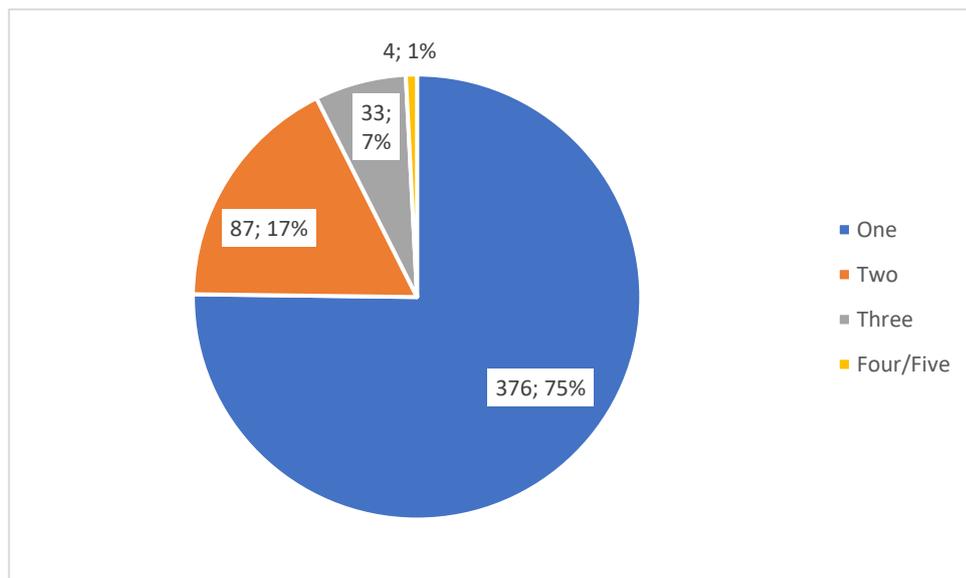

n=501

One serial postdoc clarified in the qualitative data that, "my current contract gets renewed once a year, so although I stated I have had three postdocs (contracts), it has in actual fact just been the same position". It is therefore important to also count the total number of years that respondents had spent in a postdoc position. The serial postdocs were asked to report this number of years they had spent (by the end of 2021) in all their postdoc positions taken together. Their answers ranged from 1 to 18 years, with an arithmetic average of 3 years and a median of 2. Table below presents these results in more detail:

**Table 5.2. Total number of years serial postdocs had spent in postdoc positions, by the end of 2021**

|          | N   | %    | Cum. % |
|----------|-----|------|--------|
| One      | 32  | 28%  |        |
| Two      | 27  | 23%  | 51%    |
| Three    | 21  | 18%  | 69%    |
| Four     | 12  | 10%  | 79%    |
| Five     | 10  | 9%   | 88%    |
| Six      | 6   | 5%   | 93%    |
| Seven    | 3   | 3%   | 96%    |
| Eight    | 1   | 1%   | 97%    |
| Nine     | 2   | 2%   | 98%    |
| Ten      | 1   | 1%   | 99%    |
| Eighteen | 1   | 1%   | 100%   |
| **Total**| **116** | **100%** |    |

A high standard-deviation value (2,434) reflects the skewness of the distribution in favour of the smaller number of years: 69% of the respondents reported having spent 1–3 years in postdocs positions by the end of 2021, while another 24% reported 4–6 years. The remaining 7% (or a mere 8 respondents) had spent between 7 and 18 years in postdoc positions. It was unexpected to find that the largest single percentage (28%) of the serial postdocs had spent, in total, only one year in postdoc positions by the end of 2021. From the qualitative data it was clear that some postdoc positions last only one year. Therefore, we may assume that, by the end of 2021, these 32 serial postdocs would have spent a year in their first postdoc position, while their 2022 position would have been their second one.

The relatively short length of time of some postdoc contracts also emerged as an issue for postdocs in general, and for various reasons, in the qualitative data. A minimum of a two-year contract was recommended by respondents, specifically for non-South African nationals. One reason is to reduce the number of times postdocs needed to apply for a renewal of their visitor's permit (an onerous process, as detailed in Section 5.7.3) to only once for two years. "If their renewal fails due to a failure to meet the targets, by all means, the university can inform the Department of Home Affairs to revoke the other year", the respondent added. Another respondent viewed a longer postdoc position as one possible way to address being "uprooted from strong social ties and community support from their origin country".

But a two-year postdoc position is also considered "inadequate" by one respondent, as "the time is too short for one to produce high-quality research outputs". Another respondent concurred that postdocs "are given few chances or funding opportunities to complete high-level research output before contracts conclude". A third postdoc felt that even a five-year limit "is highly problematic", because "it is not possible to build your CV enough in five years to get an academic appointment at a good institution".

The third and last question on their postdoc career trajectory requested the serial postdocs to indicate the extent to which they agree with the statement, "Poor job prospects have led to me holding more than one postdoc position since my PhD". Slightly more than half (52%) strongly agreed, and another quarter (26%) agreed, as Figure below shows.

**Figure 5.2. Serial postdocs' extent of agreement that poor job prospects had led to them holding more than one postdoc position since their doctoral degree**

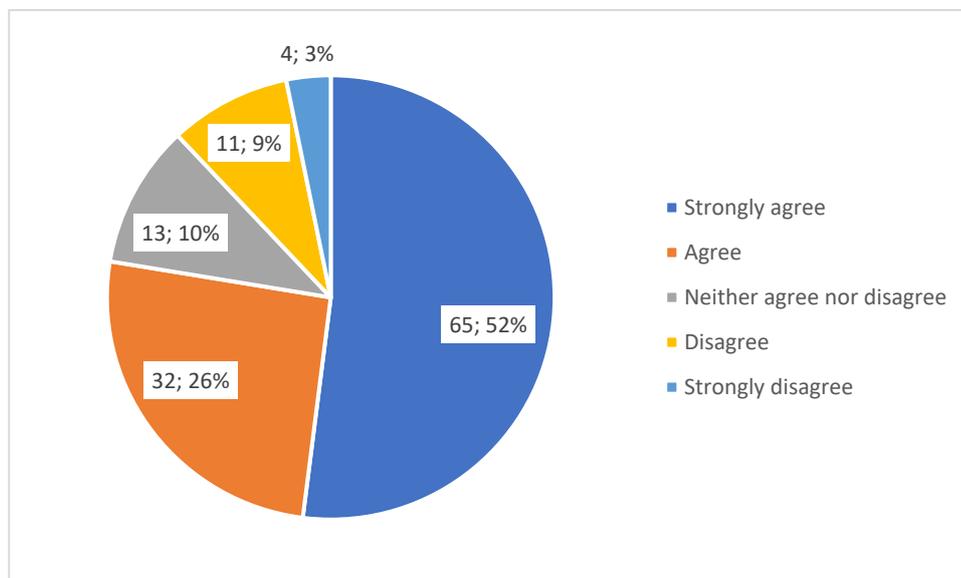

n=125 (only those respondents who had held another postdoc position right before starting the position they held in 2022)

Only 9% of the serial postdocs disagreed and a mere 3% strongly disagreed, while the remaining 10% neither agreed nor disagreed with the statement. In the qualitative data, one respondent reported feeling "forced to continue a postdoc due to the lack of job availability in my field". Another stated that they would "consider another postdoc after [their] current postdoc if it can secure a steady income and open potential positions, wherever in the world it may be". A third respondent argued that more people are driven to stay in postdoc positions because they are "forced to teach", which limits the time they can spend on research to "two hours per week".

## 5.2 Conditions of service in 2022

### 5.2.1 Remuneration and benefits

Eight questionnaire items measured various aspects of respondents' remuneration and benefits. Nearly 80 respondents also provided qualitative data that related to these issues.

#### 5.2.1.1 *Remuneration: amount, frequency and reliability of payment*

The first of the eight questionnaire items asked respondents to indicate into which one of eight categories their individual, gross annual income (at the time of the survey, and in South African Rand [11]) falls. Respondents could indicate that they preferred not to answer (13 chose this option and were excluded from the analysis). As Figure shows, the majority of respondents (60%) indicated that they received a gross annual disbursement of between R200 000 and R299 999. One quarter (24%) reported receiving between R300 000 and R399 999.

**Figure 5.3. Respondents' individual, gross, annual income in 2022**

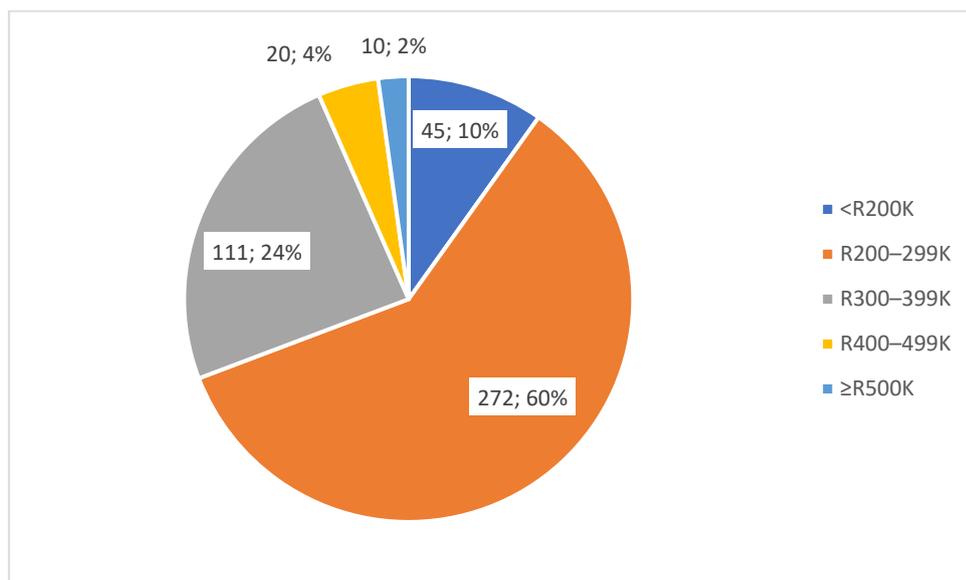

n=458

More than half of the remaining 17% of respondents (10% of the total, or 45) earned less than less than R200 000 per annum. Only 30 respondents (6%) earned R400 000 or more, and eight of these earned more than R600 000. The fact that postdocs' income (or "stipend") varies quite substantially was raised as an issue in several comments. One respondent wrote the following:

---

[11] In case respondents' income was not in ZAR, a link to a currency converter (http://www.xe.com/currencyconverter/) was provided.

> I still do not understand the remuneration difference across universities. For example, why does [one university] give us a stipend of R192 000 per annum, while other institutions pay their postdocs a good stipend between R250 000 and R350 000?

The same respondent felt that "this is abuse of postdoctoral fellows", while according to another, the "inconsistency in the stipend received by postdocs across schools and universities" is "unacceptable". Awareness of a lack of "standardisation of the postdoctoral salary" leads to feelings of resentment. "I watched with great envy my fellow postdocs under different supervisors as they collect double of what my supervisor was paying me", one respondent said. Another experienced their remuneration as "somewhat demotivating" because they "understand some institutions pay way better to their postdoctoral fellows. For example, my current fellowship pays me R16 667 per month, when I know another institution that paid R25 000 monthly about two years ago". Respondents further mentioned income disparities across fields. One respondent noted that funded fellowships "for postdocs who are clinically trained (i.e., specialists with [a doctoral degree]) are generally less well paid than PhD fellowships. It not a well-supported career path". According to another, postdocs in faculties outside of the "hard sciences" are "generally paid much less".

The survey data allow us to determine whether there is variability among the seven main scientific domains – presented in Figure above – in terms of the amount the postdocs they host, earn. For that analysis, we categorised respondents' individual, gross, annual income in 2022 – as presented in Figure above – into three categories, namely low (<R200K); average (R200–299K); and high (≥300K). As Figure below shows, the income levels of respondents in the seven main science domains do differ, although the differences are not statistically significant[12]. It is notable that no postdocs in the economic and management sciences earned a low income (although it needs to be taken into account that only 22, or 5%, of the respondents worked in this main science domain). The percentage of low-income earners is also quite low in the health sciences (6%), and in engineering and applied technologies (8%). On the other hand, the highest proportions of low-income earners are found in the social sciences (15%), agriculture (14%), and in the arts and humanities.

---

[12] Pearson chi-square=13.389; df=12; p=0.341.

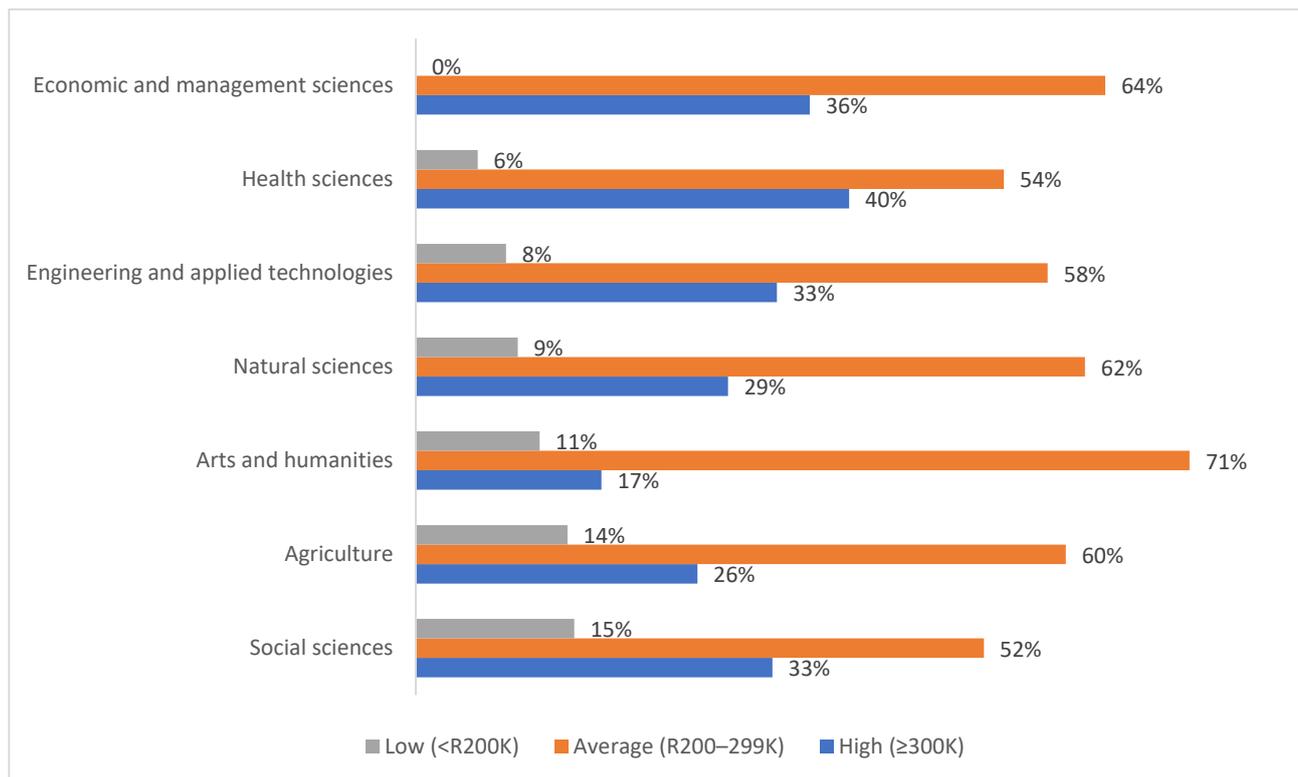

Figure 5.4. Respondents' level of income, by main science domain

n=453

When we focus on the high-income earners, postdocs in the health sciences are proportionally most likely (40%) to fall into that category, followed by those in the economic and management sciences (36%). Conversely, postdocs in the arts and humanities (at 17%) are by far the least likely (proportionately) to earn a high income, while the largest percentage of postdocs who earn an average income (71%) is also found in the arts and humanities. Therefore, there is some evidence for the one respondent's perception that postdocs outside of the "hard sciences" (especially those in the arts and humanities, but also in the social sciences) are "generally paid much less". Although this observation does not seem to apply to postdocs in the economic and management sciences, the low numbers of postdocs in that main science domain calls for caution when interpreting that result.

In the qualitative data, three respondents suggested standardisation of remuneration policies, or of the stipend amount, or at least of minimum remuneration across South African universities nationally. One suggestion was that this should equal "at least a lecturer's monthly salary". For another, "it would make more sense if perhaps postdocs were given approximately R300 000–R350 000 per annum". A respondent's observation, cited earlier, that postdocs under different supervisors earn different incomes, emerged again among the calls for standardisation: "I strongly believe" stated one respondent, "that a postdoc salary coming from a PI's [principal investigator's] personal fund should have a cut off and a decent amount should be paid to postdocs".

Calls for standardisation involved more than the value of the fellowship or stipend. A respondent felt that "more effort should be made to standardise some [financial] benefits that postdocs get across the various faculties or departments at a university", such as some "remuneration for lectures provided and financial benefits after publishing". The latter notion of additional remuneration for publishing was mentioned by three other respondents, all of whom suggested that universities should pay postdocs for "publishing articles accepted in accredited journals". One of these respondents observed a lack of standardisation between students and postdocs: "I hear that currently its only students who get paid per publication and this is not fair considering that postdocs are 'students' themselves". According to another respondent, the fact that postdocs "do not receive incentive funding for research outputs", denies postdocs "independence as researchers".

One of the respondents who called for standardisation of remuneration (but also employment) policies, based their recommendation on the belief that "postdocs are an integral component [of] national scientific scholarship", and that it would "allow postdocs to feel valued and protected from exploitation". However, contrary to the calls for standardisation, one respondent felt that postdocs' remuneration does not reflect what they observed to be "a lot of disparity between the workload of postdoc positions". Rather, remuneration remained "the same, despite these differences".

In the questionnaire, the respondents were asked to elaborate on the amount they earn, by indicating to what extent they agree with two statements. Close to 60% disagreed with the statement that the money they receive is adequate to cover their expenses and provide for reasonable leisure activities, considering the cost of living in their community.

**Figure 5.5. Respondents' extent of agreement that the money they receive is adequate to cover their expenses and provide for reasonable leisure activities, considering the cost of living in their community**

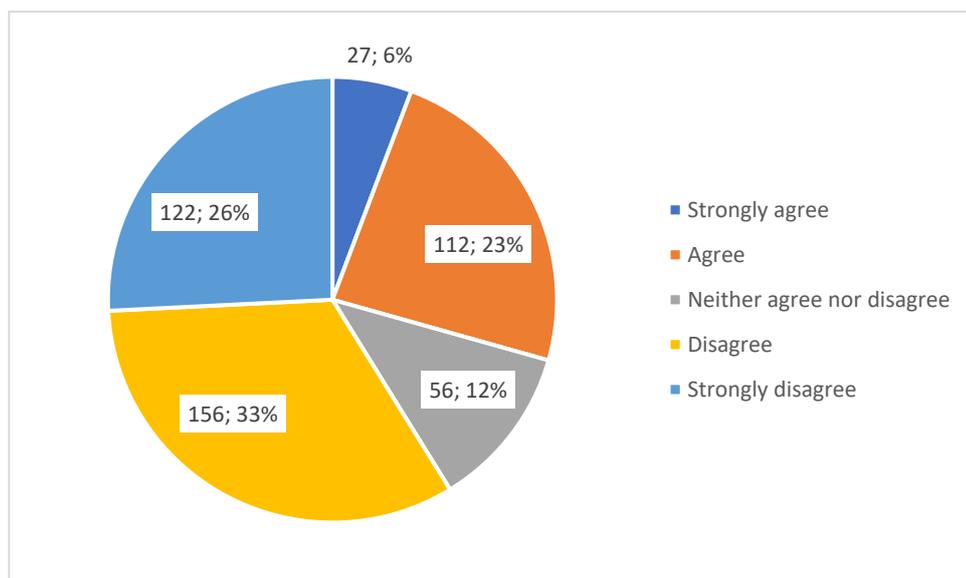

n=473

In order to determine whether there is a relationship between respondents' gross, annual income earned in 2022 and their perception of whether that income is adequate, as per the statement in the questionnaire, we compared that perception in terms of three categories of income eared, namely low (<R200K); average (R200–299K); and high (≥300K). The responses regarding the adequacy of income were also categorised, by collapsing the original categories "strongly agree" and "agree" into a single category ("agree") and applying the same logic to the opposite of the scale.

The results in Figure below show that the higher the income respondents earn, the more likely they are, proportionately, to agree with the statement that their income is "adequate to cover their expenses and provide for reasonable leisure activities, considering the cost of living in their community". Statistically, the relationship is significant[13]. Of those respondents in the low-income category, only 16% agreed with the statement. This percentage is somewhat higher (at 22%) among those in the average-income category, but it more than doubles (to 47%) among respondents in the higher-income category. It is notable, though, that even 43% of the respondents in that high-income category felt that the money they receive is inadequate.

**Figure 5.6. Respondents' extent of agreement that the money they receive is adequate to cover their expenses and provide for reasonable leisure activities, considering the cost of living in their community, by level of income**

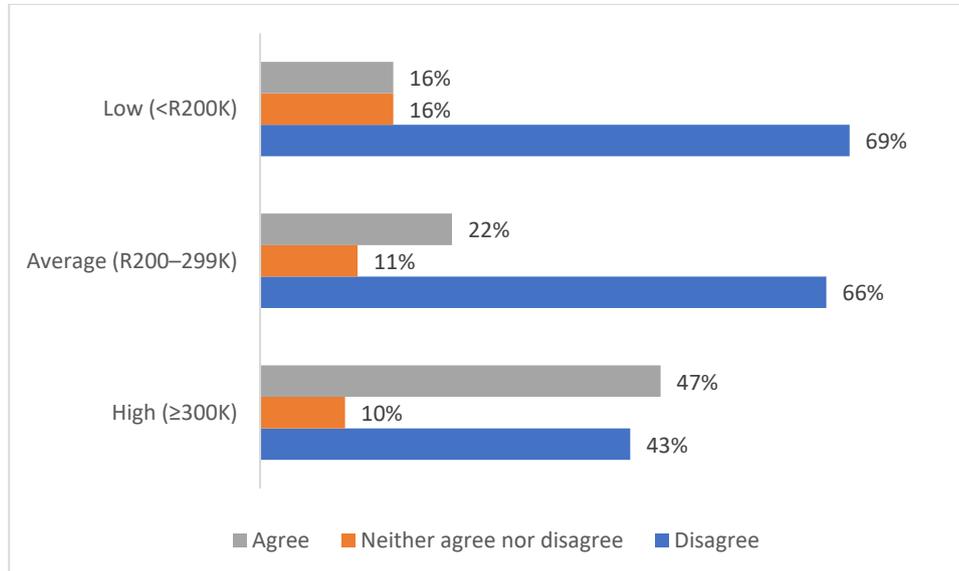

n=458

The second statement in the questionnaire that asked respondents to elaborate on the amount they earn, pertains to the ability to save the amount of money they want to, from their remuneration". Compared to responses to the first statement – on the adequacy of remuneration in general – an even greater percentage of respondents (78%) disagreed – 40% strongly so (see Figure below). As the figure further shows, only one in 10 agreed with the statement and only 2% (or 11 respondents) strongly agreed that the amount of money they earn allows them to save.

---

[13] Pearson chi-square=13.991; df=4; p<,001.

**Figure 5.7. Respondents' extent of agreement that they can save the amount of money they want to, from their remuneration**

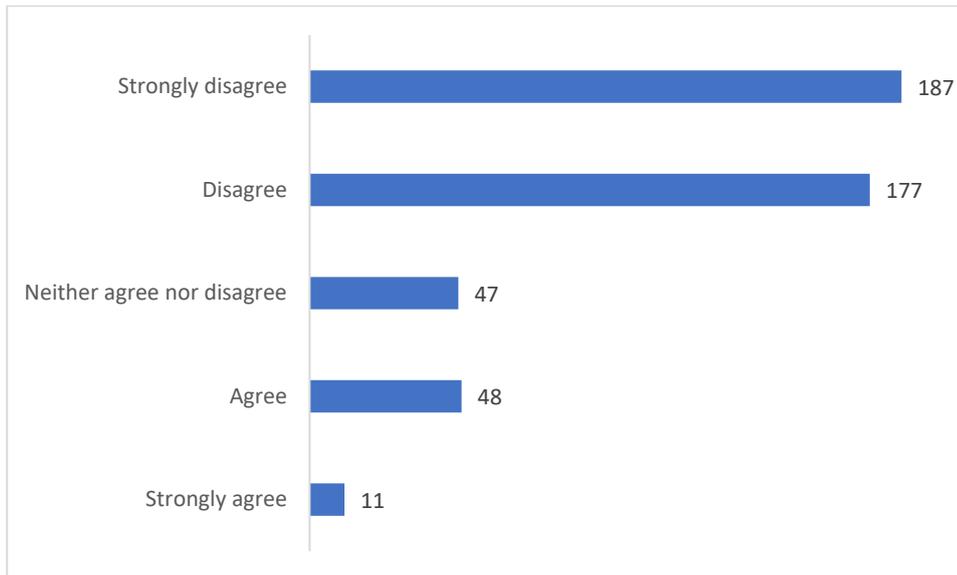

n=470

Following the same data-processing steps as for the first statement, we again determined whether there is a relationship between respondents' gross, annual income earned in 2022 and their perception, in this case, whether they can save the amount of money they want to, from their remuneration. Figure below shows in more detail the statistically significant differences[14] that resulted from our analysis. Similar to the results presented in Figure above, the lower the income level of the respondents, the less likely (proportionately) they were to agree that they can save the amount of money they want to, from that income. But the differences among, on the one hand, the lower- and average-income earners, and on the other hand, the higher-income earners, are even more stark. Of those respondents in the low-income category, only 7% agreed with the statement, and this percentage is only slightly higher (at 9%) among those in the average-income category. Again, it more than doubles (to 21%) among respondents in the higher-income category, but even 71% of those respondents felt that they could save the amount of money they wanted to, from their remuneration.

---

[14] Pearson chi-square=16.301; df=4; p=003.

**Figure 5.8. Respondents' extent of agreement that they can save the amount of money they want to, from their remuneration, by level of income**

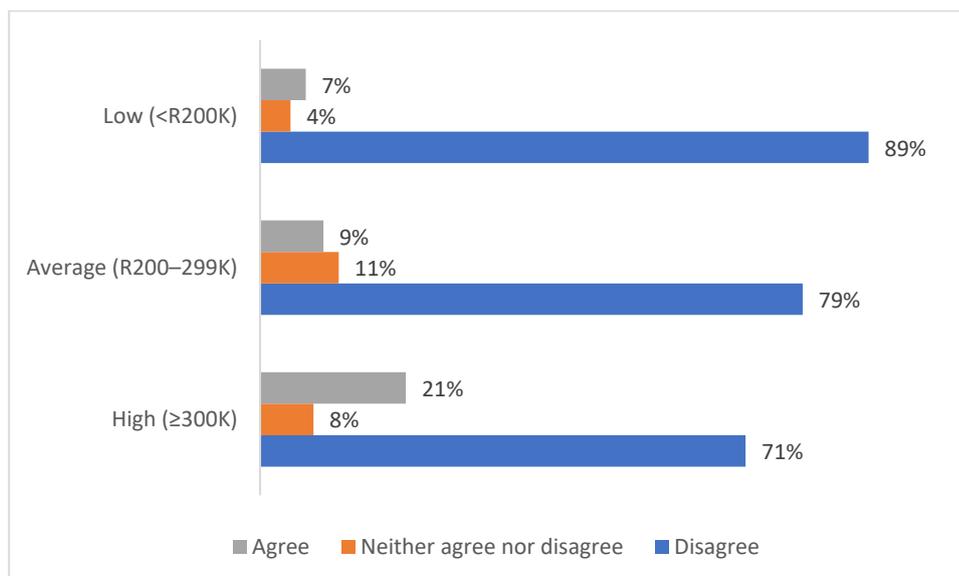

n=456

The fact that, even in the high-income category, the majority of respondents perceived their remuneration to be inadequate on these two measures, is supported by the open-ended responses. For example,

- "The pay is grossly inadequate".
- "Postdoc salary is too little!"
- "The stipend of R200 000 per annum was measly".
- "Payment is a joke".
- "You end up working as a postdoc […] earning peanuts".
- "Postdoctoral fellows are not beggars […] the remuneration given in South Africa is beggarly".

Several suggestions were made for a review of the amount postdocs receive in South Africa toward "better financial remuneration" and a "decent" income. Being "dramatically underpaid", as another described their postdoc in South Africa, raises the issue of what a postdoc income does not match. Some postdocs assessed their remuneration based on the cost of living, for example:

- postdocs "are paid a paltry stipend that is insufficient to sustain you for a month, let alone savings";
- postdocs should be "adequately compensated to attain basic living conditions";
- "the money and the cost of living are too far away from agreeing";
- "I feel frustrated when, despite my qualifications, I can barely afford to put food on the table";
- "remuneration for the position is completely inadequate for a modern, adult, life"; and
- "the pay and expectations need to be fixed to current times".

In addition to being non-aligned with "current times", postdoc remuneration was described as not increasing with inflation, which one respondent considered a "major stumbling block" in advancing their careers: "three years on the same salary bracket gets tough", they commented. A second recommended "just" to "increase the salary of postdoctoral students as per the annual inflation rate". A third respondent felt that "the remuneration system is very odd, such as having the right to, for example, negotiate my salary and to have annual increases to make up for inflation". A fourth respondent mentioned that "the remuneration package is currently small considering the inflation rate", which (combined with the lack of benefits" is a way in which postdocs "are often exploited by the institutions". That respondent added "let alone the fact that you are a holder of a PhD degree".

This last comment links to calls made by other respondents, for remuneration that aligns better with postdocs' high level of qualification, training, and skills:

- "either postdocs are the country's highest-trained researchers and compensated as such or dropped entirely from the equation";
- "the postdoc salary needs to be increased substantially", because "we are highly skilled individuals";
- "the institution itself needs to provide more support to postdocs, especially considering our qualifications; [this] could be in terms of salaries, benefits"; and
- "it takes a lot of time, money and sacrifice to get to this point: the remuneration […] must reflect that".

Other respondents' comments reflected a misalignment between remuneration and work expectations. According to one, most postdocs "are overused yet undercompensated"; another mentioned that they "do a lot more than just write publications". A third respondent compared themselves with staff – they "are expected to do similar work as staff but on minimal remuneration" – and another fourth compared themselves with master's students who "receive almost the same salary as postdocs".

In more general terms – and as with calls for standardisation of postdocs' remuneration – the sentiment of two respondents is that postdocs should be to "adequately compensated", and their funding should be "scaled up", as they are "the engines of research" who not only "increase research output of many universities in South Africa, but "produce quality research that generates millions of rands for [those] universities". One of these respondents added that "a better stipend can be a better motivation to work extra harder" and the other, that "it is prudent to benchmark with European countries on how to manage the welfare of postdocs".

A related issue is that South African postdoc positions' low rate of remuneration limits their attractiveness:

- "The national postdoc remuneration rate is […] very low lately. Institutions need to make postdoc positions more attractive".
- "The stipend might discourage many people from taking up postdoc positions".
- "Considering the low salaries, it is not attractive at all to do a postdoc in South Africa".

More specifically, a fourth respondent felt that "poor remuneration" made it "extremely difficult for South African citizens", particular ([b]lack) Africans to pursue postdoctoral opportunities, elaborating that the "postdoc system is not competitive enough and does not provide adequate incentives to attract ([b]lack) African aspiring talent to pursue this career path". A fifth applied the same argument to non-South African citizens who graduated from South African universities:

> postdoc remuneration should […] be looked into to encourage more PhD holders to be interested in coming for their postdoc programme in South Africa. As it is now, a lot of foreign PhD graduates from South African universities are not ready to pursue their postdoc here in South Africa because of this reason.

This point is underscored by two other respondents' observations, namely that "only immigration and remuneration are challenging and discouraging"; and that compensation of postdocs "should be improved for foreign candidates".

Inadequate remuneration, or "scrambling for money", in one respondent's words, has other wellness-related effects. In general terms, two respondents commented that, if the issue of "receiving more liveable or sustainable funding" were resolved, or "if the stipends [could] be augmented", "the life of postdoc fellows will be improved" and "the postdoc experience will be more beneficial and enjoyable". According to a third respondent, being "underpaid in many South African universities […] and denied […] employment benefits […] obviously leads to financial stress and reduces the ability to focus".

The need for a "better stipend" was also supported by a respondent's observation that most postdocs have families to support. Other respondents made similar comments:

- "remuneration" places "substantial pressure […] specifically [on] postdocs that have families and responsibilities";
- the "level of pay and lack of proper benefits [are] the biggest problems for postdocs who have families to support (we are no longer student-aged)";
- "the remuneration is very little and cannot support a family – most postdocs have family to look after, and that money does not allow for that"; and
- "I must save money for my family in India by cooking food".

The last respondent in the list above mentioned that such cooking "slowed" their "performance" (similar to the respondent cited above, according to whom financial stress reduces the ability to focus). One respondent "had to take on three other part-time jobs to support myself and my family, including my mother, which was exhausting". Another mentioned "not [being] significantly affected because I also give lectures. But it is very challenging for others who depend solely on the stipend". Three respondents mentioned that low remuneration, especially coupled with a lack of benefits, made them dependant on others, viz. "without the support of family, a postdoctoral position would be impossible"; "I could not survive on this pay if I didn't have financial support from elsewhere"; and "one important point is that my spouse has a permanent position which gives our family financial stability. Therefore, I can continue to be a postdoc". Contrary to most of the comments above, were a minority of positive statements, such as "I am satisfied with the money I earn as a postdoc"; "I am Brazilian. Compared to Brazil, my postdoc in South Africa has been great. Salary is 2,5 times better".

Another aspect of postdocs' remuneration that we addressed in the survey was whether they receive the pay they are due on time and without bureaucratic problems. Although the majority of respondents indicated that they do receive their pay on time, it is still worth pointing out that more than one in five indicated that this is not the case (Figure 11).

**Figure 5.9. Respondents' extent of agreement that they receive the pay they are due on time and without bureaucratic problems**

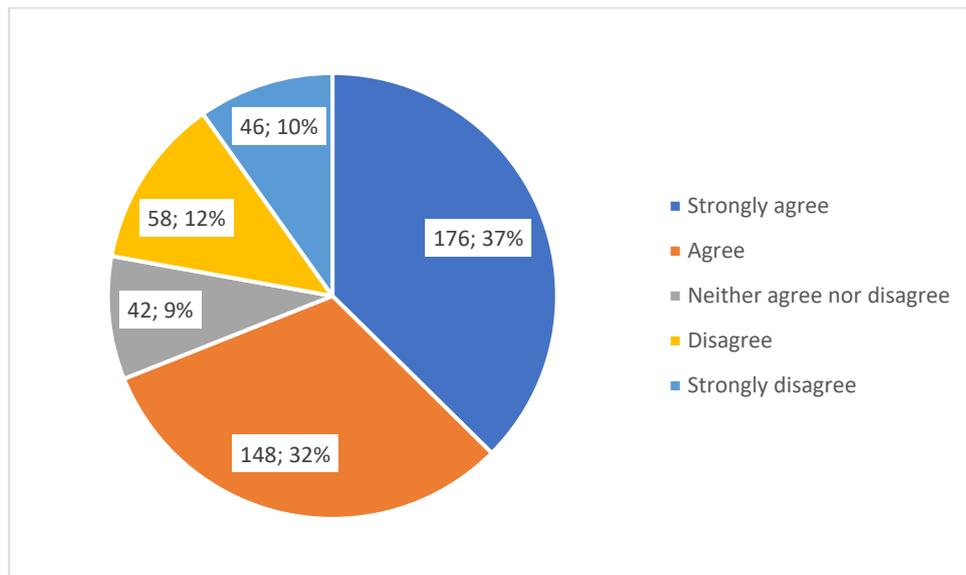

n=470

It is interesting to elaborate on the comments of those who have experienced problems with the payments of their fellowship. One had "worked as a postdoc for a couple of months in between funding without remuneration", because "adequate funding is a huge problem". Another linked non-payment to a delay in their visa approval:

> I […] was recommended for senior postdoc position that would improve my motivation through allowance, but I have not been able to register in 2023 nor receive a pay because my visa application in Nigeria since June 2022 has not been released […]. I was not registered nor paid for postdoc research from January 2020 to September 2021 when I was able to get my first-year visa. Fortunately, I [obtained] a PhD from [another South African university] in 2017, where I still do pro-bono research and publication for [an international] research entity (producing seven DHET Journal articles in 2022 separately for them), including postgraduate supervision […] I will even prefer that institutions appoint good postdocs as extraordinary professors with all the rights and privileges of research outputs and motivations [to prevent] the situation where postdocs would keep publishing and making money for the institutions without visa and emoluments.

For a third respondent, "the postdoc office was of the opinion that I am not allowed to be appointed part-time to teach and be appointed as a [postdoc]. Instead of communicating this to us at the […] faculty, the postdoc office failed to pay me for two months". A fourth respondent felt that "the uncertainty of when you will receive remuneration" places "substantial pressure on postdocs, specifically postdocs that have families and responsibilities". A fifth noted that "we are sometimes paid late, and this gives us a sense that we are not viewed as valuable human beings".

Related to the issue of receiving pay on time, is the frequency with which postdocs receive their remuneration. As Figure below shows, almost two-thirds (or 64%) received their remuneration monthly. Approximately a fifth (22%) did so quarterly, while around one in ten of the respondents were remunerated bi-annually.

**Figure 5.10. Frequency with which respondents receive their remuneration**

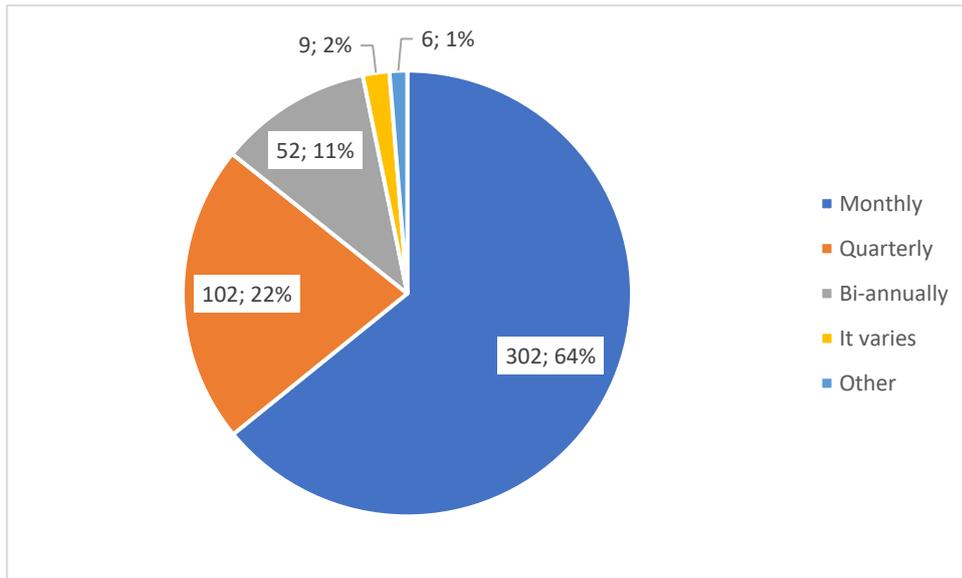

n=471

Six respondents reported unique alternatives to the listed options – i.e., "every two weeks", "annually", or "a small monthly stipend with 3 top-up payments spread over 9 months" – or provided comments ("it is supposed to be monthly, but sometime late"; "unresolved issues"; and the concerning, "I receive no payment").

We then asked the 170 respondents that had not received their remuneration monthly, whether this posed any challenges to them personally. As Figure below shows, the respondents are divided almost equally among the three response options: approximately a third of them felt that it posed "a lot" of challenges, another third "somewhat", and the remaining third, "not at all".

**Figure 5.11. Whether non-monthly payment of remuneration poses any challenges to respondents personally**

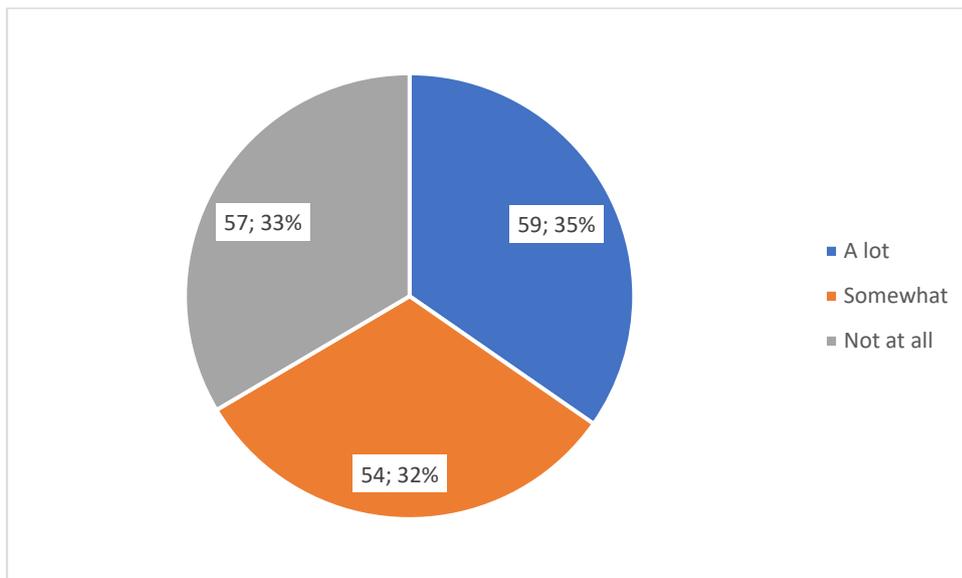

n=170 (only those respondents who had not received their remuneration monthly)

However, this still means that two-thirds of the respondents felt that non-monthly payment of their remuneration posed at least some challenges to them personally. Qualitative data on this issue included the recommendation that postdocs' stipend "should be paid on a monthly basis". The also underscored the importance of a "steady income". One respondent wistfully stated, "I will consider another postdoc after my current postdoc if it can secure a steady income […] wherever in the world it may be".

### 5.2.1.2 Employment benefits, tax-exempt status, and being treated as a student

The second-last question related to remuneration in the questionnaire listed two drawbacks of postdocs' tax-exempt status (in accordance with the relevant income-tax legislation [Republic of South Africa, 1962]), namely that they are treated as students rather than as employees (due to the bursary status of their remuneration), and that their employment benefits are limited to the minimum or are non-existent. Respondents were then asked, if given the choice, whether they would keep their tax-exempt status, or change their status to that of a tax-paying employee (noting that the latter would mean less net pay, but access to employment benefits) [15].

As Table below shows, the respondents were relatively equally divided on whether they would prefer keeping or changing the tax-exempt status of their postdoc.

Table 5.3. Respondents' preference regarding their tax-paying status

|  | N | % |
|---|---|---|
| Keep your tax-exempt status | 195 | 43% |
| Change your status to that of a tax-paying employee meaning less net pay, but access to employment benefits | 175 | 39% |
| Unsure | 65 | 14% |
| Change your status to that of a tax-paying employee with benefits, but only if pay increases | 13 | 3% |
| Change your status to that of a tax-paying employee, but with other conditions / for other reasons | 3 | 1% |
| Keep your tax-exempt status, but with other conditions / for other reasons | 3 | 1% |
| **Total** | **459** | **100%** |

One drawback of postdocs' tax-exempt status is that their employment benefits are limited to the minimum or may even be non-existent. We therefore asked respondents whether any of 12 listed benefits were available to them at their institution. Table below provides the number and percentage of respondents that indicated, for each benefit, that it was available to them at their host institution in 2022.

Table 5.4. Number and percentage of respondents for whom a benefit was available at their host institution in 2022 (n=503)

|  | N | % |
|---|---|---|
| Paid vacation ("annual") leave | 88 | 17% |
| Paid sick leave | 83 | 17% |
| Exemption from tuition fees | 71 | 14% |
| Medical scheme | 64 | 13% |
| Paid maternity / parental leave | 42 | 8% |
| Paid family or "compassionate" leave | 42 | 8% |
| Relocation costs | 25 | 5% |
| Retirement fund / Pension plan | 23 | 5% |

---

[15] Five respondents were excluded from further data analysis, as they indicated (notably) that they have the tax-paying status of an employee.

| | | |
|---|---|---|
| Transport allowance | 21 | 4% |
| Disability benefits / Workplace insurance | 12 | 2% |
| Group life insurance scheme | 7 | 1% |
| Subsidised childcare | 3 | 1% |

Even paid vacation (or "annual") leave and paid sick leave – the two benefits available to the largest percentage of respondents – was only available to 17%. Nineteen respondents indicated that benefits other than the 12 listed were available to them and half of those respondents specified those benefits. They mostly (n=5) involved the covering of costs of seminar or conference attendance (including registration and travel), but an additional R10 000 for research costs per year of contract; a data allowance; teaching assistance; or provision of an office were also mentioned. Other respondents used the opportunity to convey that they were unsure, as benefits had never been discussed (n=5), or to comment on benefit-related issues. For example, mention was made that any sick leave or leave of any kind is "subject to negotiation" with their supervisor, and that these "benefits" are therefore "inadequately institutionalised"; that one of the listed benefits – relocation costs – is only available to international postdocs, not South African postdocs; and that a R20 000 international-conference travel allowance was too low. It emerged from the qualitative data that in at least one instance an institution refused to pay a postdoc a travel allowances to attend a conference (for which their abstracted had been accepted), because they were not permanent members of staff.

Two respondents associated their lack of (at least some essential) employment benefits with them being "treated" or "considered" as "students". In another case, being treated as a student varied across employment benefits, supervisors, and institutions, again raising the lack of standardisation:

> I find it increasingly strange, even bizarre, that I am registered as a student with no health or other benefits (except a tax-exempted "reimbursement" of R17 000 per month), but then I have to apply for leave (of which I only have 15 days per annum). From what I understand, this is not everyone's experience – it is highly dependent on your institution and host. But I would be very interested in knowing what other postdocs who are in a similar position have to say about specifically applying for leave while being registered as a student. I pride myself in managing my time well and always submit my deliverables on time. Having access to only a capped number of leave days, and not being able to have more liberty in my daily/ weekly/ monthly schedule, has influenced my mental health substantially.

Being treated like a student – or being in a "pseudo-student situation", in the words of one respondent – was repeatedly raised as an employment-related issue in general. Recommendations were that postdoc positions "should be recognised as proper employment"; postdocs should be considered "as important parts of the university structure", as staff and students are, and "acknowledged for the important part that we play in the accomplishments made by the university, students, and supervisors".

One respondent recognised that they are "neither staff nor students", while another one observed that "at its core, people working in academia are students of life. If you have more answers then questions, you are in the wrong field". But nevertheless, these and other respondents recommended that postdocs should be treated more like an employee than a student. Included in their reasoning is the basic argument that postdocs "are not pursuing a degree" and have completed a doctoral degree. In addition, one respondent noted that they have "published several papers", are an "experienced researcher" and are "highly skilled", yet they "enjoy no employee benefits". As experienced individuals, another respondent argued, postdocs "need a larger playground and freedom, which allow to grow and to become independent".

A few respondents referred to the negative effects of being treated or referred to as students, and the lack of recognition this brings. For example, it "this alone makes you feel less important in the working environment"; and it "reduces [one's] self-esteem, and in most cases leads to depression and lack of hope for the future". One respondent mentioned that postdoc positions in South Africa should be "brought up to international standards". Being treated like students, for another respondent, meant that "postdocs are not valued in South Africa". Being treated as a student is an issue that arose again in respondents' assessment of their host institution (see Section 5.3.1.3 below).

Simply being a postdoc poses its own status-related challenges in the academic working environment. One respondent notes that "some academic staff just take us to be any other student", while another explained in more detail that,

> I feel that because I am in a postdoctoral position, my colleagues see me as less than or not up to scratch. Even if I have acquired more skills than some of the full-time lecturers. The fact that I am a postdoc, means that I am reaching for an impossible goal. Colleagues are patronising […] in many scenarios. This is ironic as many of the full-time staff that I work with obtained their positions after masters [and/or] finished a PhD while working in an academic role and never completed a postdoc contract.

For reasons such as these, a respondent recommended that postdocs "should be made researchers, instead of mere fellows".

Returning to the issue of employment benefits, a lack of suitable medical insurance was a particular point of concern that emerged from the qualitative data related to a lack of employment benefits. One postdoc had access to only a medical scheme, which was "below-par", and they therefore personally declined it, "due to the restrictions and lack of options". For a parent, it doesn't make sense […] to not have medical aid", and a non-South African had to "constantly think about medical-aid fees". Another respondent recommended medical benefits, but also unemployment insurance, "as in countries including Canada, the United States, etc.".

Other employment benefits that were recommended for postdocs include:

- "provision of family support, especially accommodation";
- "free food and accommodation […] at least based on merits";
- "travel allowances to attend conferences"; and
- funding to cover article-processing charges (APCs), because "insisting on and prescribing particular journals for the postdocs' publication without offering them [APCs] is unfair, given how ridiculously expensive most journals are due to currency differences".

Lack of employment benefits was also linked to postdocs' non-permanent position and led some respondents to compare themselves with other individuals in their age group. One respondent remarked that "it is frightening to think that at my age, I don't have a permanent position with benefits such as a pension fund or medical aid". These are standard things for adults". Similarly, another reported that,

> I am 32 this year: by this time, most women in South Africa already have at least one child. Unfortunately, I have no medical aid should I become pregnant, no maternity leave privileges and no housing benefits. I compare myself to other women in the same field, of the same age, and generally I feel regret that I chose to pursue a career in academia, which once was my dream job.

A third respondent compared themselves to participants in the New Generation of Academics Programme (nGAP)[16]:

> Why are people with master's degrees able to obtain permanent positions through the nGAP, while postdocs are only offered one-year contracts? Because of this, the majority of my master's students are now my seniors with important benefits, such as medical coverage. This is just wrong. A postdoc position is a scam.

---

[16] See http://www.ssauf.dhet.gov.za/ngap.html

For two respondents, a lack of employment benefits was interpreted as exploitation by their host institutions, as "it is a lot cheaper for universities in South Africa to keep postdocs without any of the standard benefits that they offer staff". The single exception to the comments on employment benefits is one respondent's view that "the postdoctoral fellowship is a steppingstone to move toward a permanent job. Therefore, it shouldn't provide too many benefits as a permanent position, otherwise fellows will be postdocs for far too long".

#### 5.2.1.3   Lack of access to financial products and services

Postdocs' tax-exempt status, non-monthly payments, and short-term contracts have another drawback that the qualitative data highlighted: lack of access to financial products and services, such as opening a bank account, applying for credit or a loan (ranging from a credit card or a short-term loan, to a car loan or home bond), renting accommodation, or even entering into a cell phone contract. The lack of a "pay slip" or "salary advice" is one part of the problem, but the underlying cause is what one respondent referred to as the "ambiguity" of the postdoc position. Postdocs "are neither students, nor employees", which is "one of the central weaknesses of the postdoc position". As postdocs are "considered as neither a student nor an employee", being a postdoc was described by another respondent as a difficult "state of limbo", which is "quite a challenge":

> The world scarcely understands what a postdoctoral fellow is. Hence difficulties are often encountered to explain why a person who claims to be 'employed' is simultaneously unable to furnish their pay slip. I think this state of limbo must be addressed as a matter of urgency.

According to a third respondent, being "neither staff nor student" comes with "great instability […] in terms of remuneration", and therefore "the position of a postdoctoral researcher within an institution needs to be revised or altered". Simply the "amount of the stipend turns some banks or suppliers off".

The lack of a "pay slip" or "salary advice" is also linked to postdocs' tax-exempt status – "grant or bursary money not being taxable, monthly income" – leading one respondent to comment that, "I would be happy to keep my tax-free status as a postdoc, provided I got some sort of salary advice. This would give me access to some useful financial products that I currently do not have access to". Similarly, another respondent felt that "postdocs prefer to get a taxed salary in order to be formalised into the financial system with banks". This was their "only grievance", but it caused "pain and suffering". Thus, "exemption from taxation is not a privilege. On the contrary, it is a handicap". "In addition", another respondent noted, "medical aid schemes also often require taxable income to calculate contributions".

Finally, the non-permanent, short-term nature of postdoc appointments makes it very difficult for postdocs to apply for and obtain loans and credit. For one respondent, this was "the biggest issue". Another remarked that, as a result, they "have to live like a student despite being in my forties", and then they recommended, "if funding is for 2–5 years, a contract for the entire period is more beneficial in providing financial security for postdocs". A third respondent concurred that "postdocs need to be offered at least two-year contracts".

### 5.2.2   Work roles and requirements

#### 5.2.2.1   Research

One of the respondents commented that they felt that the postdoc, "while being essentially a continuation of my PhD, was a lot let stress in the sense I did not have to produce a thesis. However, there was also more pressure to produce something at the end of each year". This perception is reflected in the observation that the majority (80%) of respondents were required, either formally or informally, to produce a certain number of peer-reviewed journal articles per annum, although 6% were unsure whether this was the case (Figure ).

**Figure 5.12. Whether respondents were required, either formally or informally, to produce a certain number of peer-reviewed journal articles per annum**

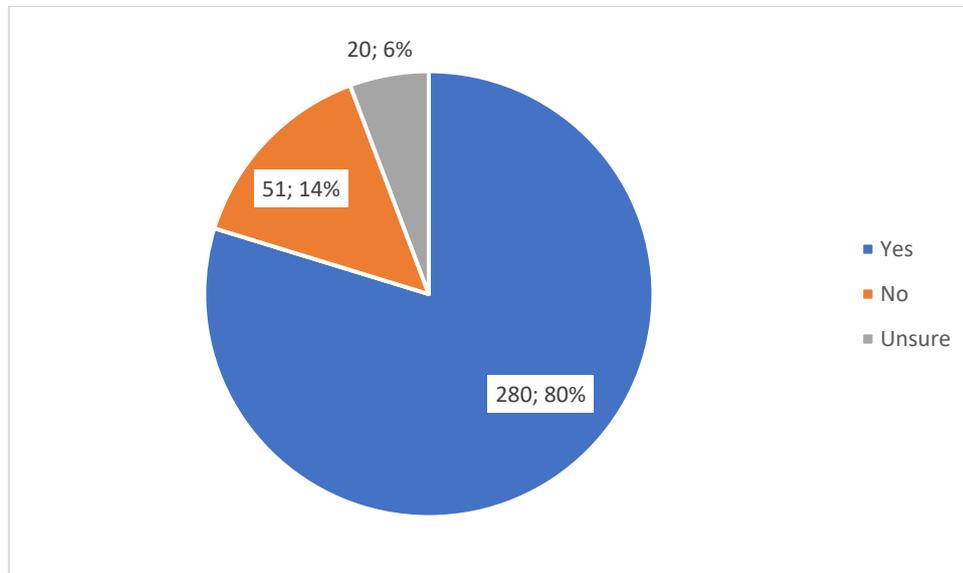

n=351 (excludes those respondents who responded to the questionnaire administered by UP, which omitted this item)

On average, those respondents were required to produce two to three[17] such articles per annum. Although the number required ranges from 1 to 12, the standard deviation (1,338) is relatively low. More detail is provided in Table , which shows that only 10 respondents (less than 3%) were required, either formally or informally, to produce more than five peer-reviewed journal articles per annum.

**Table 5.5. Number of peer-reviewed journal articles respondents are required to produce per annum**

|  | N | % |
|---|---|---|
| One | 50 | 14% |
| Two | 185 | 53% |
| Three | 57 | 16% |
| Four | 34 | 10% |
| Five | 16 | 5% |
| Six | 5 | 1% |
| Seven | 2 | 1% |
| Eight | 1 | 0% |
| Ten | 1 | 0% |
| Twelve | 1 | 0% |
| **Total[18]** | **352** | **100%** |

---

[17] Mean=2,5 and median=2.
[18] Only those respondents who were required, either formally or informally, to produce a certain number of peer-reviewed journal articles per annum

In the qualitative data, a few respondents commented negatively on the number of journal articles they were expected to publish in general, or annually. One felt there was "undue pressure on postdocs to publish a certain number of journal articles". Another two respondents mentioned that requirements (in one case, specified as publishing two articles per annum) are "problematic", as they do not take into account the lengthy "turnaround time" it takes for an article to be published. According to the one respondent, it takes approximately a year, while the other noted, "I have had articles that are under review for over a year now, with some of the editors being rude in their responses to enquiries regarding the time I can expect reviewers' responses". As a result, the same respondent observed, "some postdocs end up focusing on quantity instead of quality of research and this sometimes means sending manuscripts to not-very-reputable journals". As discussed in more detail in Section 5.3.2 below, delays in "submission to journals and therefore reaching the stipulated outcomes outlined in postdoctoral contracts" are also experienced when co-authors – who are full-time academics with myriad responsibilities – "take weeks to respond when reviewing results or drafting manuscripts and lose focus on the research".

"More realistic" requirements of the number of papers and time to publish were recommended by another respondent, "especially when you are receiving no additional assistance". This view was supported by another respondent who felt that they "have lagged without enough manpower (co-workers/students) to write and compete with time". Moreover, "the race to publish at least three articles in peer-reviewed journals can be a daunting task for a fresh postdoctoral student".

Requirements were further criticised for not taking into account a number of factors "affecting quality research". First, the coronavirus disease of 2019 (COVID-19) "is still affecting the access to laboratories and attendance of international conferences, which will aid collaborations"; secondly, the "unfortunate current spate of erratic electricity has affected the sensitivity and effectiveness of state-of-the-art equipment"; and thirdly, there are "bottlenecks of procurements" – for example, "supply of reagents and chemicals can take more than ten months". Other work roles were also reported as limiting postdocs' research output, but not as a critique of a requirement to produce a certain number of peer-reviewed journal articles per annum, and these are therefore dealt with in Section 5.6.2.5 below, in work-role conflict. Some universities seem to use a "point system […] e.g., two points as a minimum required for renewal" of a postdoc position, which was considered "exploitative" by one respondent.

The second aspect of respondents' research-related work role that was surveyed is whether they were allowed to apply for funding as a PI of a research project. As Figure shows, less than half (42%) of the respondents were allowed to do so, while the remaining 58% were divided equally between those who were not allowed and those who were unsure whether they were allowed to apply for funding as a PI of a research project.

Figure 5.13. Whether respondents were allowed to apply for funding as a principal investigator of a research project

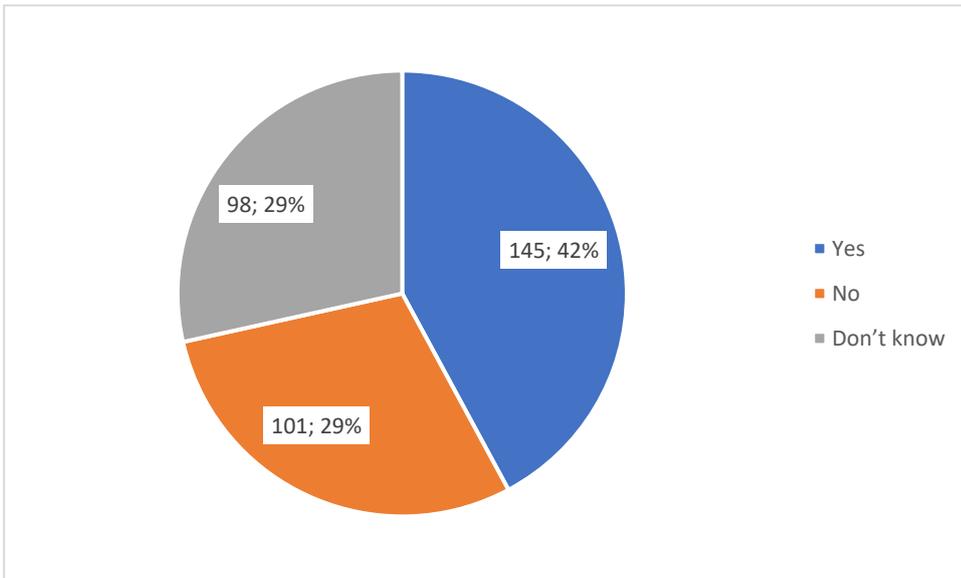

n=344 (excludes those respondents who responded to the questionnaire administered by UP, which omitted this item)

*5.2.2.2 Teaching and supervision of students*

In their 2022 postdoc position, slightly less than half of the respondents had contributed to teaching (e.g., classroom or small-group teaching) (Figure 17). Those respondents were nearly equally divided between teaching undergraduate or postgraduate teaching or doing both.

Figure 5.14. Whether respondents contributed to teaching in the postdoc position they held in 2022

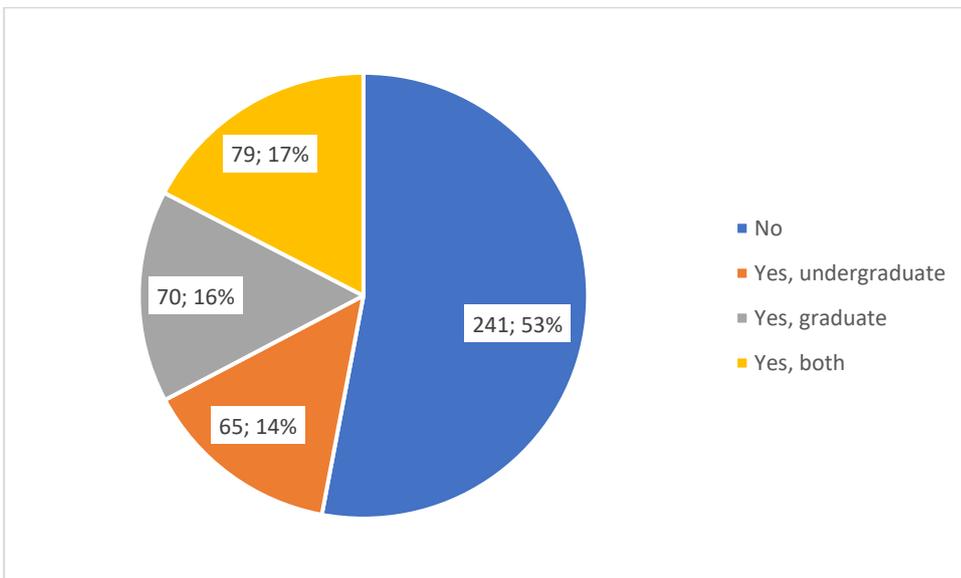

n=455

As to whether postdocs want to teach, the vast majority of the respondents (80%) answered in the affirmative, and most of those wanted to teach at both undergraduate and postgraduate level (Figure 18).

**Figure 5.15. Whether respondents want to teach**

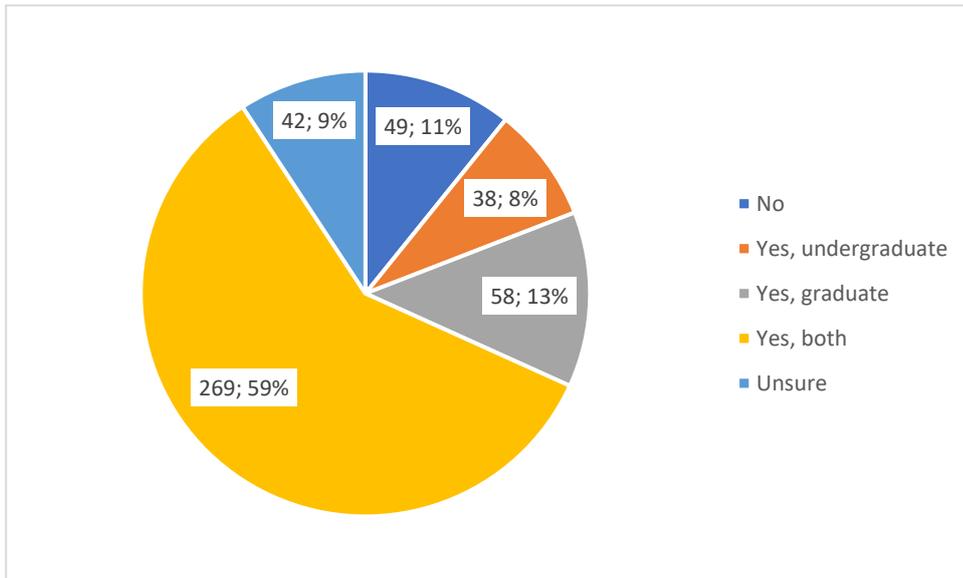

n=456

Next, we asked respondents on their supervision of students in the postdoc position they held in 2022. Slightly more than half (54%) of the respondents had contributed to formal supervision of master's students in that position. Those respondents were asked whether they were formally acknowledged (i.e., whether they appear on the students' supervision record and thesis).

The results (Figure 19) are worrisome. Although, nearly 60% indicated that they were always formally acknowledged, the more significant finding, perhaps, are the percentages who indicated that they are never (13%) or seldom (28%) acknowledged.

**Figure 5.16. Whether supervision of master's students is formally acknowledged**

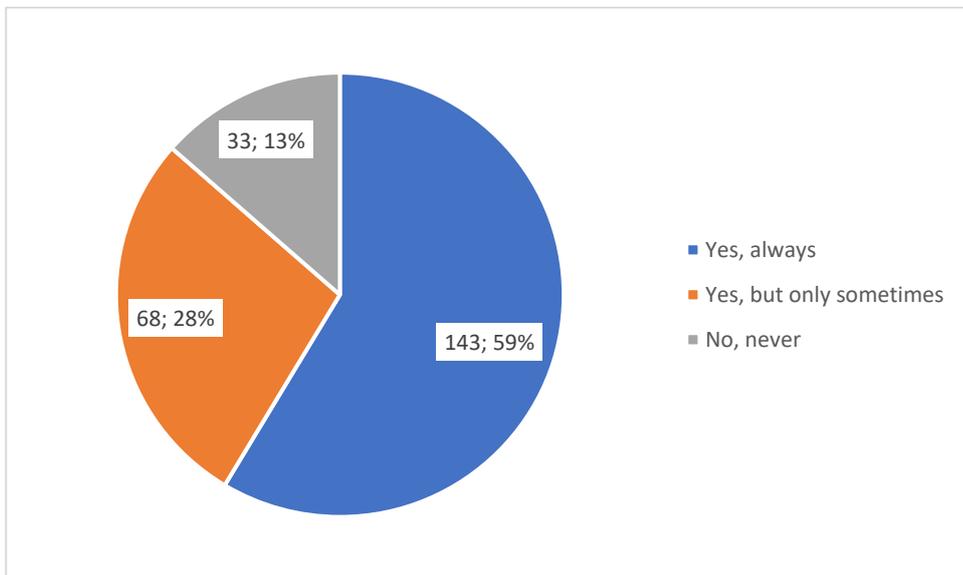

n=244 (only those respondents who had contributed to formal supervision of master's students in the postdoc position they held in 2022)

As more than half of all of the respondents had contributed to formal supervision of master's students, it is not surprising that almost three quarters (73%) of respondents are allowed to at least co-supervise such students. However, as Table shows, 50% of the respondents are only allowed to co-supervise, while 23% are also allowed to be the main supervisor. These may reflect differences in the rules regarding supervision at the university, faculty, or even departmental levels. The question also gave respondents the opportunity to specify other situations that did not fit the ones provided (i.e., the first three listed in Table below) and these 66 responses were analysed. These responses were coded and added as the last four categories in Table below.

Table 5.6. Whether respondents are allowed supervision of master's students

|  | N | % |
|---|---|---|
| I am allowed to co-supervise but not to be the main supervisor | 218 | 50% |
| I am allowed to supervise and co-supervise | 98 | 23% |
| I am not allowed to do any type of supervision | 74 | 17% |
| I am allowed to supervise, but only informally / unofficially | 11 | 3% |
| Unsure / don't know | 15 | 3% |
| I have not (yet) been offered the opportunity / requested to supervise | 12 | 3% |
| Not in contract and/or not (yet) discussed | 4 | 1% |
| **Total** | **432** | **100%** |

Most of these additional categories indicate a lack of clarity on what is allowed, but eleven respondents mentioned that they were allowed to supervise, but only "informally" or "unofficially" (including, for example, training, mentoring and other ad-hoc assistance of master's students).

As is the case with the teaching responsibilities of postdocs, we wanted to determine whether they, in fact, wanted to supervise master's students. The results (Figure 20) clear show that the vast majority (89%) does. This percentage is much higher (by 35 percentage points) than the mere 54% who contributed to formal supervision of master's students in the postdoc position they held in 2022, and notably higher (by 16 percentage points) than the 73% of respondents that were allowed to at least co-supervise master's students.

Figure 5.17. Whether respondents want to supervise master's students

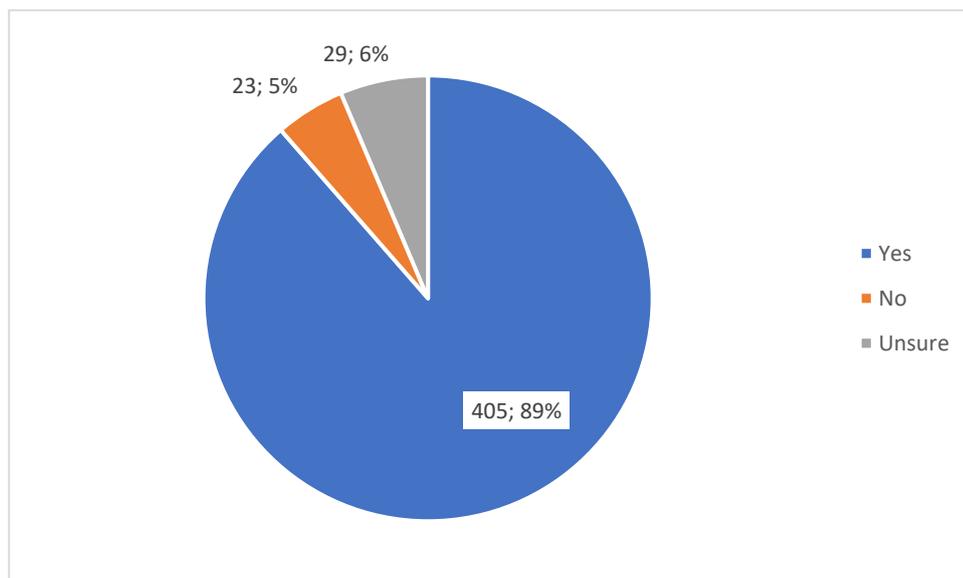

n=457

While 54% of the respondents had contributed to formal supervision of master's students in the postdoc position that they held in 2022, only 30% had contributed to formal supervision of doctoral students. Those respondents were asked whether they were formally acknowledged (i.e., whether they appear on the students' supervision record and thesis). As Figure shows, more than a quarter (28%) were never formally acknowledged for their supervision of doctoral students – twice the number that reported such as complete lack of acknowledgement for master's students' supervision. A quarter of the doctoral-supervising postdocs were only sometimes acknowledged, and slightly less than 50% were always acknowledged (12 percentage points lower than what was found for master's students' supervision).

**Figure 5.18. Whether supervision of doctoral students is formally acknowledged**

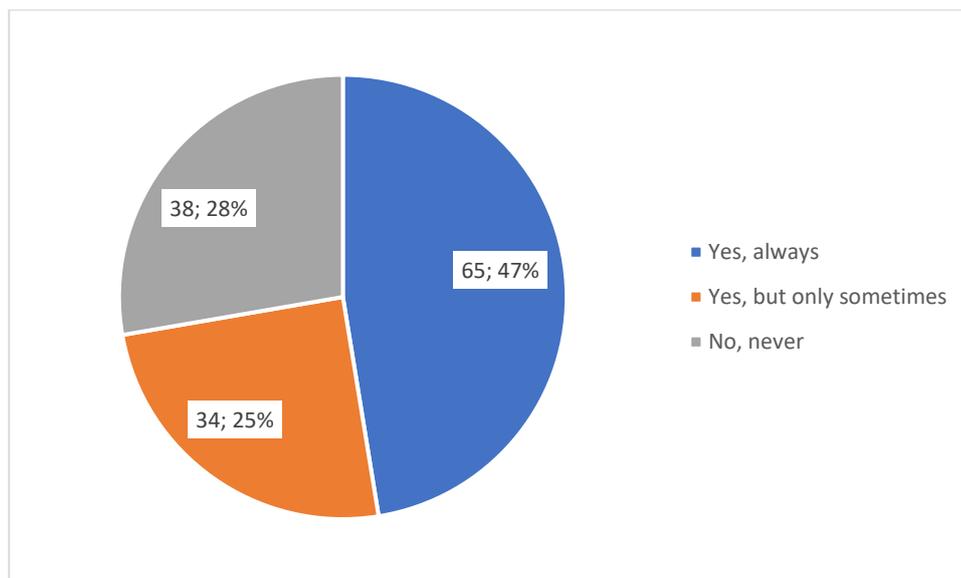

n=137 (only those respondents who had contributed to formal supervision of doctoral students in the postdoc position they held in 2022)

Less than a third of all of the respondents had contributed to formal supervision of doctoral students, but Table below shows that 58% of respondents were allowed to at least co-supervise doctoral students. Not only is this figure 15 percentage points lower than that found for master's students' supervision, but the percentage of respondents that are allowed to be the main supervisor is 37 percentage points lower than in the case of master's students' supervision. As with the same question on master's students' supervision, respondents were given the opportunity to specify other situations that did not fit the ones provided (i.e., the first three listed in Table below). A higher number of responses (88) were provided, but their analysis resulted in the same categories – and similar percentages – as for master's students' supervision. Twenty-two respondents indicated that the question was not applicable to them, as there were no (or insufficient numbers of) doctoral students to supervise, and again, those respondents were removed from the analysis.

**Table 5.7. Whether respondents are allowed supervision of doctoral students**

|  | N | % |
|---|---|---|
| I am allowed to supervise and co-supervise | 52 | 13% |
| I am allowed to co-supervise but not to be the main supervisor | 185 | 45% |
| I am not allowed to do any type of supervision | 113 | 28% |
| I am allowed to supervise, but only informally / unofficially | 11 | 3% |
| Unsure / don't know | 21 | 5% |

| I have not (yet) been offered the opportunity / requested to supervise | 18 | 4% |
|---|---|---|
| Not in contract and/or not (yet) discussed | 7 | 2% |
| **Total** | **407** | **100%** |

In one of these alternative responses, a respondent stated that they are "allowed to assist and mentor graduate students" but were told they "cannot formally supervise or co-supervise students as a postdoc", which is "bureaucratic and frustrating". Such sentiments are not surprising, considering the results presented in Figure , namely that 80% of the respondents want to supervise doctoral students.

**Figure 5.19. Whether respondents want to supervise doctoral students**

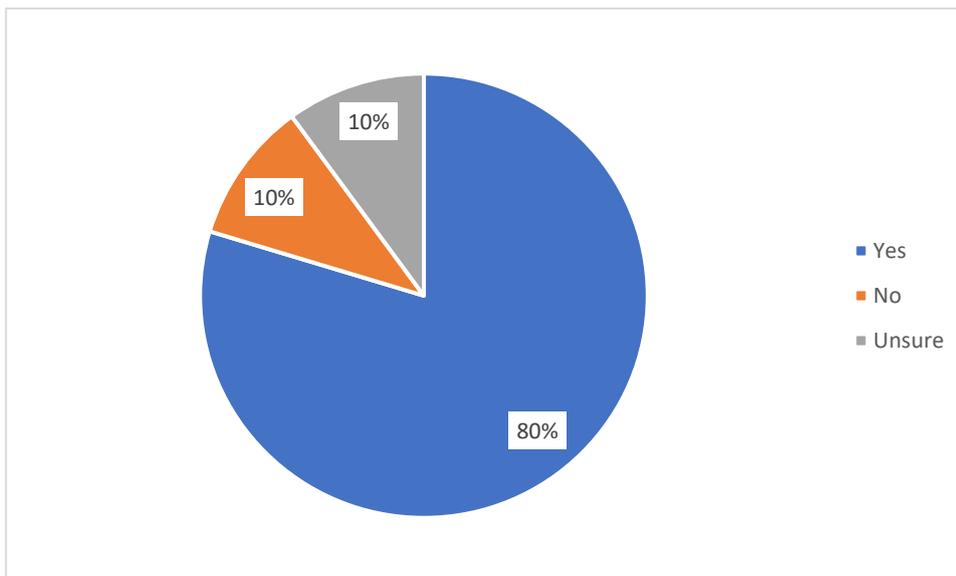

n=438

Numerous comments regarding the issue of supervision were made by respondents. As individual comments often mention both work roles – and different levels of supervision – they are discussed together here. More than half of these respondents recommended that postdocs should be allowed to teach (including marking, examination, and moderation) and/or to (formally) supervise students at both master's and doctoral level ("just like a permanent lecturing position", one noted). Indeed, one respondent felt that "it should be mandatory for postdocs to get involved in teaching, even if it means leading a laboratory", while another recommended nationally standardising postdocs' contracts to include a "minimum [of] student supervision and teaching". One wanted – as part of a "more structured postdoc experience" – to "at least have the opportunity to supervise graduate students, as it is something they would "really like to do". Other respondents critiqued the fact that, "in many instances, postdocs are not provided with the opportunity" to fulfil these work roles, or that, at their host institution, postdocs "only focus on research".

Most of these respondents agreed on the reasons for their recommendation or critique. Even if "the postdoc is primarily appointed for research", "a postdoctoral position is a developmental position", one observed. Thus, postdocs should "have the opportunity to be part of […] teaching, to and "gain experience" in these roles and to "acquire both research and teaching skills". Teaching and supervision skills were considered "very important [ones] that one has to develop during one's postdoc", particularly for a postdoc "who wants to pursue an academic career path". Such postdocs need to be allowed to teach and supervise "to grow not only as a researcher but as a person and an academic". Thus, when postdocs "are restricted in how much time [they] can commit to" supervision and teaching, they "aren't desirable candidates" for the "lecturer roles" ("the post of 'lecturer' or 'senior lecturer'") that "the majority of [them] target", as these roles or posts "require supervision and teaching experience". One respondent felt that their supervisor vehemently oppos[ing] me teaching or supervising students", illustrated that the supervisor "did his best to sabotage what career opportunities I tried to get for myself". In more general terms, "denying these skill sets to postdocs is a loss of a great opportunity to train the next generation of academics", another respondent warned.

The one respondent who mentioned that they had been given the opportunity to supervise students described the experience as "a great learning curve". For the same reasons mentioned above, two respondents recommended that postdocs should not only have the opportunity to be involved in teaching and/or supervision, but also other "faculty activities" and "relevant operations within the university".

As we have seen in the results presented above, the minority of respondents did not want to teach or supervise students. The reasons that some of them gave for this position are also worth mentioning. One respondent felt that they were "forced to teach": "not all of us want to teach, and not everyone is a good teacher", the respondent remarked, and recommended that "it's time we created research posts". Another respondent also "would prefer not to teach": "I love supervising and doing research", they state, "but teaching is not my thing". However, their department "is so short-staffed" that they "have to assist" with undergraduate and honours-level teaching, even if postdocs "aren't expected to" do so. Similarly, a third respondent noted that the "departmental activities" they were "encouraged and even pressured to assist in" exceeded their "postdoctoral contract stipulations". These activities included not only "increased teaching hours [and] supervision demands", but administration/assistance" as well – all "without acknowledgement".

One respondent indicated that one could find oneself in a slightly precarious position:

> if you accept, it interferes with your research or supervision, but if you refuse, the university might see that as a sign that you are not willing to go beyond the scope of your task agreement for "the greater good". You therefore accept the request to teach, not because you want to, but because it could possibly help your future job prospects.

Concern was also raised by the respondent who felt "forced to teach", about the fact that this teaching left very little time for what they referred to as "the luxury" of research "in the two hours per week that are free". According to the respondent, this situation is "driving more people to stay in postdoctoral positions". A fourth respondent did not want to "seek […] out" supervision experience, even though they know that permanent academic jobs require such experience. According to them, supervision "is an on-the-job fulfilment. In other words, when I receive full-time employment, I will be offered training in supervision and not before, so why should I seek it out and potentially do a bad job?" The last of the five respondents provided a reason why they did not want to undertake supervision specifically at the honours level (although "it's an expectation in most schools"), as "in many institutions, [it] doesn't count towards promotions or as 'supervision', so [it] isn't attractive". These qualitative data also hint at various reasons why respondents took a postdoc position, which is the topic of the next main section.

### 5.2.3 Other contractual issues

A large majority of the respondents (92%) had a memorandum of agreement (MoA) or other similar contract (e.g., a letter of appointment) with their host institution. Nine respondents (2%) were unsure, and the remaining 6% answered in the negative. Those respondents who responded in the affirmative were asked whether they were able to negotiate the terms and conditions of their MoA or contract to the extent that they would have preferred. As per Figure 14, more than two thirds (68%) indicated that this was not at all the case.

Figure 5.20. Whether respondents were able to negotiate the terms and conditions of their memorandum of agreement or other similar contract to the extent that they would have preferred

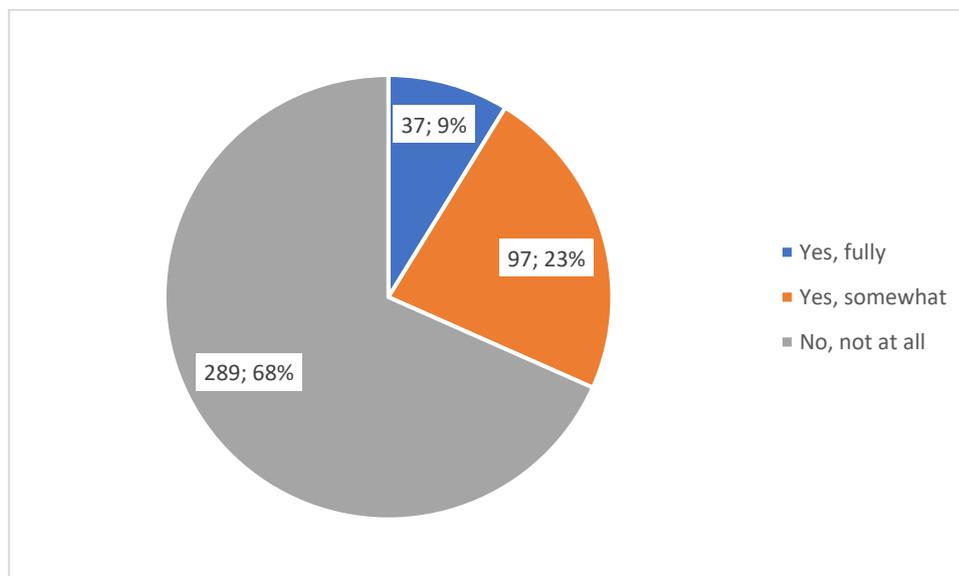

n=423 (only those respondents who had an MoA or other similar contract with their host institution)

Only 9% of the respondents reported that they were fully able to negotiate the terms and conditions of their MoA or contract to the extent that they would have preferred.

## 5.3    Reasons for taking the 2022 postdoc position, and its assessment

In Section 3, on postdoc career trajectories, we reported that close to 80% of serial postdocs agreed that poor job prospects have led to them holding more than one postdoc position since obtaining their doctoral degree. But we also wanted to determine – for all postdocs – what the primary reason was (i.e., only one option could be selected) why they chose to take the postdoc position they held in 2022. The vast majority of respondents (90%) cited the positive aspects of having a postdoc, i.e., improving future employment prospects, developing a research portfolio, gain more skills and experience in their fields of specialisation, and the like. A small minority of 9% indicated that they had no choice, as they could not find a different and suitable employment position.

Table 5.8. Primary reason respondents took the postdoc position held in 2022

|  | N | % |
|---|---|---|
| To enhance my future employment prospects within a university or research institute | 119 | 27% |
| To develop my research portfolio through focussed research | 111 | 25% |
| To gain additional research training / experience in my doctoral field | 74 | 17% |
| I was unable to find a different, suitable position | 65 | 15% |
| To gain research training / experience in a different field of research | 38 | 9% |
| I feel it is a necessary step to obtain a desired permanent position | 33 | 8% |
| **Total** | **440** | **100%** |

Further insight into the quantitative responses was provided by the open-ended responses. Some referred to that the fact that postdoc positions are a "good to kick start" for an academic career"; "a great opportunity for emerging scholars"; "an extremely important career step and helps with personal career development". "Doing a postdoc is still good and relevant, even with the current economic climate and social movement towards leaving academia" was another comment. Even a respondent who felt that working in a postdoc position was "not desirable", noted that "fellowships are designed to train prospective academics for a permanent role in a university setting".

One respondent recommended that, because "postdoc opportunities are fantastic", they "should be built upon, not neglected". Similarly, a second respondent recommended that postdoc positions "should be prioritised", but because "postdocs are a valuable asset to any university". A third respondent even recommended that,

> postdoc positions should be offered yearly by all public universities! PhD graduates who excelled with a minimum of five publications within three years or less, should be provisionally given the offer to do a postdoc – doing so will encourage PhD research outputs.

### 5.3.1 Satisfaction with the postdoc position

We also assessed respondents' level of satisfaction as far as seven aspects of their current postdoc position. Figure below clearly shows that the percentage of satisfied and dissatisfied respondents differs quite markedly for the first five aspects. Opportunities to work on interesting prospects, and to interact with high-quality researchers from other departments and institutions, are the aspects about which the greatest percentage of respondents expressed their satisfaction. In the qualitative data, one respondent mentioned that "collaboration opportunities" were "excellent" and two other respondents commented as follows:

- "I am so thankful for the opportunities I have had as a postdoc to work on many different projects and with external stakeholders".
- "I work on a number of wonderful, different projects with my mentor that have opened up numerous networking and skills development opportunities".

On the other hand, on the last two aspects respondents are very similar in their responses. These are the two aspects that involve host institutions' provision of support (workplace-sponsored training opportunities in skills that postdocs need; and career-development opportunities, assistance or advice). It is also on these two aspects that the lowest percentages of respondents expressed satisfaction.

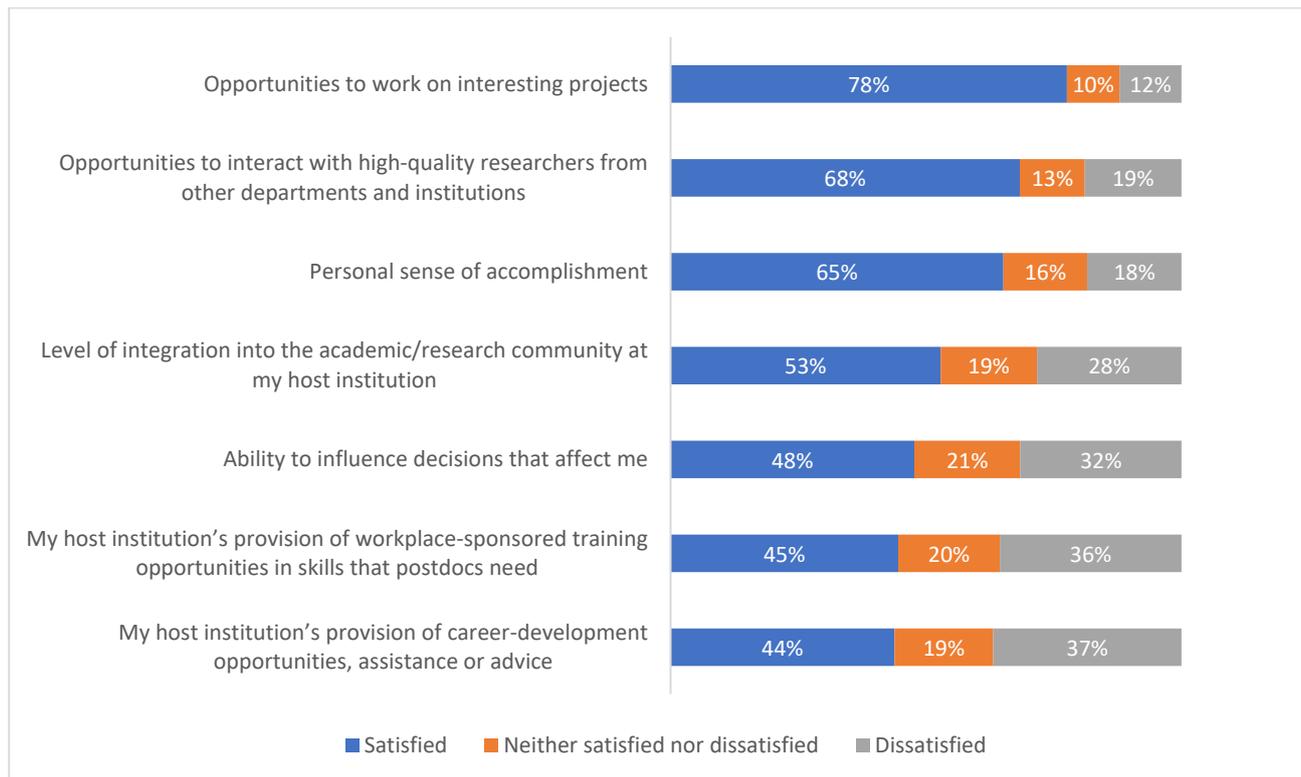

Figure 5.21. Respondents' level of satisfaction with seven aspects of the postdoc position they held in 2022, arranged from highest to lowest percentage of satisfied respondents

n=422–439

The qualitative data provided insights into respondents' satisfaction with some of these aspects of the postdoc position they held in 2022, as discussed in the following sub-sections.

### 5.3.1.1 Level of integration into the academic/research community and ability to influence decisions that affect postdocs at their host institution

According to one respondent,

> Very little interest from other staff members was shown towards the postdocs (three) in both years, who had to find their own feet and assist each other, which was isolating. Only towards the end of the second year, after making continuous efforts to engage with other academics in the department, did I manage to form a connection and affinities with them […]. I thought more could be achieved in terms of publication support and interdepartmental events/activities and found I formed close relationships with the admin and support staff.

Another respondent reflected (also negatively) on their previous postdoc, which was also in South Africa, stating that "I was generally excluded from academic life". The lack of "induction" at some institutions was mentioned: "staff get inductions, [while] postdocs get an invitation to a welcome meeting once a year, which does not really help in any meaningful way with regards to daily activities on the job", one respondent commented. Another agreed that

> there was very little "induction" to starting a postdoc at my current institution. My level of understanding of the support services at the university was completely dependent on my [supervisor]. It was lacking in structure to introduce a new postdoc to the university and basic functions, like the library and finance team.

One recommendation was for "better mentorship and introduction into [a] postdoc position and what the requirements entail […] in order to feel less overwhelmed, especially in a new research field". Another was that, especially management (i.e., heads of schools), "must […] create a conducive environment so that [postdocs] feel they are also colleagues". On a positive note, one respondent "forged many new relationships and connections (also joining the Postdoc Committee at UCT)". Such committees may increase postdocs' ability to influence decisions that affect them at their host institution, thereby addressing perceptions in the qualitative data that postdocs are "often forgotten in institutional decision making"; and, relatedly, that there "is limited representation of postdoc researchers' needs at the top". Most of the qualitative data involved an assessment of the host institution, which the following sub-sections address, after which we consider respondents' general assessment of their postdoc position in 2022.

*5.3.1.2 Host institution's provision of training and career development*

On a positive and general level, respondents mentioned that they were "very glad to have done [their] postdoctoral research" at the host institution, because they "learnt at a lot". For another, "the postdoc journey with [their host university] was one of my finest education ever", "in terms of research" (and specifically research-ethics training). One university's graduate school for postgraduate students allowed a respondent to "develop a number of skills", so that now, "I can boldly affirm that I am a better person when it comes to issues on research, compared to when I first arrived in [South Africa]".

However, by far the majority of the qualitative data reflected negative assessments of host institutions, on a number of issues related to training and career development. One respondent wished their host institution "were clearer about job opportunities that could come as a reward of postdocs' hard work [as] such clarity would go a long way in motivating most postdocs to publish". Another was more critical: "being a postdoctoral fellow should prepare one with the needed research skills as well as guaranteed hope of being hired, but the reverse has been the case". In fact, one respondent noted, "any professional skills, academic growth has been the result of my own initiative and taking on additional work to obtain extra skills. The institution overall has not provided much support, guidance, opportunities for networking and job opportunities".

Added to this critique was a perception that "you are meant to boost the research base of the university just as a tool, without prospects of growth". This perception of postdocs being exploited by their host institutions, especially for their research output, is reflected in two further statements. "The university cares more about more publications of papers than mentoring some of us into this process", one respondent observed, while in another's view, the postdoc "programme is centred around the progression of the host and not the postdoc. In simple terms, postdocs are only being exploited for the growth of other individual researchers and students".

That respondent noted that "there is no concrete plan to develop" postdocs, and another agreed that "there is no clear plan on how to help developing our careers e.g., helping with supervision, teaching and others. No proper training in new skills". Thus, the respondent recommended, "universities need a clear plan how to accommodate postdocs and train the lab leaders how to deal with this workforce", because "most South African universities have no clue what a postdoc is and what their role at the university is. It feels like we are just there". Another respondent agreed that "there is need for the university schools that a postdoc is attached to be educated on what a postdoctoral fellow means and the contributions done during the tenure – especially the management, thus head of schools".

Some respondents expected their institutions to "provide more support to postdocs, especially considering [their] qualifications". Such support was conceived as providing "opportunities for future prospects and more training opportunities, as well as exposure"; and addressing "major stumbling blocks" that "remain in advancing [their] careers". One respondent called for a more fine-grained approach by host institutions:

> those in charge of making [the] postdoc experience and career opportunities better are less interested in listening to individual experiences or challenges and focus more on the lowest denominators (postdocs with poor doctoral training) for the collective failure of postdocs to make a successful transition.

Two suggestions referred to the need for "platforms" for postdocs to network across universities, e.g.,

> Postdoctoral collaborations and networking should be encouraged. This will be advantageous as the postdoc charts his or her career path. For instance, if there's any opportunity for visiting or exchange programmes, even if it's unpaid or with minimal stipends from the foreign host institutions, postdocs and sponsors should avail postdocs the chance to engage on those platforms.

Specific training needs that emerged from the qualitative data are "student supervision, teaching methods, and leadership". In addition, "workshops on the application for NRF rating and grant proposal writing" were considered "of paramount importance to postdocs". One the need to be trained for teaching, one respondent mentioned that "not being groomed in teaching" in their previous postdoc was the reason they "left after a year of being there" (also see Section 5.2.2.2 on reasons why respondents wanted to be given the opportunity to teach).

### 5.3.1.3 Other qualitative responses on the host institution and/or its subsidiaries

In addition to the topics from Figure that were further illustrated above with qualitative data, there were other issues with host institutions that emerged from an analysis of the open-ended responses. Positive comments included being supported, ranging from observations that "the university has done a lot of work toward improving the life of postdocs in terms of providing support on multiple fronts, to "superiors" at the host institution "showing great concern and being very supportive of [the respondent's] situation", and simply being "surrounded by amazingly supportive people". The respondent who expressed gratitude to the superiors, noted that, "professionalism should be humanist professionalism for postdocs: the majority of them are battling with very challenging circumstances beyond what you see". Feeling "at home" and "enjoying" oneself was another aspect that a respondent mentioned.

Comparative to these "softer" criteria, the quality of the research environment was somewhat underplayed in the qualitative data. One respondent mentioned that their host institution "is the best in terms of technical knowledge"; another appreciated the "highly-motivated, research-driven environment" that "supports individual goals", and at third recognised the value of working with "some great researchers", who were also "caring, inspiring and effective at their jobs". However, the qualitative data included many more negative assessments of respondents' host institution. Also, these were often mentioned as contrary to the positive experiences of respondents' supervisors or immediate research environment. For example,

- "My institute is great […]. However, at the university level things are not great".
- "My institution at large is not as supportive and engaging" as the environment in which "I am fortunate to work".
- "The institution overall has not provided much support, guidance, opportunities for networking and job opportunities. These have been created by my supervisor".

Respondents sometimes communicated their perceptions of postdocs being "rarely valued", "undervalued", "mistreated"; "exploited", not "at all supported" by, or "often invisible to", a host institution. Some of the reasons for these perceptions are financial, especially the need for a monthly salary (also see Section 5.2.1.1) and employee benefits (also see Section 5.2.1.2), while others cite a lack of "career-growth opportunities" at the institution. The ambiguous status of postdocs – i.e., being treated as students, rather than staff (or as neither) by their host institutions – again emerged strongly in the qualitative data:

- "The university groups us with students in their admin system".
- "Most admin staff and some academic staff just take us to be any other student".
- One of the disadvantages of being a postdoc in South Africa is that "we are often treated as students even though many of us supervise PhD students".
- "Having done a PhD and published several papers, I am an experienced researcher and am highly skilled, yet I am treated like a student".
- "We are not students but, in some universities, we are treated as such".
- "We are not students; we are not staff. Even HR [human resources] cannot place us in a position that does not cause problems in most cases. I will cut it short here because I can talk about this forever!".

The results of being treated as students (albeit "glorified students" according to one respondent) rather than staff, are that postdocs are "inherently […] put on a lesser footing than other employees and encourages their supervisors to treat them as disposable". Similarly, for another respondent, it was "very disappointing […] as it […] harms the integration into the university staff system. Moreover, it keeps me thinking that I am still a student – no more a qualified professional authority by myself. It degrades my professional self-esteem". Another respondent

> found it a challenge that in my institution postdocs hold no status – it is made clear to us that we are neither student nor staff. We are expected to do similar work as staff but on minimal remuneration and without the respect of having access to certain resources.

The lack of resources referred to by this respondent and others, highlights the implications of being treated as a student. According to the qualitative data, some postdocs are not provided with basic facilities, such as offices to work or operate from; staff cards; institutional email addresses; or access to staff Wi-Fi and computer programmes, such as Adobe Pro and a full [MS] Teams account, "which would help us significantly in doing our work more effectively". "Having clear policies on what staff-like administrative rights are available would make life simpler" was one recommendation, as the respondent "has access to a staff portal but none of the processes (leave-application processes, etc.) are actually functional". Another respondent even felt that they "have to compete with students for research resources".

More general recommendations include that postdocs "should be given more recognition" and "treated as PhD holders, not just like undergraduate students" at their host institutions. "Universities should [also] thoroughly think about this and need to come up with solutions", and "the postdoc needs to be standardised across institutions". One respondent stated that "something needs to be done regarding postdocs' positions in an institution. As it stands, a postdoc is neither here nor there concerning being a student or staff member. Yet, their contribution towards the institution is rewarding for the institution". Another felt that "postdoctoral fellowships could be beneficial to both parties if there could be recognised, full employment".

Respondents' perceptions of a "lack of parity" between the benefits they offer a host institution, and the lack of support or other contributions (e.g., funding or job security) offered in return by the institution, were quite frequent. It was noted by two respondents that institutions provide facilities and other resources for research, but the general argument is that postdocs are "taken advantage of by hosts, [because] the balance between postdoc contribution versus host contribution are completely unfair". One respondent observed that particularly those postdocs who are "soft-funded" are being exploited, and for another, "it is clear that some universities prefer to keep postdocs as students in order to get around their BEE [black economic empowerment] requirements for employees".

The unreciprocated contributions postdocs feel they have made, include teaching and supervision, but also research (e.g., postdocs are "push[ing] research forward" are "necessary to produce research" for their host institutions). Especially the publication of journal articles, for which a host institution receives subsidy from the DSI, was noted. One respondent mentioned having published such articles over the previous ten years, while another (the "soft-funded" respondent) had published over 40 articles over the previous seven years. The perception of a "lack of parity" or exploitation of postdocs with regard to publication of articles in particular, was described by the latter respondent as "a pyramid scheme", while another respondent described postdocs as "frequently the cash cows of universities". A third respondent felt that host institutions "seem to see postdocs as their ticket to achieving a certain number of publications that they would ordinarily not be able to get from their permanent academic staff and postgraduate students". In a similar vein, one recommended that,

> All positions need to be re-evaluated annually. To remain competitive (the university must survive in the QS and ranking), the university must assess the performance of permanent staff and terminate those who do not contribute to research. The research output should be based on factors such as the papers-to-students ratio in the group and publication in reputable journals.

"They are more interested in my research output, not my personal academic career", was another comment. Perceptions such as these led one respondent to the decision that they "will cease to write journal articles while continuing to enjoy [their] teaching and supervisory work".

Administrative issues at host institutions were also mentioned. One respondent bluntly stated, "the university is administratively incompetent", while a second described "the admin" at their host institution as

> extremely poor – I have not once been able to get through to someone on the phone, emails take weeks to get a response, and no one is ever in the office either, so it is just impossible to get anything done.

A third respondent noted that "opportunities for networking and job opportunities" that were created not by their host institution, but by their supervisor, had "been hard won against bureaucratic ineptitude". A fourth was more forgiving, stating that "having efficient university administration assistance" is an issue that, if resolved, would make "the postdoc experience [...] more beneficial and enjoyable". A fifth respondent described in detail how their experience specifically with the postdoc office at their host institution "has been extremely poor". In summary, "the postdoc office was of the opinion that I am not allowed to be appointed part-time to teach and be appointed as a [postdoc]. Instead of communicating this to us at the [...] faculty, the postdoc office failed to pay me for two months". The respondent felt that "experiences such as these result in researchers leaving an institution" and that, therefore, that university "will not reach its 2040 mission [...] if it does not prioritise the people doing the research".

Another more specific difficulty "with support from the administration", according to a respondent (also based on their "discussion with other colleagues"), included problems with banks – such as "not [being] able to use our funding inside and outside of South Africa", but also with their visa. "Most of the international postdoc regret the behaviour of staff, and services" in this regard, they note. Another respondent also felt that "the university is not willing to commit in assisting with employment documentation", such as the processing of work- or permanent-residence permits, or critical skills permits.

Recommendations from three respondents in this regard were that host institutions should:

- "take [it] as their responsibility to use [their] international offices" to address the fact that "most postdocs, [who] are foreign nationals [...], are not treated well regarding the immigration issue";
- "liaise with the Department of Home Affairs on behalf of postdocs who desire to come along with their families"; and

- "fight for immigration renewal when it relates to postdocs".

The last recommendation on this topic is based on the following experience and argument of one respondent:

> Save for the stipend, I really think we are treated worse than normal students (undergraduate to PhD). Here is why. As a "normal" student, you only register once a year. In the case of postdocs, if my contract ends from, say, April to March, I have to first register for the new year in January and then apply for [immigration] renewal again at least three months before the current contract lapses. This means I need to register in January and apply for renewal in January. **If** the application is approved, I would have to register again (I guess), in April. This is an ordeal that normal students do not have to endure […] How can a postdoc focus on their research if there is a cloud of uncertainty hanging over their immigration status and their renewal status with the university?"

As is the case with the positive assessments, the quality of the research environment offered by host institutions rarely featured in the qualitative data. In fact, there was only one such a comment, namely that "sometimes one would find out that a certain equipment is not working at that moment they want to use it". According to the respondent, their host institution "needs lab managers to take care of materials and equipment. It would save time, especially for new students/staff".

#### 5.3.1.4  Satisfaction with the postdoc position as a whole

As Figure  below shows, respondents' level of satisfaction, overall, with the postdoc position they held in 2022 is quite high. Two-thirds, or 66% are either somewhat or completely satisfied, while only a fifth (or 17%) are either somewhat or completely satisfied. This ranks the level of satisfaction with the postdoc position, overall, higher than the level of satisfaction with five of the aspects (the last five listed) in Figure  above.

**Figure 5.22. Respondents' level of satisfaction, overall, with the postdoc position they held in 2022**

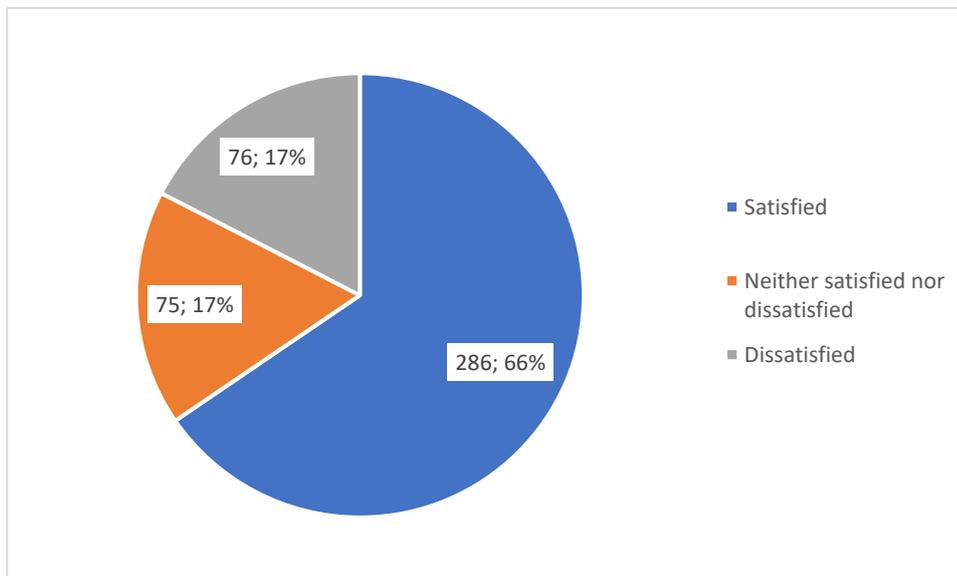

n=437

The qualitative responses provide additional rich data on why respondents provided **positive** feedback on their experiences with their postdoc position. One stated that "the postdoc journey has been fair", but most comments were more 'expansive', for example, "I am overall very happy with my postdoc"; "the postdoc experience is certainly good and rewarding"; "I think the postdoc experience in general has been very beneficial to me"; and "the postdoc opportunity […] improved my life and given me opportunities I did not even dream about having".

A postdoc position in South Africa in particular was positively assessed in general terms, ranging from the extremely positive statement that "South Africa is truly a heaven's paradise", to more moderate ones, such as, "it has been a great opportunity to have come to South Africa for my postdoctoral fellowship"; and "my postdoc experience in South Africa is satisfactory".

The individual responses provided further detail on which aspects of their postdoc position the respondents found satisfying, in addition to the seven ones included in the questionnaire (see Figure ). Two respondents mentioned the flexibility a postdoc position offers: "I love the flexibility of postdoc life […] I love academia" was one's comment, while another was "satisfied with the time flexibility that I am granted to conduct my work". A third respondent explained that,

> this postdoctoral position has been a good time for me to explore academia without the pressure of probation. While it sometimes feels like I am stagnating a little, I am reminded that it is a valuable and privileged position to set up my career.

Finally, the following comment reflects another, different perception of how the postdoc position may be constructed by postdocs and supervisors alike:

> When I received my award letter […] for my postdoctoral research fellowship, I shared the news with one of my PhD supervisors; she said: "The postdoc is the cherry on the top of the PhD journey". Having completed my two-year postdoc in 2022, I can say with confidence that she was correct.

General statements of a more **negative** orientation were less frequent, but insightful. Two respondents provided very detailed descriptions of why they "personally felt that [their] postdoc experience in South Africa was a nightmare", or why "currently, working in [a postdoc] fellowship position is not desirable". Both apologised for the "long-winded and negative comment", or "the negativity, but this is the transparent truth of my experience as well as the sentiments of my peers". Their negative experiences ranged across the topics of this section and were captured in other relevant sections, but their conclusion of one of these respondents is of further relevance here, namely that "worst-case scenarios" [of which they consider themselves to be one] can develop within" the postdoc system in South Africa. The respondent takes "comfort that I am not alone in these thoughts, as researchers such as P. Kerr[19] have discussed the various ways in which our postdoc model in SA is problematic". For another respondent, similarly, "it appears that the postdoc position in South Africa is poorly conceived as a whole". A third felt that "an experience which should be characterised as creative and enjoyable [was] actually negative", and a fourth commented that postdoc positions "in South Africa aren't desirable and [are] avoided".

Other, general calls were for postdocs to be "treated better" and to be given "more opportunity compared to those who only did a PhD or master in academia". From their negative experience as a postdoc, one respondent concluded that, although "research and development are very critical for economic growth, […] government is lacking the will power to adopt locally produced knowledge".

---

[19] P. Kerr has published three articles on this issue:
- Kerr P. 2022. Academic pipeline or academic treadmill? Postdoctoral fellowships and the circular logic of 'development'. *South African Journal of Higher Education*, 36(3):72–90. https://doi.org/10.20853/36-3-5080
- Kerr, P. 2022. Career development or career delay? Postdoctoral fellowships and the de-professionalizing of academic work in South African universities. *British Journal of Sociology of Education*, 43(4):550–565. https://doi.org/10.1080/01425692.2022.2045902
- Kerr, P. 2021. Are we in a parallel pipeline? Bringing the casualisation of academic work onto the South African higher education agenda. *Transformation: Critical Perspectives on Southern Africa*, 105:26–51. https://doi.org/10.1353/trn.2021.0005

### 5.3.2 Assessment of the supervisor

In addition to respondents' assessment of the postdoc position in general, their assessment of their supervisor was probed by the questionnaire, and with the results on this topic, we conclude our main section on the position postdocs held in 2022. When asked to rate their level of satisfaction with their supervisor's overall performance (as their supervisor), the responses were very positive, with 70% indicating that they were either "very satisfied" or "satisfied". Having said this, it is still noteworthy that (as Figure below shows) 16% of respondents indicated some degree of dissatisfaction with the performance of their supervisor.

**Figure 5.23. Respondents' level of satisfaction with their supervisor's overall performance as their supervisor**

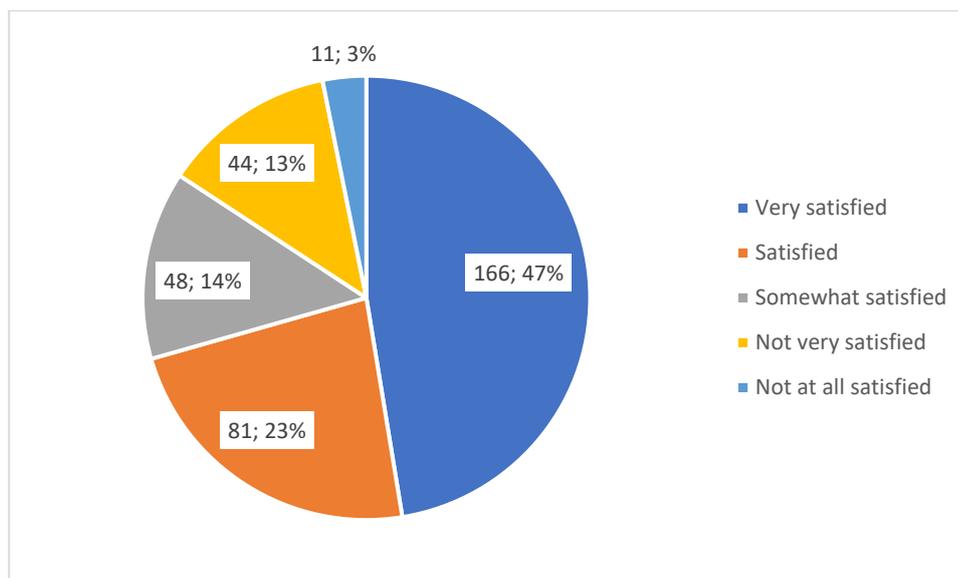

n=350 (excludes those respondents who responded to the questionnaire administered by UP, which omitted this item)

In the qualitative data, respondents' general comments on their supervisors were mostly positive, for example, "I am beyond happy with my supervisor"; or my supervisor is an "excellent mentor". Often, respondents' positive experiences of their supervisors were compared with negative experiences of other aspects of their postdoc. Two respondents felt that their supervisor was the only or primary reason why they were "grateful for [their] experience as a postdoc", or why their "postdoc experience was a very positive one". A third described their relationship with their supervisor as "the highlight of my postdoc fellowship" and a fourth respondent, who was very critical of their host institution, mentioned that their "issue" was "more with the universities not the supervisors".

Next, we analysed respondents' extent of agreement with seven statements concerning their supervisor. The five response categories (ranging from strongly agree to strongly disagree) were recoded into three, by collapsing the original categories "strongly agree" and "agree" into a single category ("agree") and applying the same logic to the opposite side of the scale. One of the statements was phrased with a negative orientation, namely, "My supervisor demands to be first or senior author on all the journal articles we publish". For comparison with the other, positively-oriented statements, its codes were switched around during recoding and the statement was rephrased (in square brackets).

Figure below shows that the highest percentages of respondents agreed that their research work and scientific contributions are valued by their supervisor (87%); and that their supervisor understands that they have family and/or personal obligations (84%). Relatively speaking, the lowest percentages of respondents consider their supervisor to be a mentor and few agreed that they had learned much from their supervisor about how to succeed as a scientist. However, at 75% and 72%, respectively, these percentages are still high.

**Figure 5.24. Percentage of respondents who agreed with seven statements concerning their supervisor, arranged from highest to lowest percentage of agreeing respondents**

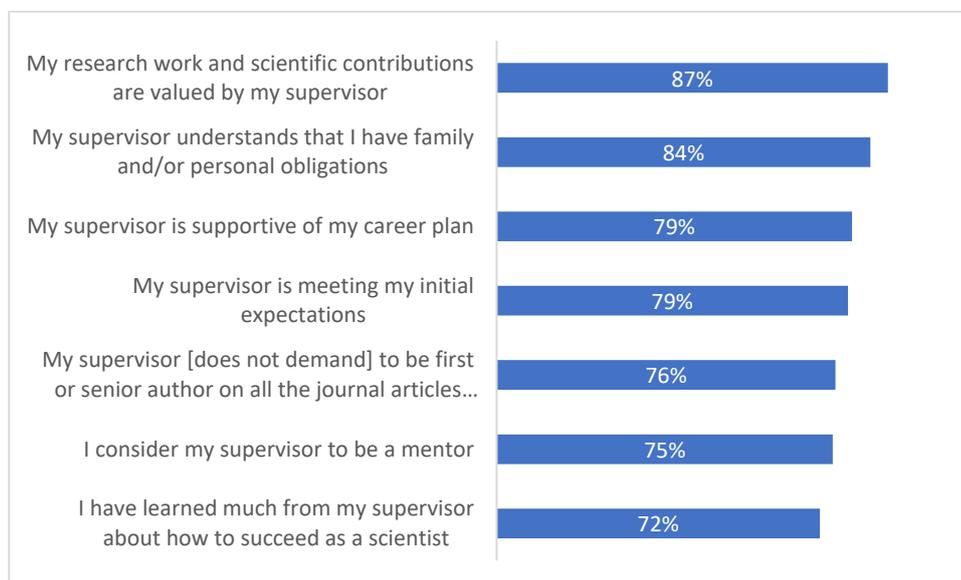

n=433–437

In the qualitative data, we identified aspects of their supervisors – other than those listed in Figure above – that respondents experienced as positive. One respondent "really appreciate[d] the opportunity that I was granted by my supervisor; he does everything to make sure I enjoy my work". Two respondents mentioned the support that they had received from their supervisor. Another described their supervisor as "an incredible individual and she and I just happened to be a good fit", while being supervised by a female PI was central to a respondent's decision to pursue postdoctoral research:

> I started my postdoc under a female PI at [a South African university], and she was wonderful, supportive, and inspiring. She is since deceased. I am now also under the supervision of a female PI, and her management style drives our team to deliver our best. I cannot underscore the importance of what representation of women, especially in engineering, has meant to me, a female, pursuing research that is meaningful to me, and delivers social impact.

Two respondents expressed gratitude for, or felt privileged to have had, "the opportunity to be mentored by experts in [their] field"; or "an erudite and established scholar as host". One of them elaborated that, "I always see my host as someone I still have a lot to benefit from in terms scholarship in the field. I really look forward to continue working with him even after completing my postdoctoral programme".

As the results in Figure above show, 75% of the respondents considered their supervisor to be a mentor. Some of the areas in which one respondent's supervisor had mentored them, included "critiquing a research problem and methodology. He also encouraged me to target high impact journals (Scopus Q1 or Q2) as outlets for possible publication".

A related theme that emerged from the qualitative data is that working with the supervisor was "intellectually rewarding" for respondents, or that they had worked on "a number of wonderful, different projects" with their supervisor. In the latter case, these projects had also "opened up numerous networking and skills development opportunities" for the respondent. Similarly, appreciation was expressed for "the avenues [supervisors] open for me, the networks I build with them and their colleagues", and "opportunities for networking and job opportunities" that were created by their supervisor (and not, the latter respondent emphasised, by their host institution).

In addition to a supervisor being an "excellent […] researcher, mentor, and teacher", their excellence in administration was also highlighted by one respondent. In the case where the supervisor met all these expectations, the same respondent concluded that "to be his student is a blessing […] I was boosted to work for my supervisor who offered me this postdoc offer based on trust that I will serve him well". Similarly, being given the "space […] to excel" by their supervisor, led one respondent to do exactly that – "especially towards my long-term goal to be a consultant (I accepted an offer with an excellent firm for 2023 – very happy here)". "I consider myself lucky to have had a great supervisor", the respondent concluded.

However, some respondents were not as positive in their assessment of their supervisor or postdocs' supervisors in general. The most frequent theme was the perceived 'exploitation' of postdocs by supervisors. Respondents would indicate that "most postdoc positions are created to help PIs gain promotion", while such positions "should rather allow for personal career development". Similarly, "being a postdoctoral fellow under a PI who only wants to climb the academic ladder has been a limiting factor", in one respondent's experience.

Exploitation by supervisors can take other forms as well. For example,

> Postdocs need to pay rent and eat, and supervisors are good at saying, "there is no other funding, we can only pay so much, else we can find someone else who is willing to do research for that amount". So, we end up taking much less than we are worth, as something is better than nothing.

And "even the nicest" supervisors "'abuse' postdocs by dangling [the] carrot of a permanent job, which actually only exists for <10% of postdocs". "Some supervisors are slave masters", one respondent noted. Other respondents agreed and provided more detail. According to one, "many supervisors use postdocs for grunt work and churning out articles, which is a waste of expertise developed over many years". Another explained in more detail that they were "burdened by other duties forced onto me by my supervisor, which I am neither acknowledged nor compensated for, and yet I'm still expected to produce (at a minimum) two first-authored publications per year". An alternative view was that postdocs "become slaves" because "these days most [of them] are more experienced than their mentor and as a result, [they] take over the project". The need for a system or structure to be put in place "to check supervisors from abusing or exploiting the postdocs" and, in general, for "clear standards of how postdocs work with supervisors in general", was identified.

Hindering postdocs' academic career opportunities was mentioned by two respondents. According to one, their supervisor "did his best to sabotage what career opportunities I tried to get for myself (for example, he vehemently opposed me teaching or supervising students)". For another, "the assumption that postdocs are working towards an academic career hinders any possible assistance from supervisors who have connections to industry partners for permanent employment opportunities".

Other negative attributes of supervisors that emerged from the qualitative data included "control", "exclusion", "discrimination", "verbal harassment", and lacking the "skill of being a mentor" (as do "most academics in higher positions", based on one respondent's experience).

Contrary to the negative experiences described above, we find that the respondents' positive assessment of their supervisor in the first set of closed-ended questions (see Figure above) is maintained in other results from closed-ended questionnaire items. In total, more than two thirds (69%) were either very satisfied (49%) or satisfied (20%) with the amount of communication with their supervisor. As Table below shows, the percentage of satisfied respondents is only slightly lower – and the percentage of dissatisfied ones only slightly higher – with regard to the amount or level of supervision received.

Table 5.9. Respondents' level of satisfaction with two aspects of supervision they received from their supervisor

|  | No. of valid cases | % very satisfied | % satisfied | % somewhat satisfied | % not very satisfied | % not at all satisfied |
|---|---|---|---|---|---|---|
| Amount of communication with supervisor | 437 | 49% | 20% | 19% | 8% | 4% |
| Amount/level of supervision received | 437 | 44% | 22% | 16% | 14% | 3% |

The majority of negative comments from respondents related to these two aspects. Three respondents spoke about receiving "very little" mentorship, guidance or structure from their supervisors, a "distant and unavailable" supervisor, or "begging for supervisor's time". Similarly, two respondents described their supervisors as "extremely" or "very" "hands-off". The lack of supervision received was variously ascribed to supervisors' "workload" (combined with an "intensely erratic, but kind, personality"), or their lack of understanding of "the technical aspects" of the respondent's work. An interesting and rather unexpected result from the qualitative data analysis is that a lack of supervisory structure was not always interpreted in negative terms. For one respondent it meant being "given a huge amount of freedom", while for another, this "setup is rather privileged", as it gave them "the opportunity to learn highly transferable skills for […] industry, opening doors to future jobs that I was previously unqualified for".

At the same time, one of the "hand-off" supervisors also wanted "to be involved in most things (without saying it)", which brings us to the issue of communication with supervisors. The same respondent reported that they "have repeatedly said that […] what I work on, is a collaborative effort that requires excellent communication between supervisor and trainee; however, this seems to fall short".

In Section 5.2.2.1 above, concerning the research-related work role and requirements of the respondents, we reported that only 42% of the respondents were sure that they were allowed to apply for funding as a PI of a research project (see Figure ). We investigated this issue of the research-related autonomy of the respondents further, specifically in relation to the amount of control they had, compared to their supervisors, over decisions about five aspects of their research.

As Figure below shows, this issue varies across the five aspects that we included in the questionnaire. Only 52% reported that the writing of grant proposals was completely or mostly under their control, which aligns with the results reported in Section 5.2.2.1 above. On the other hand, close to three quarters (73%) of respondents reported the writing of papers to be completely or mostly under their control. As Figure shows in more detail, this percentage decreases quite substantially for determining the authorship of papers, and for planning new projects (at 59% each); and for choosing collaborators (57%).

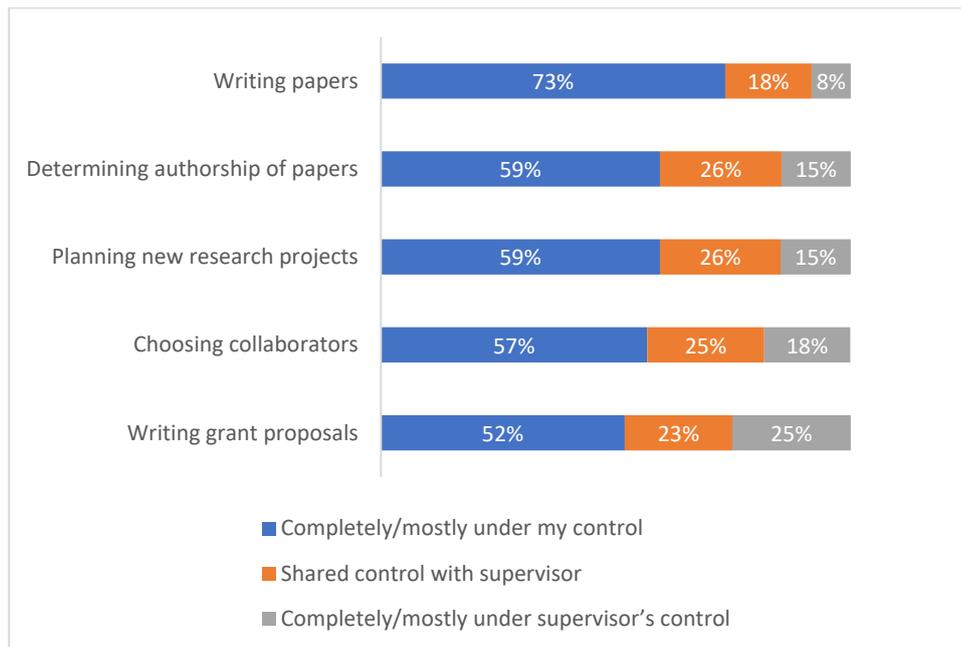

**Figure 5.25.** Amount of control respondents, compared to their supervisors, had over decisions about five aspects of their research, arranged from highest to lowest percentage of completely/mostly under the respondent's control

n=347–350

Although close to three quarters of respondents reported the writing of papers to be completely or mostly under their control, the qualitative data raised the issue of some respondents not being "allowed to submit a paper for publication before a supervisor can review it". The issue has two related aspects, the first of which is postdocs' perceived lack of independence. For one respondent it means that "postdocs are treated same way as postgraduate students": "how do you train one to be the best when you don't allow them to fly on their own and learn from their mistakes?" the respondent asks. The response they "get when asking, is that [the supervisors] are well-known and respectable scientists and cannot allow postdocs to submit imperfect work to journals, with their names on it. It will mess up their image".

The second, more pragmatic aspect then arises "in a case where the supervisor is too busy" to review all the papers, meaning that "only one or no paper can be published in a year", because the postdocs "have to queue for the supervisor's attention with 10 other students in the lab". As a result, one respondent says, "I have a pile of prepared manuscripts waiting for the supervisor to finish reviewing theses and dissertations. In this sense, then, the respondent feels that "some supervisors act as barriers to the success of postdocs in their labs".

For another respondent, not only to supervisors, but co-authors in general, delay submission to journals and therefore reaching the stipulated outcomes outlined in postdoctoral contracts". Such co-authors "take weeks to respond when reviewing results or drafting manuscripts and lose focus on the research because, as "full time academics" they "are encumbered by a myriad of responsibilities. Heavy teaching load, participation is multiple committees, supervision of master['s] and PhD candidates and administration of modules and courses". Although the respondent "sympathise[s] with the demands of the job", in their experience the supervisors' "commitments come first. Therefore, any research outputs aren't priority. They are in a permanent position. If they don't publish in a given year, they won't lose their job".

The perception above, that supervisors "act as barriers to the success of postdocs", extends beyond supervisors' insistence to review postdocs' papers before submission. In the experience of one respondent, some

> are restricting career development of fellows through some restricting rules. Also, a lot of micromanagement from a [supervisor] does not allow a fellow to perform at their best. Fellows are not allowed to make decisions on important matters of their research.

Respondents who experienced the opposite of such restrictive "micromanagement", relayed it in positive terms, such as being given "the liberty to carry out research" or, in general, "the space […] to excel". One respondent even described having a "very hands-off" supervisor as a privilege, as "this has given me the opportunity to learn highly transferable skills for the data science industry, opening doors to future jobs that I was previously unqualified for". With regard to control over planning new projects specifically, one respondent noted that how "enriching" a postdoctoral opportunity is, "depends very much on your PI and a balance between imposed and co-created projects/programmes".

## 5.4 Discrimination and harassment

Our next main results section deals with issues of discrimination and harassment, which were covered in four questionnaire items. These items related to awareness of gender discrimination and racial discrimination, as well as asking about actual experiences of such forms of discrimination.

First, respondents were asked about **gender discrimination**, and specifically whether there are any differences between the treatment of male and female postdocs by their supervisors and colleagues. Although nearly two thirds of the respondents (64%) indicated that they had not experienced such gender discrimination in their environments, it is noteworthy that a substantial quarter of respondents indicated that they neither agreed nor disagreed (see Figure ) with the statement. Although a small percentage, it is still cause for concern that 9% agreed that male and female postdocs are treated differently by their supervisors and colleagues.

**Figure 5.26. Respondents' extent of agreement that there are no differences between the treatment of male and female postdocs by their supervisors and colleagues**

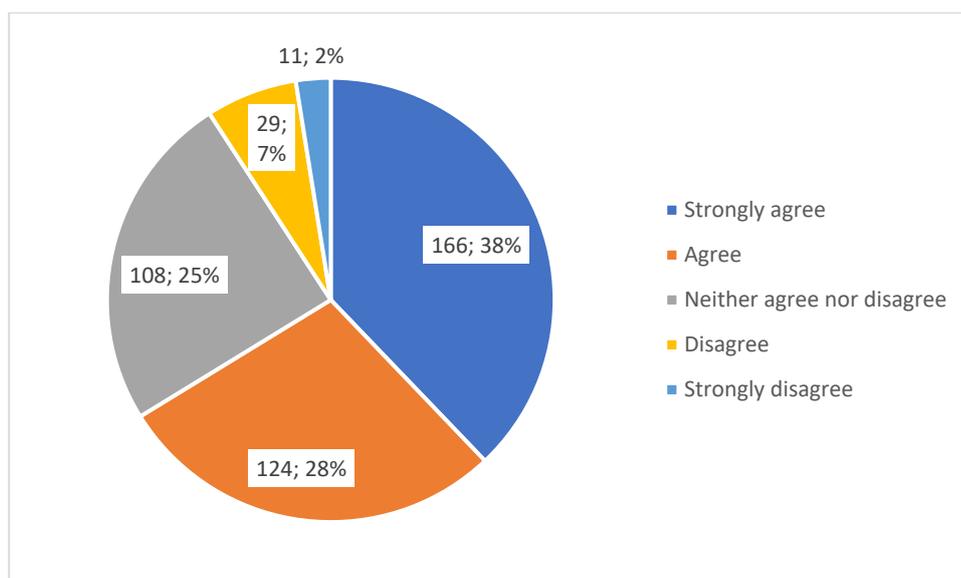

n=438

A similar questionnaire item concerned **racial discrimination**, by asking respondents to agree or disagree with the statement, "There is no discrimination against postdocs based on their race or ethnic background". Only 55% agreed that this was the case. Again, approximately a quarter (24%) neither agreed nor disagreed (see Figure ), but around one in five respondents (21%) – double the percentage found with regard to gender discrimination – either disagreed or disagreed strongly with the statement that there is no discrimination against postdocs based on their race or ethnic background.

Figure 5.27. Respondents' extent of agreement that there is no discrimination against postdocs based on their race or ethnic background

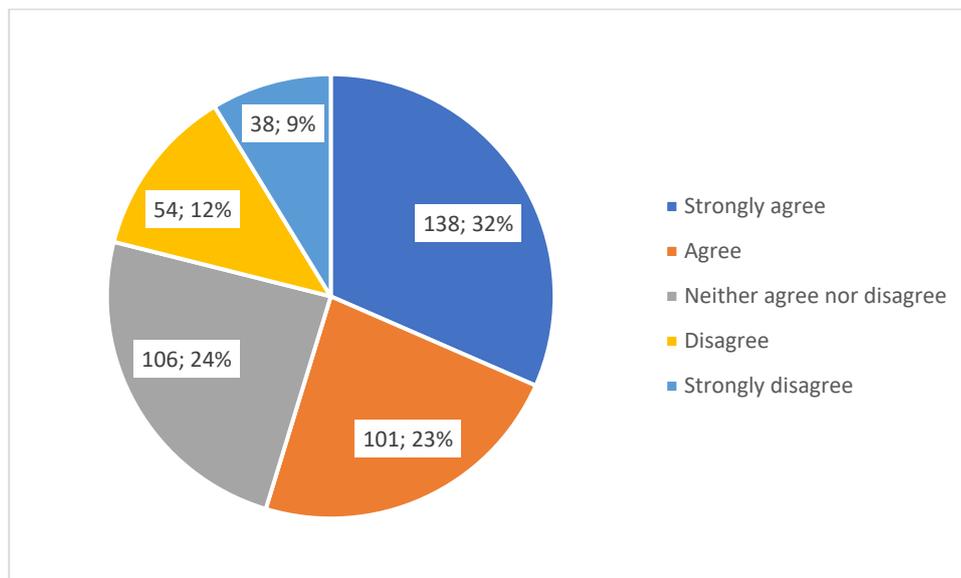

n=437

The third questionnaire item on the issue asked respondents directly whether they themselves had experienced discrimination or harassment in the postdoc position they held in 2022. Only 16% reported in the affirmative. However, when asked – in the last question on the issue – whether they had observed discrimination or harassment during the postdoc position they held in 2022, almost a quarter (24%) answered "yes".

In the qualitative data, some respondents provided more detail on their experiences of discrimination based on gender, race or ethnic background. Two respondents reported experiences of sexism and/or racism, viz., "colleagues are […] sexist in many scenarios"; and "at my previous postdoc (also in South Africa) […] I was sexually harassed and witnessed both racism and sexism". A few other respondents referred to perceptions, experiences, and/or opinions of discrimination based on their white race, mostly their male gender, and/or their older age, in relation to the granting of postdoc or permanent positions. One respondent provided a detailed account of their and their fellow postdocs' experiences:

> At the moment, there is a high level of bias against white candidates for both postdoc positions and permanent appointments. There is also an "Africanisation bias". My (now) fellow postdoc did not initially receive the position, as she could not adequately explain to the interview panel how her research promotes the Africanisation of higher education, which I feel is unfair. […] my postdoc colleague and I face insurmountable hurdles set in place by employment equity. I've heard of colleagues from our discipline who are currently on their second or even third postdoc positions, at different universities, with massively impressive publication records, but no prospects of permanent employment.

Other, shorter responses on this topic are as follows:

- "I decided to pursue a postdoc position as there were no other options available to me at the university. The positions available for lecturing in my department were earmarked as BEE positions".
- "The biggest challenge for me is that I am not sure there is a place for me in academia in South Africa, given [that] I am a white male".
- "Because I am a 59-year-old white man, I have no expectation of securing an academic post".
- "As a middle-aged, white woman, finding employment in my field is impossible regardless of how much knowledge I have".

Another respondent observed that, "the questions about racism and discrimination may have a preconceived notion that it applies to non-white students only. In reality, they (white people) are by far the most discriminated against in academia in South Africa". A respondent who self-identified as a black South African, felt that "the standard of research is dropping because these days is not about quality anymore, but about reaching the number of females, black people, Indian people, white people, etc." Three respondents seemed to agree, although they were less specific about the basis for the discrimination. One commented that, "unfortunately, at my university, the [permanent] positions are not filled with quality", while the second explained that "the transition from postdoc to finding a permanent position at a South African academic institution is extremely difficult. This is partly related to few opportunities to apply for, and prejudiced decisions on candidacy for available positions". The third respondent noted such discrimination in the private sector, in the field of molecular biology and biotechnology, as there is a lack of such work in South Africa, "especially if one does not meet the correct demographic".

Three respondents noted unspecified discrimination or bias as challenges that, in two cases, in part "propelled" them to "seek better employment opportunities in countries around the world", or to leave their previous postdoc "after a year of being there". We will return to this topic in Section 5.6.2.1, on respondents' career preparation, where we report that of discrimination or bias was by far the most frequently selected option by respondents as the greatest, single challenge for their personal career progression. We also consider respondents' perceptions and experiences of discrimination against non-South African nationals specifically, in Section 5.7.4, on the perceived lack of job opportunities for non-South Africans, including perceptions of xenophobia.

## 5.5   Contributors to a successful postdoc experience

Extending the focus beyond the postdoc position respondents held in 2022, we asked them how important they deemed twelve factors or attributes as contributing to a successful postdoc experience. The set of questionnaire items did not produce much variability in the responses. As Figure below shows, the majority of respondents (90% or more) considered all but one of the factors or attributes as either extremely or very important. The figure does, however, show the relative importance of the factors or attributes, with quality research almost unequivocally considered as extremely or very important, followed by the security of knowing that one's training will push one's career forward. However, in the qualitative data one respondent mentioned that, as a postdoc specifically in South Africa, they "have often felt […] uncertain as to whether doing a postdoc is a good career decision".

Rated relatively low, in comparison, are supportive colleagues and training (at 90% each). But the exception concerns the personal lives of respondents: as Figure  shows, only 71% of respondents perceived accommodations made by a spouse, partner, and/or their family to be extremely or very important in contributing to a successful postdoc experience.

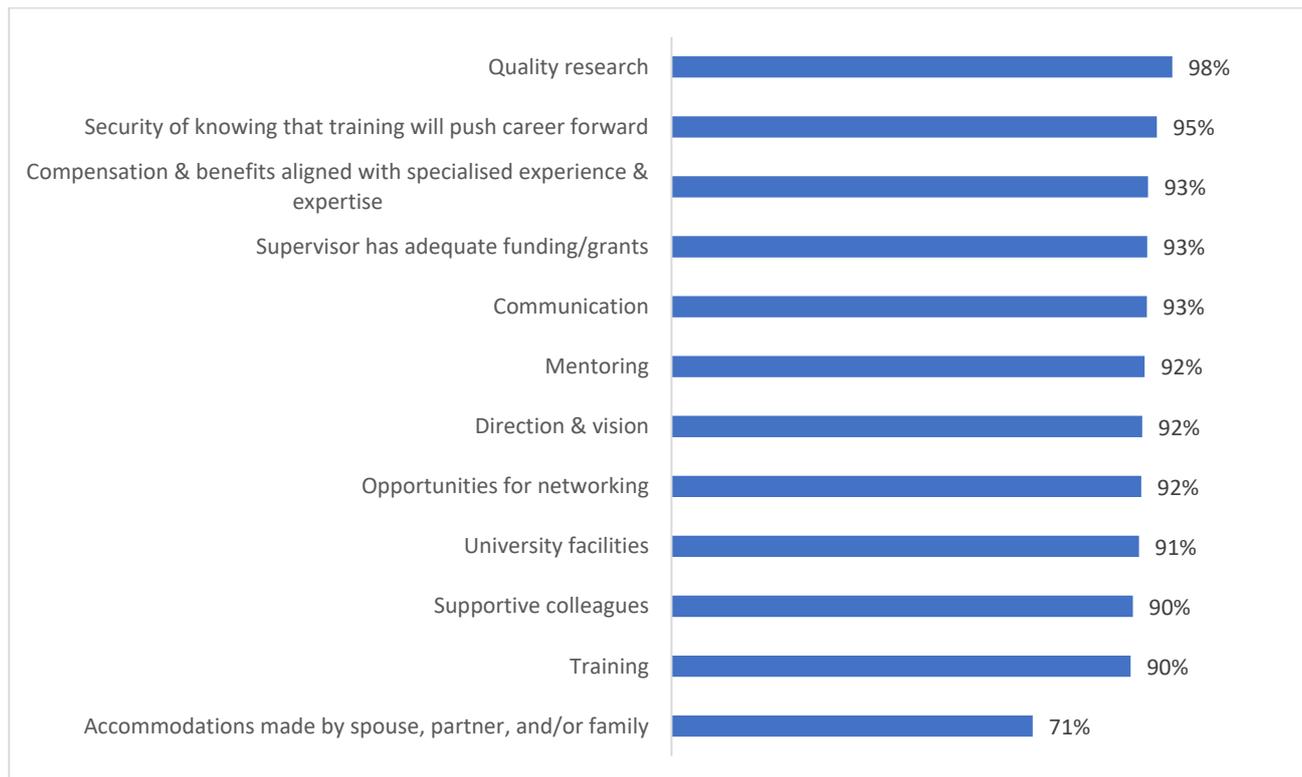

Figure 5.28. Percentage of respondents who deem 12 factors/attributes extremely or very important in contributing to a successful postdoc experience

n=424–429

The qualitative data identified another factor, which does concern respondents' personal lives, namely flexibility. According to one respondent, a postdoc position "should be flexible […] to allow work[ing] remotely and still deliver[ing] quality output". Another felt that "working from home should not be viewed as a researcher unwilling to be present", as

> mothers, and parents or caregivers in general should be allowed to manage their time to get the work done around their care needs. If the work is being done, then the appearance of their face in the office should not be necessary. Coming into the office adds incredible strain to situations where a carer is needed, and this hinders the work needed doing.

Mentoring was observed to be "very important at this level of someone's academic journey", but another respondent noted that, among postdocs, "there is sometimes a fear of asking for assistance", as "having a PhD results in the assumption that you should know what you are doing".

## 5.6 Future career expectations and preparation

### 5.6.1 Career plans

Three quarters of the respondents hoped to pursue an academic career, although another 17% were unsure, as Figure shows.

**Figure 5.29. Whether respondents hope to pursue an academic career**

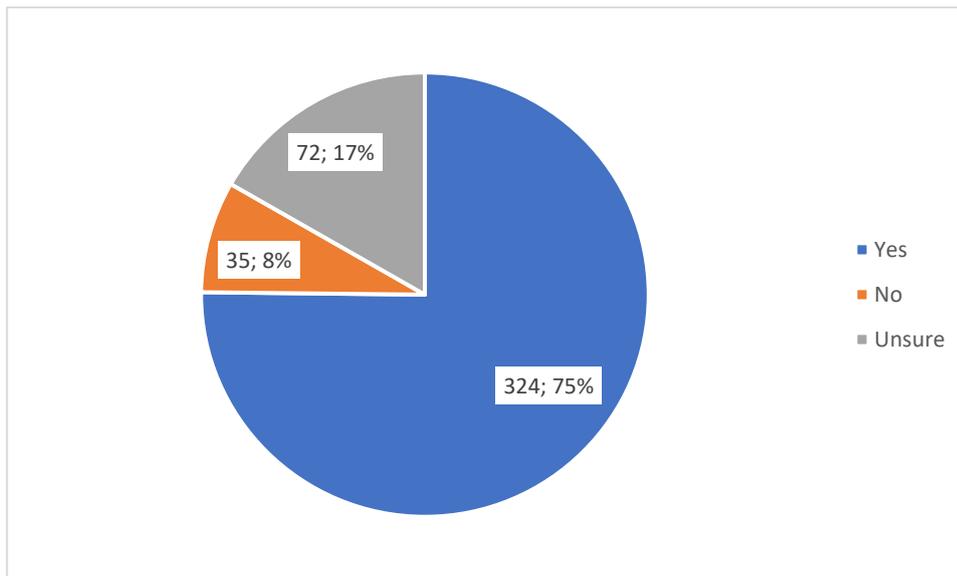

n=431

Those 324 respondents who were sure that they wanted to pursue an academic career were requested to specify the role in which they would most prefer to work (in their future, long-term career plans). As Table shows, only five respondents had not yet decided, while the majority (79%) preferred a role that combined research and teaching (i.e., a professor or "tenure track" position). It is notable, however, that 16% were hoping to pursue an academic career that involved primarily, or only, research, compared to the 3% that preferred a role involving primarily, or only, teaching.

**Table 5.10. Role in which respondents who hope to pursue an academic career would most prefer to work**

|  | N | % |
|---|---|---|
| Research and teaching combined | 254 | 79% |
| Primarily or only research | 51 | 16% |
| Primarily or only teaching | 9 | 3% |
| Other[20] | 3 | 1% |
| Not yet decided | 5 | 2% |
| **Total[21]** | **322** | **100%** |

The 35 respondents who hoped to pursue a non-academic career were asked the same question, i.e., "When thinking of your future, long-term career plans, in which role would you most prefer to work?" As Table shows, the majority (10) indicated research, followed by smaller numbers of respondents preferring other roles.

**Table 5.11. Role in which respondents who hope to pursue a non-academic career would most prefer to work**

|  | N |
|---|---|
| Research | 10 |

---

[20] "Primarily research and teaching and still take consulting work from clients in industry"; "research and clinical work"; and "working for international research think tanks".
[21] Only those respondents who hoped to pursue an academic career.

| Professional practice | 7 |
| Research management / administration / policy | 6 |
| Consulting | 6 |
| Science communication / Scientific publishing or writing | 4 |
| Research and other[22] | 2 |
| **Total[23]** | **35** |

Next, we move from a survey of plans and preferences to respondents' job-seeking behaviour. The majority (61%) of respondents reported that they had applied for permanent positions in the 12-month period before the survey. Close to 90% of the respondents reported that the positions they applied for are at least somewhat related to their current research, and – As Figure shows – close to half (48%) strongly agreed that this was the case.

**Figure 5.30. Whether the position/s the respondents applied for are related to their current research**

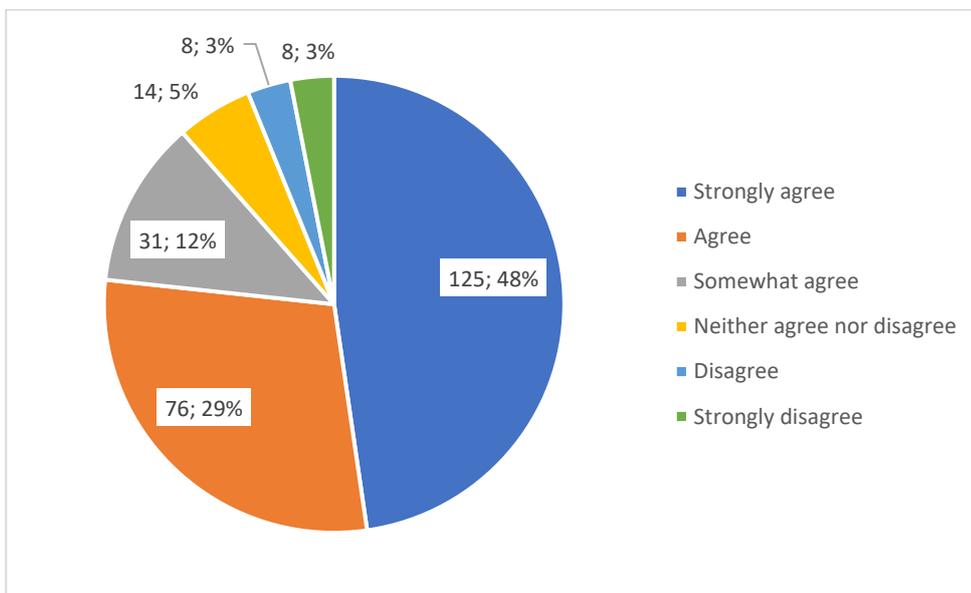

n=262 (only those respondents who applied for any permanent position/s in the 12-month period before the survey)

All respondents – regardless of whether they had applied for any permanent position/s in the 12-month period before the survey – were asked to assess the job market in their field. Four percent were unsure how to respond and excluded from the next analysis. As Figure shows, only 4% considered the job market in their field to be excellent, and the perception of less than a quarter (24%) was that it was "good". Another 27% reported the job market in their field to be fair, but the most frequent response (45%) was "poor".

---

[22] "Hybrid role including research and other 'non-academic' skills"; and "research, curating and management".
[23] Only those respondents who hope to pursue a non-academic career.

**Figure 5.31. Respondents' perception of the job market in their field**

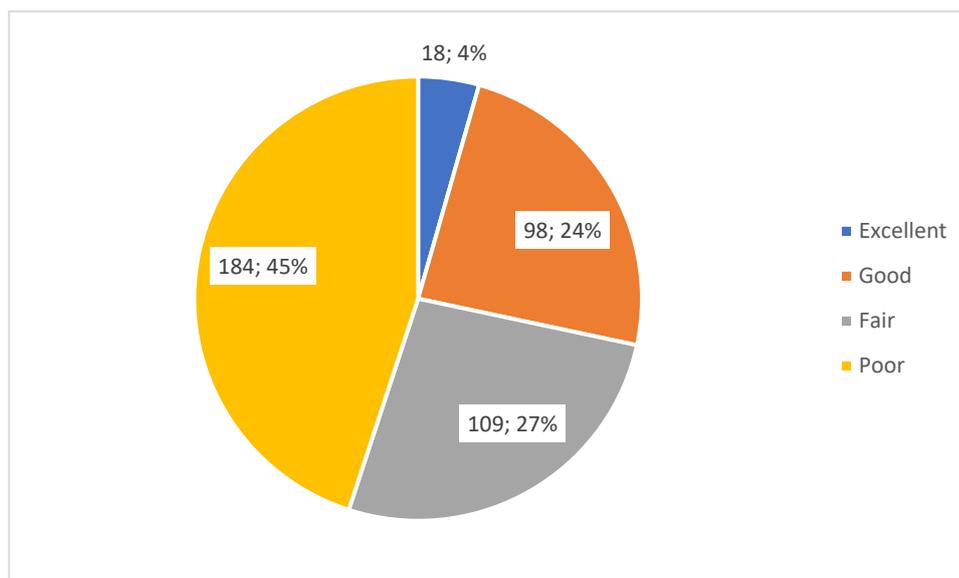

n=409 (excluding those respondents who were unsure)

We conducted the same analysis for only those 306 respondents who were sure that they wanted to pursue an academic career and had a clear perception of the job market in their field. The results were slightly more positive, but still, only 5% considered the job market in their field to be excellent, 26% considered it to be good, 28% to be fair, and 42% to be poor. In the qualitative data, respondents expressed mostly negative perceptions of the job market in their field. Many of those are related to the lack of job opportunities for non-South African postdocs, and these are analysed in more detail in Section 5.7.4 below. Others were cited in Section 6, when they pertained to discrimination based on nationality, race or ethnic origin, and gender. The remaining comments are considered here.

Job prospects for postdocs in South Africa were described by one respondent as "abysmal" for "most" – a statement quantified by another respondent according to whom "a permanent job […] actually only exists for <10% of postdocs". According to a third respondent, opportunities are "extremely limited", even "for career researchers […] who manage to secure their own funding".

According to other respondents, the problem is not limited to postdocs but applies to "PhD graduates in South Africa at large". Respondents linked the lack of job opportunities for doctoral graduates to those graduates, such as them, taking a postdoc position. One believed that "the fact that someone with a PhD will go ahead and take on a postdoc role is already a reflection of how bad [that] job market is". According to another, the government "constantly campaigns for people to pursue PhD" but has "no plan in place to create industries for [the graduates] to work in", thus they "end up working as a postdoc on a yearly contract, earning peanuts". The respondent concludes that they "cannot in good faith encourage any person to register for a PhD because the return on investment is terrible".

A similar notion, which was raised by two respondents" is that many degrees "groom" individuals for positions that either do not exist, or are limited – not only in universities, but also in government and industry. According to the first of these two respondents, "many degrees groom us for such positions (for state work) but then there are no positions for these specialist skills – neither at a state level nor a university position". The second agrees that, because "universities are in the business of degrees", academics "'groom" undergraduate and honours students to study further. I wonder if they realise that, following graduation, opportunities are limited – not only in academia but in industry". Having specialist or "niche skills" were also viewed as potentially problematic by a third respondent, because of either "the inability to transfer highly specific skills, or the inability of potential employers to see the benefits of those skills in their organisation".

For the other two respondents, the issue is more general, namely the inability of the country to see the benefits that doctoral graduates and postdocs can offer. According to one, "South African institutions do not take cognisance of scarce skills developed and needed in the country, neither at a state or university level". The other felt that "the only factor that keeps one in the field is the interest and passion to effect change and make a difference", which led them to argued further that,

> as academics, we impact the social fabric of the country. Idealistically, we are problem solvers. If society doesn't value research and this contribution, why place oneself in an inferior position? As human beings, we all need security and appreciation, a sense of job satisfaction. In sum, postdoctoral positions in South Africa aren't desirable and avoided.

Above, we cited respondents who perceived job opportunities to be extremely limited, or even non-existent, regardless of sector. Also, three respondents felt that the there was a lack of jobs both in the academic sector and in the private sector, or for doctoral graduates in industry. But the majority focused in their comments on the limited availability of positions specifically in the academic sector in South Africa. As 75% of the respondents hoped to pursue an academic career (see Figure above), this is not surprising. Most of those respondents referred to the lack of permanent (or "tenure track") positions, although one mentioned the lack of "semi-permanent positions" as well. was ascribed by one respondent to inadequate funding and – aligned with the earlier comments in this sub-section, on an oversupply of doctoral graduates – "high numbers of PhDs", particularly "in the sciences".

Host institutions' appointment policies and practices were also relevant, according to two respondents. One feels they have "absolutely no hope of permanent employment in [their] current institution", explaining that

> it is not the department's fault, but that of the university. Almost half of our senior staff members (mostly full professors) have retired over the past two years, but the university has made no effort in replacing them. They promised us that positions will be advertised, but nothing has happened yet […] Despite the fact that our university is in dire need of postgraduate supervisors (due to increasing MA and PhD registrations) they don't seem to have any plans towards employing postdocs as permanent supervisors any time soon.

The other respondent understood their host institution's appointment practices to be a response to the short-term nature of the funding that may be secured by a postdoc:

> Labour law means that if I secure three years' research funding, my institution has to advertise the position and make it a permanent position, but the funding is not permanent funding and they might have to give the job to someone else, even though I secured the funding, so South African institutions are reluctant to host me and support me in building a research agenda and new coursework material. Please, please, please look into this further, because it will definitely affect other career researchers!

The result of a limited number of permanent academic positions, for one respondent, means that "the transition from postdoc to finding a permanent position at a South African academic institution is extremely difficult" (the other being "prejudiced decisions on candidacy for available positions" – a point already raised in Section 5.4 above). And, according to another respondent, such positions may be open to them "if I were willing to relocate, but I have a family". Thus, although they "would love to stay in academia", they are "feeling very discouraged about their career". A certain measure of desperation was also evident from wishes that were expressed for "you" to "offer me an academic position in any institution, even if it is for only three years, as a start for career progression"; or for "get[ing] networked with a university that could take me in as a lecturer in [my field]". Another simply stated, "I am open to any available opportunities in academia". The hope was also expressed that

> the essence of the collective feedback from this survey is to make decisions that will create a pathway for persons that have undergone postdoc training at South African universities to be integrated into academia so that they can also train and motivate individuals interested in academia or research-only.

As only 8% of the respondents did not hope to pursue an academic career (see Figure above), negative perceptions of the job market for non-academic positions were less frequent. In fact, such non-academic positions – also outside of South Africa – were viewed as the only alternative for two respondents. For the first, "the uncertainty of career opportunities in academia in South Africa has prompted me to consider non-academic jobs outside [of] South Africa, but within the broader field of research". The second respondent's view also reflects the feeling of discouragement evident in a previous comment:

> It saddens me that after nearly two decades of study in my field, and trying my best to build an academic career, I will have to find employment elsewhere […] I'm at a point where I want to say, academia is a farce, and I want out, and that saddens me immensely.

Field of study was relevant, according to two other respondents, in determining job opportunities outside of academia. One respondent mentioned that the "hard sciences" in general offer more such opportunities than "for postdocs in some other faculties". Another was more specific, stating that, "with the biotech industry booming, we won't be choosing to be postdocs associated with universities for much longer". However, according to a third respondent, there is a lack of private-sector work in South Africa in the field of molecular biology and biotechnology, "especially if one does not meet the correct demographic" (also see Section 5.4).

### 5.6.2  Challenges for career progression, and career preparation

The questionnaire asked respondents about the greatest, single challenge for their personal career progression. The option that was by far most frequently selected (by 28% of the respondents) was "discrimination or bias". Clustered in the range on 13–17% are competition for funding, respondents' desire to stay in academia, and lack of appropriate networks or connections. The remaining respondents are distributed in relatively small (<10) percentages across other challenges. In this regard, it should be noted that an option to specify alternative, greatest challenges to the ten listed in the questionnaire was provided, of which 47 respondents availed themselves. Those responses were analysed and added to the existing response options, viz. "lack of (permanent) employment opportunities"; "lack of opportunities (in field), incl. lack of funding", "lack of support for career development", "work-role conflict", and "personal health".

Table 5.12. Respondents' perception of the greatest[24] challenge for their personal career progression

|  | N | % |
|---|---|---|
| Discrimination / bias | 115 | 28% |
| Competition for funding | 68 | 17% |
| My desire to stay in academia | 54 | 13% |
| Lack of appropriate networks / connections | 51 | 13% |
| Unwillingness / inability to relocate | 24 | 6% |
| My desire to leave academia | 19 | 5% |
| Unwillingness / inability to sacrifice personal time / time with family | 17 | 4% |
| The economic impact of COVID-19 | 16 | 4% |

---

[24] Respondents could only select one option.

| Lack of relevant skills | 14 | 3% |
|---|---|---|
| Lack of (permanent) employment opportunities | 7 | 2% |
| Language skills | 5 | 1% |
| Lack of opportunities (in field), incl. lack of funding | 4 | 1% |
| Lack of support for career development | 3 | 1% |
| Work-role conflict | 2 | 0% |
| Personal health | 2 | 0% |
| Other[25] | 4 | 1% |
| **Total[26]** | **405** | **100%** |

One of these challenges for postdocs' personal career progression, namely lack of relevant skills, was further investigated with other questionnaire items. In addition, the qualitative data provided more insights on this challenge, as well as on some of others in Table . In the remainder of this section on the career preparation and challenges for career progression, some of the challenges are used to structure these additional results according to topic.

### 5.6.2.1  Discrimination or bias

The fact that "discrimination or bias" was selected by the highest percentage of respondents as the greatest (single) challenge for their personal career progression, was somewhat unexpected. It was higher than the percentages reported for all the indicators in Section 5.4, which deals with respondents' perceptions, experiences or observations of regarding discrimination based on gender and race or ethnic background. Based on our analysis of the qualitative data, which included many references to discrimination or bias against non-South Africans, we surmise that respondents referred here not only to discrimination based on gender, race or ethnic background, but also to discrimination or bias based on nationality. Those comments are analysed in more detail in Section 6.4.3, on respondents' reasons to leave South Africa, as they mostly relate to that topic.

### 5.6.2.2  Lack of funding

Based on what one respondent had observed over the previous five years of their postdoc, "funding for your position" is one of the "critical factors". Other respondents' experiences reflect this observation. Some simply stated, "there is a lack of funding in South Africa", or "adequate funding is a huge problem". Providing more detail, one respondent identified "the lack of grant opportunities [for research funding] that postdocs can apply for" as one of the "major stumbling blocks [that] remain in advancing our careers". Similarly, for another there was, specifically in academia, "a lack of support and funding for postdocs to make or attempt to make the transition from postdoc to researcher". The qualitative data further show that factors perceived as influencing the availability of funding opportunities include the following:

- field (e.g., they are "not readily available" "for postdocs who are clinically trained", such as a specialist with PhD);
- competition (especially when applying "in a personal capacity as a postdoc");
- ageism ("after 35, opportunities dwindle to very few grants"); and

---

[25] "A failing ANC government"; "competition: public health research is currently a very competitive field"; "international conferences"; and "possible relocation internationally".
[26] Excludes 24 respondents who selected the "none" response option.

- nationality ("the lack of funding opportunities for foreigners in research in South Africa is a significant drawback").

The result of a lack of funding opportunities for foreigners, according to this respondent, "exposes scientists to exploitation and makes it difficult to create an independent scientific identity". On the other hand, a respondent with a South American nationality, who compared their postdoc in South Africa to their country, described it as "great", partly because there are "much more funding opportunities in my field".

### 5.6.2.3 The impact of coronavirus disease of 2019

In their comments, four respondents mentioned effects of COVID-19 on their postdoc studies. One simply stated that "opportunities were lost during the pandemic". For another, "COVID-19 is still affecting the access to laboratories and attendance of international conferences which will aid collaborations", which makes it an even more "daunting task" for a "fresh postdoctoral student" to publish a required minimum of "three articles in a peer-reviewed journals". For two respondents, COVID-19 led to a lack of integration into the host institution. The first one mentioned that COVID-19 "created a lot of issues around engagement and separated" the postdocs from other staff members "for a long time". When they "finally could connect on campus, it was a bit awkward". Moreover, as "very little interest" from those staff members "was shown towards the postdocs (three) in both years", the postdocs "had to find their own feet and assist each other, which was isolating". Similarly, the second respondent ascribed to COVID-19 some of their experiences of a lack of "structure to introduce a new postdoc to the university and basic functions, like the library and finance team". Consequently, their "level of understanding of the support services at the university was completely dependent on" their supervisor.

### 5.6.2.4 Skills development

Only 14 (or 3%) of the respondents perceived a lack of relevant skills as the greatest challenge for their personal career progression. The skills those respondents perceived as lacking, are listed in Table below.

**Table 5.13. Relevant skills, the lack of which 14 respondents perceived as the greatest challenge for their personal career progression**

|  | N |
|---|---|
| People management / leadership skills | 5 |
| Statistical skills | 5 |
| Interpersonal / communication skills | 4 |
| Specific experimental techniques | 3 |
| Computational skills | 2 |
| Time- or self-management skills | 2 |

Respondents were also requested to assess the postdoc training they had received up to time of the survey, focusing on three aspects. As Figure below shows, the respondents' assessment was mostly positive regarding the extent to which their postdoc training had prepared them for achieving their career goals (very well, or at least well, according to 69%).

Figure 5.32. Respondents' assessment of three aspects of the postdoc training they had received up to the time of the survey, arranged from highest to lowest percentages in the category "well".

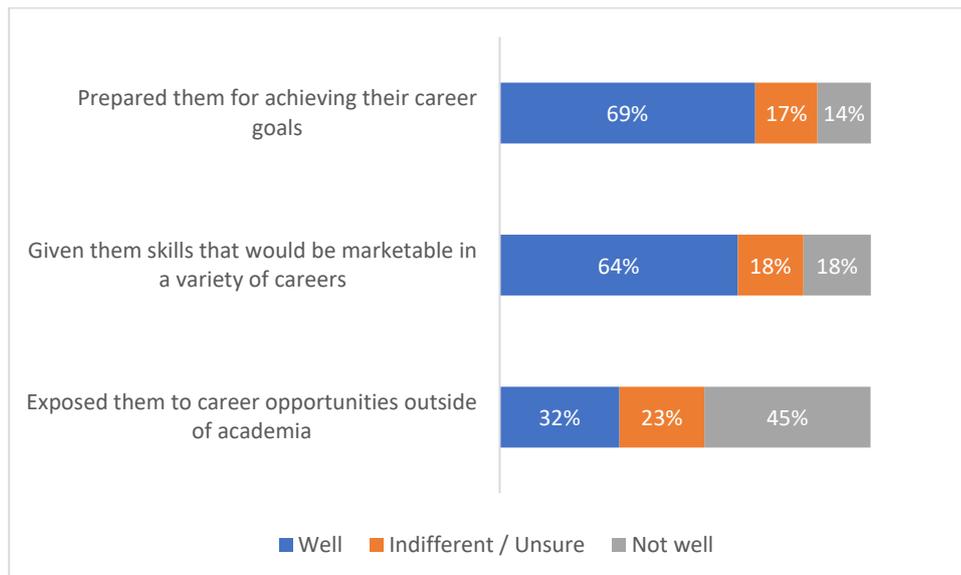

n=411–415 (excludes 11–16 respondents who indicated that they had not been in a postdoc position long enough to judge)

In the qualitative data, skills developed during the postdoc position were assessed by three respondents in positive terms. One was "extremely satisfied by the knowledge I gained during my postdoc tenure", while another (a non-South African national) reported that their "postdoctoral experience in South Africa has enhanced" specifically their "research skills and networking abilities". A third found that their postdoc

> was just a minor extension of my PhD skill set. I did continue work post-PhD in my field of interest, but I also branched out with other work. I also got the opportunity to supervise students which was a great learning curve.

Conversely, two respondents were critical of the extent to which postdoc training had prepared them for achieving even job security, by stating that "a postdoc position does not provide job security"; and while it "should prepare one with guaranteed hope of being hired, […] the reverse has been the case. There is no hope".

The need for greater job security in general emerged strongly in the qualitative data. What one respondent referred to as "the precarious situation many postdocs face", results from contracts that are renewed "once a year", according to one respondent, with "no guarantee that that the university will renew my contract once I have completed the first year", according to another; or simply, by "not receiving long-term commitments", according to a third. The perceptions are further that the lack of job security does not reflect that "it takes a lot of time, money and sacrifice to get to this point", and it does not take into account performance. "I am currently intending to publish about 10 peer reviewed articles this year, yet I know that such hard work will never translate into gaining a full-time position", one respondent noted.

As to the results of "this reality", the respondent stated that it "can be very discouraging and entrap one into looking for job security, as opposed to doing actual research". This "debilitating" effect of job insecurity was also noted by a second respondent, who suggested that "we need to rethink the postdoc position and create a path to integrate us into research positions in universities or research institutions". A third respondent highlighted the "substantial pressure" of job insecurity, specifically on "postdocs that have families and responsibilities". Job insecurity also becomes more pronounced in the last year of postdocs' fellowship. One respondent felt that, at that stage, "universities should not let us go, especially when we have spent more than two years with them". To address job insecurity, respondents recommended that "the postdoc programme needs to have well defined career path", "stability and a structured promotion framework".

Returning to the results in Figure above, specifically how well postdoc training has prepared respondents for achieving their career goals, qualitative data on the perceptions and experiences of respondents with regard to how well they qualify specifically for a permanent academic position are relevant. One noted that the experience they gained during their postdoc is "not really considered as working experience in the job market since postdocs are regarded as students and not employees". Similarly, another questioned "the point of a postdoc". That respondent had developed teaching and research skills during the previous "ten years as a tutor, demonstrator, teaching assistant, guest lecturer and researcher", but was told after an interview for a lecturer position, "we were impressed with your qualifications, but we are looking for skills and experience". For a third respondent, a "generalist" in their field, the issue was that "extensive experience in the chosen field" is required, as positions seem to become "increasingly specialised".

A fourth respondent, with "two years full-time teaching experience", was shortlisted for only "one permanent academic job out of the six [they had] applied for". They ascribed this to their lack of supervision experience and having no single-authored publications (a requirement in their field). However, they argue that supervision experience "is an on-the-job fulfilment. In other words, when I receive full-time employment, I will be offered training in supervision and not before, so why should I seek it out and potentially do a bad job?". With regard to the latter, the respondent is "trying to achieve this requirement during my postdoc". The last of the five respondents agreed that lack of supervision experience (but also lack of teaching experience) means postdocs "aren't desirable candidates" for lecturer positions. However, they argue that postdocs "are restricted in how much time we can commit to these activities". While activities such as honours supervision are expected of them, but do not "count towards promotions or as 'supervision'". Contrary to the experiences and perceptions narrated above, one respondent was seemingly "overqualified" for an academic position:

> I recently applied for a permanent position in our department (junior lecture position). After a very long time, it was given to someone with only an honours degree. Very shocking and very disappointed that I'm seen as overqualified. How am I supposed to get a permanent position?

Figure above further shows that 64% of respondents felt that the postdoc training they had received gave them skills that would be marketable in a variety of careers. However, a notable deviation in Figure is that only 32% of respondents felt that they have been exposed "very well" or "well" to career opportunities outside of academia. These results align with those produced by a more general question on how well prepared for non-academic career opportunities the respondents considered themselves. Figure below shows that approximately a third (34%) of respondents felt well prepared for such career opportunities, while around half (49%) considered themselves somewhat prepared, and 17% felt not at all prepared for non-academic career opportunities.

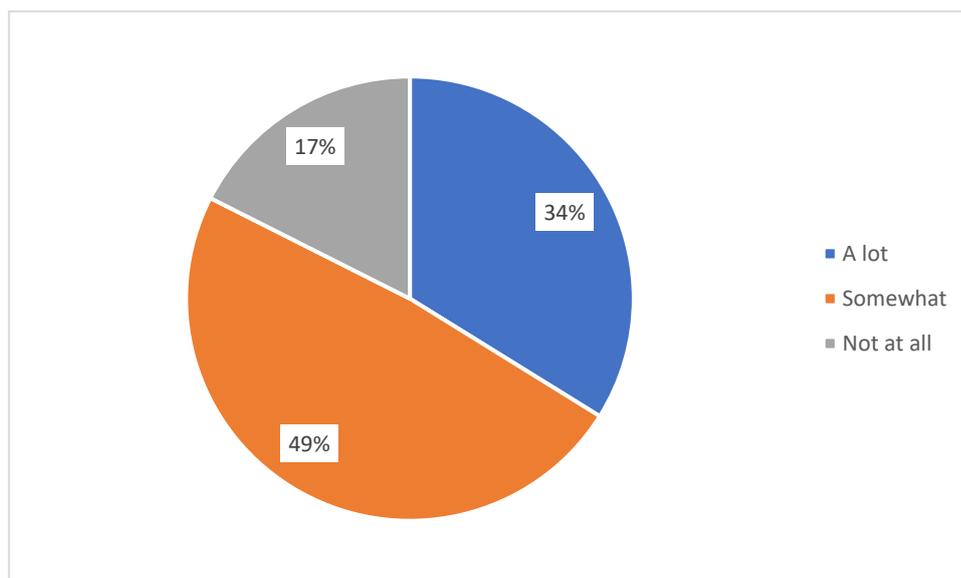

Figure 5.33. How well prepared for non-academic career opportunities respondents consider themselves

n=428

The qualitative data provide some insights into respondents' perception that they are ill-prepared for non-academic career opportunities. One, who had "applied unsuccessfully for several jobs outside of academia" realised that "industry does not view my years of research as relevant work experience", although he had "done a PhD and published several papers, [is] an experienced researcher and […] highly skilled". According to another, "We aren't trained for industry. Even with a PhD, industry positions will be limited to entry/junior level". A third respondent agreed that "non-academic jobs are either very junior or require management experience that I don't have and don't have an opportunity to gain". Contrary to these views is a respondent's suggestion that postdocs' "overqualification" (for industry) may be a cause, while another disagreed with even the notion of offering training to postdocs that is "essentially geared towards preparing oneself for a career **outside** academia". To support their position, the respondent asks the rhetorical question, "Why would someone do a postdoc if they were not considering an academic position? Trainings should be targeted solely on improving the research skills of those conducting postdocs, or not at all".

The qualitative data provided some suggestions to better prepare postdocs for non-academic career opportunities, and these ranged from creating "opportunities for connections to industry, to meet postdocs", to host institutions "arrang[ing] for industrial placement for postdocs to enable them to network and consider careers outside of academia". These suggestions align with an observation that "fear of academics to integrate into industry" plays a role in the lack of employment opportunities for postdocs in that sector. Finally, one respondent mentioned that "training should allow for registration with" the Health Professions Council of South Africa.

### 5.6.2.5   Work–role conflict

In two cases, postdocs' work in laboratories, including training, left "insufficient time for publication" (and for "funding-application writing"). One recommended that "the postdocs who work in a wet lab daily with massive lab activities should have one or two days off-lab, only to focus on paper and grant writing, supervising and classes". It was further noted by the other respondent that academic staff who were not "physically involved in research anymore" had "very little understanding of this need for time" for research.

As reported in Section 5.2.2.2, and especially in the results of an analysis of the relevant qualitative data, the majority of postdocs are of the opinion that postdocs should be allowed to teach and/or supervise students, as experience and development of skills in these roles would benefit those of them who want to pursue an academic career. However, at least two respondents noted that teaching responsibilities leave very little time for (or "interferes with") research and/or supervision. Contrary to this view, another respondent felt that postdocs "are restricted in how much time we can commit to" supervision and teaching. Therefore, they "aren't desirable candidates" for the "lecturer roles" ("the post of 'lecturer' or 'senior lecturer'") that "the majority of postdocs target", as these "require supervision and teaching experience". In more general terms, the "great expectation that postdocs should be working 24/7 to prove themselves" was mentioned by one respondent. If they do not, then they "are not deemed worthy to be a researcher" and "are severely criticised".

### 5.6.2.6 *Personal health*

Although only two respondents selected "personal health" as the greatest challenge for their personal career progression, in the qualitative data some respondents explained how their postdoc position caused personal-health issues. One respondent felt that both their "mental and physical health has markedly deteriorated", while another reported "a host [of] mental problems", including "burnout, depression, [and an] ED [eating disorder] relapse".

Such personal-health issues are ascribed to various features of their postdoc positions, some of which have already been highlighted in earlier sections. One respondent mentioned a number of features:

> I am feeling exploited and deeply unhappy. Having done a PhD and published several papers, I am an experienced researcher and am highly skilled, yet I am treated like a student, enjoy no employee benefits, and have no career-growth opportunities. I am burdened by other duties forced onto me by my supervisor, which I am neither acknowledged nor compensated for, and yet I'm still expected to produce (at a minimum) two first-authored publications per year […] I have applied unsuccessfully for several jobs outside of academia.

Another respondent also felt that one of these features, namely being treated like a student, "in most cases leads to depression and lack of hope for the future". Feeling burdened by duties and expectations aligned with another observation, namely that "having access to only a capped number of leave days, and not being able to have more liberty in my daily, weekly or monthly schedule, has influenced my mental health substantially". A factor that specifically affects foreign nationals "emotionally and psychologically" is that they "are not treated well regarding the immigration issue" (see Section 5.7.2 below for more detail), and this prevents them from focussing "on their assignment".

But the qualitative data, including the quote above, focused mostly on the negative effects of job insecurity on respondents' personal health. These effects may be direct or indirect. In terms of the former, job insecurity of the postdoc position itself caused anxiety. In an extreme example, a "contract gets renewed once a year", and this has been the case three consecutive postdocs (contracts). The resulting "uncertainty causes […] great anxiety". In addition to postdoc positions being "unstable", there is also "a lack of surety of permanent positions following" the postdoc, with the result that "my peers and I feel anxiety and pressure daily". Lack of employment prospects after the postdoc position "depresses greatly", another respondent stated.

As to the indirect effect of a lack of job security on respondents' personal health, the strategies they follow in an attempt to address the issue play an important role. One respondent, who feels there are "no prospects of permanent employment", has "been doing non-academic freelance work in the meantime, keeping one foot in the door in terms of future employment prospects. This 'moonlighting' has placed undue stress on my personal life and health". Another ascribed their mental-health problems to the fact that,

> at the start of my postdoc, I tried extra hard to be indispensable, because that is your reality in a market where so many people are looking for the same position. I put in extra work: worked on other projects not on my contract, supervised many students, lectured, and presented workshops.

For the respondent quoted above, the direct result of these experiences, and their accompanying mental-health issues, were that "my productivity reduced […] making me more anxious about my situation, until I became completely paralyzed". Similarly, another respondent spoke about their postdoc position affecting their "mental health to the point where I could no longer function".

One respondent cited a lack of "support for postdocs specifically" in regard to mental and physical health. Another suggested "counselling and/or therapy services", although they recognised the need for "a permanent position and greater job security" as the main, underlying issue. Relocating to another university for that purpose, would involve leaving "family and pets behind" and "being human, that will probably make my anxiety skyrocket and lead to a bunch of other mental issues".

Ultimately, in the case of all three of these respondents, a recognition that (as one stated) "the job is just not good for my health" led them to terminate their postdoc position, "for the sake of my health and happiness", and/or to cease pursuing an academic career. However, that decision also leads to regret, e.g., "that I chose to pursue a career in academia, which once was my dream job".

## 5.7 Migration from and to South Africa

### 5.7.1 Plans to leave South Africa

Approximately a third (34%) of respondents planned to leave South Africa upon completion of the postdoc position they held in 2022, while the same percentage did not, and the remaining 32% were unsure (Figure ).

**Figure 5.34. Whether respondents planned to leave South Africa upon completion of the postdoc position they held in 2022**

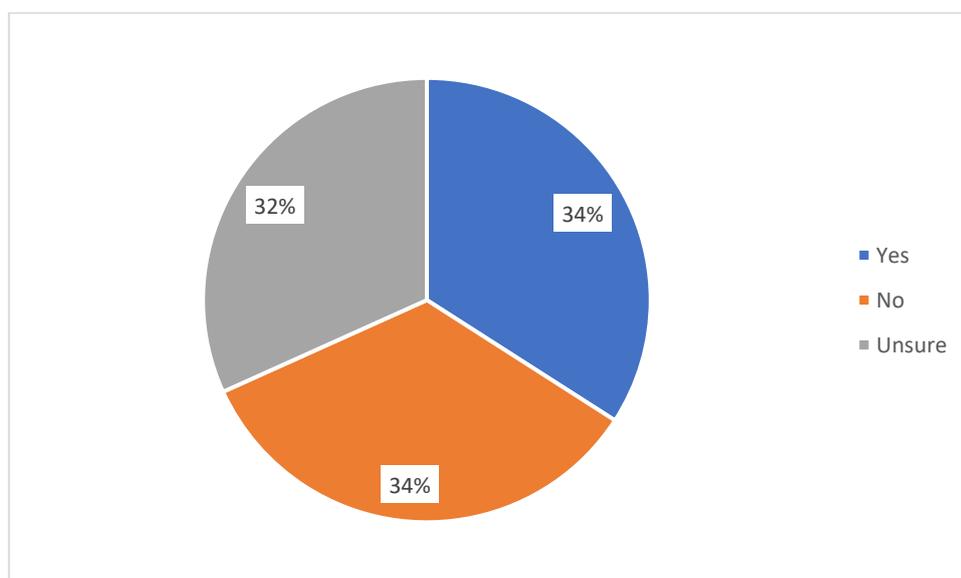

n=428

As to the countries outside of South Africa in which respondents planned to pursue their career, approximately half identified either the United States (26%) or Canada (23%), while a further 11% planned to pursue their career in the United Kingdom.

Table 5.14. Country outside of South Africa in which respondents planned to pursue their career

|  | N | % |
|---|---|---|
| United States | 40 | 26% |
| Canada | 36 | 23% |
| United Kingdom | 17 | 11% |
| Netherlands | 7 | 5% |
| Germany | 6 | 4% |
| Nigeria | 4 | 3% |
| Kenya | 3 | 2% |
| Switzerland | 3 | 2% |
| United Arab Emirates | 3 | 2% |
| Zimbabwe | 3 | 2% |
| Australia | 3 | 2% |
| Cameroon | 2 | 1% |
| Ethiopia | 2 | 1% |
| India | 2 | 1% |
| Saudi Arabia | 2 | 1% |
| Spain | 2 | 1% |
| Uganda | 2 | 1% |
| Other[27] | 17 | 11% |
| **Total** | **154** | **100%** |

When we categorised all the countries selected by world region, we found that the majority –exactly half of the respondents – planned to pursue their career in North America. Close to one-fifth (18%) selected a European country (most often the Netherlands or Germany, as Table above shows) and another 11%, the United Kingdom. A similar percentage (12%) planned to pursue their career in an African country, most frequently in Nigeria, Kenya or Zimbabwe. Smaller percentages (<5%) of respondents planned to do so in the Middle East (mostly in the United Arab Emirates or Saudi Arabia), Asia (especially India), Oceania (mostly Australia) and in South America.

---

[27] On respondent each selected the following 17 countries: Albania; Belgium; Botswana; Brazil; Burkina Faso; Chile; China; Czech Republic; Denmark; Finland; France; Iran; Portugal; Romania; Rwanda; Sweden; and the Bahamas.

Figure 5.35. Region outside of South Africa in which respondents planned to pursue their career

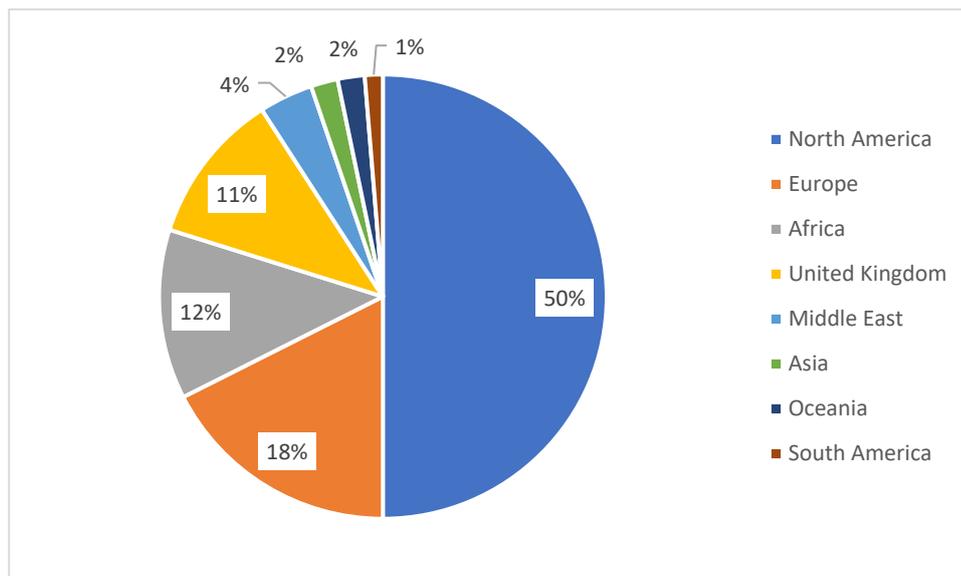

n=154 (excluding respondents who planned to pursue their career in South Africa)

As to those respondents who planned to remain in South Africa, the qualitative data provided some insights on their reasons, albeit with some caveats. One respondent "would like to stay and work in South Africa" "if given the opportunity". Another "would only consider leaving South Africa as an absolute last resort and only if I can first secure employment. I would first seek employment in South Africa, however, as "this is where I prefer to live, work and play". A third respondent mentioned two positive aspects of remaining in the country:

> I can see the application of my studies much faster in South Africa. And since I am in the African continent, I can find international collaborators easier than national collaborators and that helps my career to expend in the international platform.

However, that respondent also felt they "cannot leave" partly because their "spouse has a permanent job here" [but] "the electricity crises, crime, and the instability of South Africa always make me question why I still live here". Another respondent "felt a great sense of insecurity during the country's vandalism in 2021, beside the theft and killings across places in Gauteng", although they did not mention that as a reason to leave South Africa.

The remaining questionnaire items concerning respondents' migration plans were posed only to those respondents who planned to leave South Africa upon completion of the postdoc position they held in 2022 (n=146). The most frequently-selected option (by approximately half the respondents, or 51% – see Table below) was "better or more job opportunities elsewhere". As one respondent later commented, "I am a job seeker, so where the opportunity comes, I will go". A second explained that they "didn't stay back in South Africa because I was already employed as an academic staff in my home country (Nigeria)". A third felt that "opportunities for career researchers, even those who manage to secure their own funding, are extremely limited in South Africa. I am being forced to seek hosting internationally, even though I am committed to research activities in southern Africa".

Table 5.15. Reasons[28] why respondents who planned to leave South Africa upon completion of the postdoc position they held in 2022 (n=146), intended to do so

|  | N | % yes |
|---|---|---|
| Better/more job opportunities elsewhere | 75 | 51% |
| Prefer to stay but cannot because of immigration rules / visa issues | 52 | 36% |
| Financial reasons | 37 | 25% |
| Variety of experience / exposure to new people and ideas | 35 | 24% |
| Lack of support (e.g., funding) for R&D in South Africa | 34 | 23% |
| Personal reasons | 27 | 18% |
| Better facilities/technology/researchers elsewhere | 23 | 16% |
| Other reasons | 31 | 21% |

Table above further shows that more than a third (36%) of the respondents preferred to stay in South Africa, but could not, because of immigration rules or visa issues. A quarter of the respondents selected financial reasons (25%) and 18%, unspecified personal reasons. Slightly less than a quarter (24%) were seeking variety of experience, or exposure to new people and ideas. A similar percentage – still almost a quarter (23%) – planned to leave South Africa because of lack of support (e.g., funding) for R&D in South Africa, and another 16% cited better facilities, technology and/or researchers elsewhere.

A fifth of (or 31) of respondents indicated that reasons other than those listed, informed their intention to leave South Africa. Of those, 29 provided reasons[29] that were further coded. Four were added as "visa issues" to the second, existing reason (concerning immigration rules) in Table above. The remaining other reasons (specified by 25 respondents) for planning to leave South Africa, are summarised in Table below. Nine mentioned xenophobia, including discrimination against or hostility towards non-South Africans. Another six referred to lack of opportunities (mostly job opportunities) for non-South Africans, followed by three who cited lack of opportunities because of their race (mostly specified as white).

Table 5.16. Other reasons why respondents who planned to leave South Africa upon completion of the postdoc position they held in 2022, intended to do so

|  | N |
|---|---|
| Xenophobia (discrimination against / hostility towards non-South Africans) | 9 |
| Lack of (job) opportunities for non-South Africans | 6 |
| Lack of (job/funding) opportunities because of race | 3 |
| Other reasons[30] | 3 |
| **Total** | **25** |

As the questionnaire generated a large amount of qualitative data on some of the issues in these two tables, the results of an analysis of these data are presented in the following sub-sections. However, it should be noted that these data did not always pertain directly to these issues as reasons for planning to leave South Africa.

---

[28] Respondents could select all options that apply.
[29] Excludes 5 respondents who responded to the questionnaire administered by UP, which omitted the option to specify other reasons, and one respondent who stated, "I prefer to stay but I have no job after completion of my postdoc after 4 years".
[30] "To pursue another postdoc abroad"; "my field is understudied in South Africa"; and "In my observation of others it seems an academic career in South Africa would be stressful".

### 5.7.2 Immigration rules

Immigration rules, policies or law were variously described by respondents as being "unfavourable", "harsh", or "unfriendly" from the perspective of "foreign postdoc fellows". Immigration law was considered by another respondent as "one major problem with the postdoc programme in South Africa", and for another, immigration is one of only two aspects of the postdoc that are "challenging and discouraging" (the other being remuneration). Not only is immigration law "affecting postdocs", mostly by preventing them from obtaining a permanent position, but "immigration and rising insecurity" are noted as "major reasons for considering relocation upon the completion of a postdoc". As another respondent explained, immigration policies lead many postdocs to plan "not […] to settle in South Africa to consolidate and build their career profiles". Rather, they "opt to go to developed countries where their skills and potential could be given the enabling environment to thrive, consolidate and grow". Similarly, unfavourable immigration rules were viewed as a signal that "postdocs are not valued in South Africa". Taking a broader view, immigration law is also "discouraging a lot of PhD holders from coming for their postdoc in South Africa", according to one of these respondents.

Three respondents perceived recent changes in terms of immigration rules, namely they "have not been favourable in recent years" and "are becoming increasingly harsh", or that there has been a "recent manipulation of visa politics to keep foreigners away from South Africa". Another referred to "the new rule regarding foreign researchers", which has "greatly changed [their unspecified] plan". Similarly, a fourth respondent explained that "the new immigration policies" was one reason why they "don't have a choice but to leave" South Africa, even though the decision "has been really difficult" and "it is tragic for us as members of the South African society and for the country after investing many resources in training us".

Two other respondents were more specific regarding the changes in immigration policy that impacted them. One noted a recent change with regard to the critical-skills visa, which means they "need to jump through all sorts of hoops to get" that visa, even though they obtained a doctoral degree from a South African university. These changes, the respondent opined, "have made South Africa much less attractive for non-residents". The second respondent noted the recent "cancelling", by the Department of Home Affairs, of the permanent-residency permit for highly skilled and qualified graduates, which, in their view,

> has led to the mass exodus of South African-trained doctoral graduates who were hanging around doing postdoctoral research, in a hope of job consideration. They have now [exported] their skills to other countries that place a premium on their skills, in exchange for stable immigration statuses and better job promises. If these policies are sustained, it is only a matter of time that the country will not only lose relevant skilful people but become unattractive to prospective graduate students who could become assets for the future. A change of policy is urgently needed.

Another respondent agreed that policy change is needed, specifically less strict immigration rules for postdocs, as "there is need to make good use of postdocs as […] an important resource. The knowledge, the energy, the zeal must be put to good use, rather than let them fly away with the gained knowledge and skills".

### 5.7.3 Visa-related issues

Three of the four respondents who highlighted visa issues as a reason why they planned to leave South Africa upon completion of the postdoc position, referred to the fact that the visa "doesn't allow [an] accompanying spouse to work in South Africa"; or (more frequently) that family members are prevented from visiting the postdoc. According to one respondent, their family was refused a visiting visa twice [and] "the reason the Department of Home Affairs gave, was that I had limited funds to take care of one wife and a three-year old son. They indicated that my family would become a public charge.

This issue led another respondent to conclude that "South Africa is hostile to [a] family being together". In the comments at the end of the questionnaire, this visa constraint was raised again, although not necessarily as a reason for planning to leave South Africa. According to one respondent, the fact that they "cannot bring my family to South Africa because of visa issues" is "the main challenge with a South African postdoc", which "affects [their] ability to focus and leads to high turnover". For another, "it is disheartening when some postdocs work for months away from their families because of visa constraints that impede their families from visiting".

Another visa-related issue that emerged as a reason why respondents planned to leave South Africa, is delays in obtaining a visa. Again, this is an issue that emerged quite strongly in the qualitative data, where lengthy delays in the approval of visas were often described in detail. The importance of the issue was highlighted by a respondent who "will be very glad if this could [be covered] in the outcome of this research survey", noting that "most international postdoctoral fellows" at their host institution "were delayed with the outcome of their renewal visa for 2021 and 2022".

Specific home countries that were mentioned in regard to delays in the processing of South African visas are Ghana and Nigeria. As to the extent of the delays, the Ghanaian respondent reported that "the visa-processing period goes beyond eight weeks". A second respondent had applied for their visa in November 2022, and by the early-February 2023, there was still "no hope of getting it". A third had already applied in June 2022, but by early-February 2023, it had "not been released". Other respondents reported visa processes that have taken more than a year, "with no conclusion" or "hope of getting it".

As to the effects of delays in the processing of visa applications, a respondent mentioned how it "makes it difficult for applicants to meet institutions' deadlines". In fact, one respondent reported that "many have failed to take up their postdoc offers owing to delayed processing of appropriate visa"; and a second "almost lost [their] postdoc place". A third respondent "was not registered nor paid for postdoc research from January 2020 to September 2021 when I was able to get my first-year visa" and again, had not been "able to register in 2023, nor [had they] "received pay, because [their] visa application […] has not been released".

The same respondent noted that, in 2022, they had "produced seven DHET-accredited journal articles for [a South African university] and [were] recommended for a senior postdoc position", which brings us to reported effects of visa delays on postdocs' research output. "I couldn't complete my research due to the failure of the Department of Home Affairs to renew my permit. So, an unfortunate and wasted effort", one respondent noted. For another, the effect they had observed was more indirect: because "some fellows were still outside of South Africa", this "hindered their research from being conducted" in the country. In response to this situation, some of them "focused only on review papers instead of scientific data, i.e., lab, face-to-face interviews, quantitative data, etc.". However, the respondent felt that such delays impact negatively on South Africa's scientific-publication output and its "ranking globally". Similarly, another opined that the issue "needs to be looked into for South Africa to continue having good research base locally and internationally", warning that "a lot of foreign PhD graduates from South Africa universities are not ready to pursue their postdoc here in South Africa because of these reasons".

Recommendations made by two of the respondents to address the issue of visa delays were that the relevant "authority at the Department of Home Affairs in South Africa" should "intervene and allow international postdoctoral fellows with the approved visas in due time". More specifically, another felt that "the Department of Home Affairs needs to be better informed by the DHET in relation to the importance of expediting visa processing for postdocs".

In addition to the specific issue of delays in the processing of visa applications, a number of respondents described more general terms how difficult it is to obtain a visa, and the "challenging" or "cumbersome" nature of the application processes. In the words of one respondent, "a South African visa is now like gold moulded in heaven", while another viewed "permit or visa approval" as "the greatest challenge with postdoctoral training in South Africa for foreign students". However, according to another respondent, it is South African embassies who experience the processing of a visa for postdocs as a "major challenge". One simply stated that the Department of Home Affairs "is frustrating me".

For two respondents, visa issues signal a lack of dignity and respect:

- Postdoctoral fellows [...] should be treated with dignity in South Africa. It frustrates a researcher after he is faced with pressures of research, he/she is still faced with visa issues.
- Academics are not burdens. They contribute one way or the other to research and development. Academics should be given [a] visa and treated with respect.

The process of renewal of one's visa or immigration permit was also described a "onerous" and is further complicated by host institutions' deadlines referred to earlier. One respondent sketches the situation as follows:

> if my contract ends from, say, April to March, I have to first register for the new year in January and then apply for renewal again at least three months before the current contract lapses. This means I need to register in January and apply for renewal in January. IF, the application is approved, I would have to register again, (I guess), in April.

It is therefore not surprising that the policy of having to renew one's visa or permit every year was criticised by respondents. According to one, it creates "a precarious situation many postdocs face", with detrimental effects on their research:

> as a foreigner, I have to constantly think about [...] applying for an immigration permit at the end of each year in order for me to register the next year [...]. How can a postdoc focus on their research if there is a cloud of uncertainty hanging over their immigration status and their renewal status with the university?

Another respondent focused on the effects in terms of social integration:

> The nature of postdoc positions require that fellows relocate from their respective countries. This relocation involves fellows being uprooted from strong social ties and community support from their origin country, which unfortunately are not available in a new country like South Africa.

The recommendation of that respondent was "to increase the minimum duration of postdoc positions to two years and encourage special migration status for fellows willing to relocate to South Africa, whether for a long period or permanently". The other respondent felt that "universities should fight for immigration renewal when it relates to postdocs", but similar to the recommendation above, suggested that universities

> should try to have postdocs for at least two years, so that international students can apply once for two years and, if their renewal fails due to a failure to meet the targets, by all means the university can inform the Department of Home Affairs to revoke the other year.

Universities were also called upon to support foreign-national postdocs that are "placed in impossible situations by the Department of Home Affairs". "Some support" in this regard "would not be misplaced", a respondent remarked, as those postdocs "are frequently the cash cows of universities".

Some recommendations on a national level were made. According to one respondent, postdocs "should be given a work visa and not visitor's visa", while a second suggested that "postdocs should pay taxes to qualify them for a work permit". For a third respondent, what needs to be addressed is one of the "central weaknesses" of the postdoc position, namely its "ambiguity", as being "neither students nor employees […] has **direct** consequences on our visa requirements and ability to obtain permanent residency status". This view aligns with a fourth respondent's suggestion "that postdoctoral positions in South Africa be […] given proper employment status that the Department of Home Affairs should recognise".

### 5.7.4 Lack of (job) opportunities for non-South Africans[31] and perceptions of xenophobia

According to Table above, only six respondents cited a lack of (mainly job) opportunities in South Africa for non-South Africans as a reason why they planned to leave South Africa upon completion of the postdoc. However, many more commented on this issue in the qualitative data. Three stated that

- there are "no career opportunities for non-South African citizens";
- "after a postdoc, there's no future for foreigners in South Africa"; and
- "affirmative hiring is making it impossible for foreigners to stay in academia".

Being non-South African was also perceived as "the main reason I haven't gotten a permanent position", or why, for four other respondents it was "difficult" to obtain jobs in the country, specifically "in the academic system". Another respondent was slightly more optimistic, stating that "as a non-South African, opportunities seem to be tricky to come by". Thus, even though "the idea of staying in academia is all set, […] the decision to stay in or leave South Africa hinges on openings that avail themselves".

Most respondents were also at pains to highlight that the lack of job opportunities for foreign nationals pertained despite the fact that they have, for example, permanent residency; "have been working hard towards improving research in South Africa", "put in effort into [their] research career at a South African institution"; and worked "extremely hard […] to make it […] "to do our job". Without permanent residency, the situation is more dire, according to one respondent:

> you will not even be considered for permanent appointment, irrespective of having over 60 DHET-accredited journal papers, co-supervising two PhDs [at a South African university] and supervising five master's students (Gambia and Nigeria) as well as three doctoral examinations [for two South African universities] and two master's students [for two South African universities].

Thus, as one respondent noted, "it is difficult for foreign researchers to get hired in South Africa, even when you are overqualified for the position that you apply for". But others noted that they were also more qualified and experienced for the positions they applied for, than their South Africans peers who obtained them, were:

- "people with lesser qualifications and profile get the job ahead of me because they are South Africans";
- "positions and opportunities [for] which I qualify for [were] filled by people with a lot lower qualifications and experience";
- "when opportunities come, you are being told you are not South African so you cannot have such opportunities; then they bypass you to give them to a South African student you are mentoring".

---

[31] In this section, they are defined as individuals without a South African passport or permanent-residency status in the country.

- "even with four years' postdoc experience, my department can only make me a tutor while they hire South African citizens with master's degree to lecture. […] One department that I worked for […] sacked all the non-South Africans, including us who had PR [permanent residency], and they hired honours students to lecture".

"The humiliation is too much. Every day, you are reminded that you don't belong here", the last of the respondents cited above, concluded. It was interesting to note, then, that two respondents were supportive of reserving job opportunities for South Africans:

> South African citizens should be the first priority for any postdoctoral position available in any South African universities or research institutions, like in Botswana, where the Botswana citizens are prioritised for all postdoctoral and job positions in their country.

The second felt that "it's fair to pick South Africans before any other country", but argued that, "when excellence is involved, that should not be waved aside at the expense of country of origin". Another non-South African agreed, by stating that "unfortunately, at my university, the positions are not filled with quality. I have numerous publications, and national and international grants (including Marie Curie) but still I am not being appointed. This demotivates me most of the time".

Demotivation or discouragement was an effect of discrimination against foreign nationals that three other respondents also noted. One had stopped applying for "positions [that] remain vacant and are re-advertised again and again […] because I see it as wasted effort going forward. I direct my energy elsewhere". The second explained how

> the bias against foreign nationals in the workplace and job placements is alarming and very discouraging for me. I have little motivation to continue to do my best as I already know that I literally have no chance to be absorbed into the system in South Africa.

The third was "discouraged" – by their experiences of what they perceived to be job discrimination against foreign nationals – "from taking up master's and doctoral supervision, as well as teaching", at a South African university. Two other respondents described their experiences as "very sad […] to realise that it does not matter how hard we do our job, the opportunities are not there"; and "heart-breaking [because] my academic career seems to be at a dead-end post-postdoc!"

While at an individual level, policies restricting the employment of foreign nationals are clearly "counter-productive for the growth in the career of foreign postdoc fellows in South Africa", the results of such "nationality discrimination", in one respondent's words, have broader implications:

> South Africa is losing out hugely […] especially for highly-skilled workers such as postdocs. It is more disheartening to note that despite that South Africa has used taxpayers' money to train international doctoral students and have equipped them, they cannot be retained because the policies of higher education institutions and government around their employment are not favourable. This leads directly to the loss of skills and manpower and a huge economic loss to South Africa.

To explain the source of their experiences, some respondents referred to broader political forces, such as "the informal negligence of foreign manpower" and, as mentioned in the previous section, "the recent manipulation of visa politics to keep foreigners away from South Africa". Still on a national level, but more specific to the HE sector, were "the unspoken 'admission' policies for permanent university staff members in the country". It is noticeable that all three these respondents felt that the causes could be traced less directly to "official" policy than to practices guided by more informal norms.

On the institutional level, one respondent's perception was that institutional policies restrict "the employment of foreign nationals", with the result that those institutions "hesitate to absorb foreign postdocs" even those "with permanent residence status in South Africa". "Often universities overlook people like us", a Zimbabwean national with permanent residency stated, remembering that,

> at one time I applied for a permanent position as a lecture at some university here in South Africa and they refused to take me because I am foreigner. I pray and hope that one day someone out there will see my worth and give me an opportunity to work and do great things in my career.

It is not surprising, considering the experiences conveyed above, that terms such as "bias" and "discrimination" were used by some foreign postdocs to explain those experiences and their resulting lack of career progression. Other respondents were more critical, stating that "foreigners like me are no longer welcome for junior posts, e.g., lecturer or senior lecturer positions"; and "the South African employment/academic environment is too hostile to African migrants".

This issue brings us to perceptions of xenophobia. According one respondent, the country "is being torn apart" by "academics' xenophobia". "Academic xenophobia should be ended", another argued, as "the university exists as a site of pluralversalism and tolerance". A third respondent described their host institution as "institutionally xenophobic", for the following reasons:

> The mainstay of the university's postdoc strength is hard-working Zimbabwean, Ghanaian, Kenyan, Nigerian students [who are] prepared to work hard for very little. Yet the university overlooks their contribution and hires mediocre black South Africans as junior lecturers, even straight out of their PhD without postdoc experience. [The university] is systematically replacing talent with mediocrity.

Xenophobia – or what two respondents more aptly referred to as "Afrophobia" – was experienced by one respondent, both while studying for their doctoral degree, and after completion, as they "would not get a permanent teaching position in a South African university", even though they are a permanent resident in the country. The phenomena of xenophobia and Afrophobia were also perceived by six respondents as extending beyond job opportunities in the academic sector, as the following quotations reflect:

- "South Africa is presenting [as] a discriminating country with the loss of the spirit of 'Ubuntu'".
- "The current political climate in South Africa is very unwelcoming to foreign nationals".
- "There is a very thin section of the South African society that is not Afrophobic".
- My host institution, "like South Africa in general, is institutionally xenophobic".
- "The effect of xenophobia is still much pronounced, [both] within the university and the society, therefore one must continually watch over one's shoulder to move freely in towns, malls, and campus".
- "In South Africa, Africans are very much discriminated against. Afrophobia is very much institutionalised".

## 5.8  Summary of key findings and recommendations

The survey produced evidence that universities require to inform interventions that will ensure full harnessing of the potential of postdocs as an important group of emerging scholars in our country. In this last section, we summarise that evidence and make recommendations for the HE sector and possibly other sectors to manage their postdocs in an equitable manner.

As a background to the key findings, it is important to note that eight universities hosted 5% or more (up to 20%) of the postdocs, and together, hosted 80% of them. These universities, ordered from highest to lowest percentage of postdocs, are: UP, UJ, UCT, SU, UFS, RU, UNISA and Wits. Thus, almost two-thirds of the postdocs were hosted by a traditional university, and less than 10% by a university of technology. Secondly, the largest percentage of the postdocs work in the natural sciences (37%), while the social sciences (18%), health sciences (16%) and agriculture (11%) are relatively well-represented.

The majority (75%) of the postdocs' career trajectories aligned with its "standard" conceptualization, as they were in their first postdoc position, taken directly after their doctoral graduation. Most postdocs took the position that they held in 2022, in order to deepen their training and experience for future employment as researchers, not because it was other options were unavailable. Among the minority of "serial postdocs", most (70%) had held only one previous postdoc position, but holding up to four such previous positions was reported. On average, the serial postdocs had spent 2–3 years in a postdoc position by the end of 2021. It is notable, however, that 24% had spent 4–6 years (and one even 18 years) in a postdoc position. For close to 80%, poor job prospects had led to them holding more than one postdoc position. More than 80% of the serial postdocs had held a previous postdoc in South Africa.

A large majority (92%) of the postdocs had an MoA or other, similar contract (e.g., a letter of appointment) with their host institution, but more than two thirds (68%) of them felt unable to negotiate the terms and conditions to the extent that they would have preferred. Longer duration of postdoc contracts (at least two years) was recommended, and for various reasons, including a reduction in renewals of visitors' permits; increased financial security and access to credit; and an increase in the time available to complete one's research and produce the required outputs.

Almost 60% of postdocs earned R200 000–R299 999 per annum, but among the remainder there is much variation in income. For example, one in ten earned less than R200 000, while 6% of the postdocs earned double that amount, i.e., R400 000 or more. The income levels of postdocs differ depending on the main science domain in which they work, but the differences are not statistically significant. Perceived variation in the annual income of postdocs was strongly criticised by the postdocs, and standardisation, also of other financial incentives (e.g., for research outputs), was recommended.

Close to 60% of the postdocs perceived their income as inadequate, and almost 80% felt they were unable to save the amount they would prefer to. The lower the income that postdocs earn, the more likely they are to hold these perceptions, but especially the latter perception of the inability to save. However, even 43% of the postdocs who earn a high income (≥R300 000 p.a.) felt that their income is inadequate, and 71% of them felt they were unable to save the amount they would prefer to. A host of qualitative data support these results. Those data further convey the perception that a low rate of remuneration limits the attractiveness of postdoc positions, and that it has negative effects on the wellness of postdocs, particularly those that have families to support. It is therefore not surprising that financial reasons were cited by a quarter of the postdocs as a reason for them planning to leave South Africa upon completion of their position. Recommendations were made by the postdocs for an increase in remuneration to match not only the rising cost of living and inflation, but also postdocs' high level of qualification and work expectations.

More than a third (36%) of the postdocs did not receive their remuneration monthly, and for two-thirds of them, it posed at least some challenges to them personally. In addition, more than one in five (22%) did not receive their remuneration on time and/or experienced bureaucratic problems in this regard. These issues arise because of various factors, such as inadequate funding, delays in visa approvals, or poor communication from host institutions on regulations. Recommendations were therefore made for a steady income that is paid on a monthly basis.

Less than half (43%) of the postdocs were unequivocal in their preference for maintaining their tax-exempt status (thus remaining non-employees, with limited or no employment benefits). Having a tax-exempt status (of a student) limited their access to financial products and services, and being treated as a student in general, especially by host institutions, was a broader issue of concern raised repeatedly by the postdocs.

Some benefits were available to the postdocs, but only to a minority. The two benefits available to the largest percentage (17%) was paid vacation (or "annual") leave and paid sick leave, followed by exemption from tuition fees and a medical scheme (the remainder of the benefits were available to less than one in ten of the respondents). A lack of suitable medical insurance was a particular point of concern that emerged from the qualitative data related to a lack of employment benefits. There also seems to be variation in institutions' covering of research costs, in particular conference attendance, leading again to postdocs' recommendations for standardisation.

The majority (80%) of the postdocs were required, either formally or informally, to produce a certain number of peer-reviewed journal articles per annum (2–3, on average). These requirements were viewed by some postdocs as unrealistic for various reasons, such as work-role conflict, but also delays in feedback from co-authors (including supervisors) and/or during the peer-review process, thus partly informing recommendations for longer postdoc contracts.

The percentages of the postdocs who wanted to contribute to teaching or the supervision of master's students and doctoral students are 80% or higher, and consistently much higher than the percentages of those who were given the opportunity to do so. By far the largest difference (of 50 percentage points) between preference and practice is found with regard to doctoral supervision. The percentages of the postdocs who want to contribute to the supervision of postgraduate students is also higher than those who are allowed to at least co-supervise such students (73% on the master's level and 58% on the doctoral level). It is safe to conclude from these results that the majority of postdocs see teaching and supervision as key elements in their preparation for, ultimately, a career in academia. A large amount of qualitative data, including recommendations that postdocs should be allowed to teach and/or supervise students, support this conclusion.

It is concerning that relatively large percentages of the postdocs who contributed to "formal supervision" did not (always) receive acknowledgement for doing so: 41% in the case of master's supervision and 53% in the case of doctoral supervision did not appear on the students' supervision record and thesis. Having said this, it is important to keep in mind that universities differ in terms of their rules with regard to whether postdocs may be formally involved in the supervision of master's students, through an appointments process, or only informally. In the latter case, formal acknowledgment is not possible.

Two-thirds of the postdocs were either somewhat or completely satisfied with the position they held in 2022. However, the highest percentages of dissatisfaction were recorded for host institutions' provision of support in terms of training and career-development. The qualitative data support these and other negative assessments of host institutions, while also showing how they lead to perceptions of the institutions as uncaring or even exploitative. A number of recommendations were made by the postdocs to address their needs for various forms of support, especially from relevant administrative divisions in host institutions.

Compared to their satisfaction with their host institution, the postdocs' level of satisfaction with their supervisor was much higher, according to both the quantitative and qualitative data. Having said this, it is still noteworthy that 16% of respondents indicated some degree of dissatisfaction with the overall performance of their supervisor. Similar percentages of postdocs were dissatisfied with the amount of supervision and communication their supervisor provided. Mentoring postdocs on how to succeed as a scientist was also assessed relatively negatively, while perceptions of exploitation and similar issues were conveyed in the qualitative data, accompanied by recommendations for standards to guide the relationship between postdocs and their supervisors.

Approximately half of the respondents were sure that they were allowed to apply for funding as a PI and felt they had more than shared control with their supervisor over the writing of grant proposals. The percentage that had such a level of control over other aspects of their research was also relatively low (<60%), with the exception of the writing of papers (73%). Even then, the submission of those papers is often (perhaps justifiably so) under the control of a supervisor, but this may lead to frustrating delays in publication, according to some respondents.

It is concerning that 16% of the postdocs had themselves experienced discrimination or harassment in the postdoc position they held in 2022, and almost a quarter (24%) had observed discrimination or harassment in that position. Most concerning is that, for 28% of the postdocs, discrimination or bias was the greatest, single challenge for their personal career progression, and the percentage is much higher than for any other challenges. Such discrimination or bias was most frequently highlighted in the qualitative data by non-South Africans from the rest of the continent, specifically in relation to job opportunities. In this regard, postdocs often described South Africa, or its universities, as "xenophobic" or "Afrophobic". With regard to the other forms of discrimination, the postdocs were twice as likely to report that postdocs are discriminated against based on race or ethnic background than gender (21 and 9%, respectively). Interestingly, discrimination against whites, rather than blacks, and against males rather than females, was mentioned most often in the qualitative data.

Three-quarters of the postdocs were sure that they wanted to pursue an academic career, and almost 80% of those would most prefer to work in a role that combines research and teaching. Many (61%) of the postdocs had applied for permanent positions (in the 12-month period before the survey) that were, for almost 90%, at least somewhat related to their current research. Only 28% perceived the job market in their field to be good or excellent, while the most frequent response (45%) was that it is poor. Among those postdocs who wanted to pursue an academic career, the perception was slightly more positive (a difference of 3 percentage points).

Perceptions of poor job market were often linked to a lack of job opportunities in South Africa for non-South Africans, particularly those from other African countries, which, as already mentioned, is linked to a perception of "national" discrimination or bias as a major challenge for postdocs' personal career progression. Competition for, or a lack of funding, also emerged quite strongly as such a challenge, in both the quantitative and qualitative data. Notably, almost a quarter (23%) of the postdocs felt that lack of support (e.g., funding) for R&D in South Africa contributed to their decision to leave the country on completion of their position. For 16%, better facilities, technology and/or researchers elsewhere was one reason for them to leave.

A lack of relevant skills was a career-related challenge for very few postdocs, which is further supported by the fact that, for the majority, their postdoc training had prepared them for well for achieving their career goals and had given them skills that would be marketable in a variety of careers (69% and 64%, respectively). But regardless of their training and skills, a postdoc position does not seem to provide them with job security, especially if they are not South African nationals. The lack of job security has negative effects on postdocs' personal health, both directly and indirectly. Some of the postdocs made recommendations to better define and structure postdocs' career paths.

The postdoc position does not prepare the majority of postdocs well for a non-academic career. Only around a third (32%) of the postdocs felt that their position had exposed them well to career opportunities outside of academia, and a similar proportion (34%) felt well prepared for such career opportunities. This may be a response to the fact that the minority of postdocs hope to pursue non-academic careers, but may also create the intention, among two-thirds of respondents, to pursue a career in academia, despite the fact that (1) more than 40% of them know the job prospects in that sector are poor, and (2) for 13%, the desire to stay in academia is the greatest challenge for their personal career progression (the third-highest percentage recorded for all of the challenges). The qualitative data provided some recommendations to better prepare postdocs for non-academic career opportunities.

A third of the postdocs were sure that they will be leaving South Africa upon completion of their position, mostly (78%) for North America, Europe or the UK. For half (51%) of them, the perception of better or more job opportunities outside of the country played a role. According to more than a third (36%), immigration rules or visa issues contributed to their decision, although they preferred to stay. The qualitative data strongly support these results and show how immigration rules are preventing non-South African postdocs, especially those who are nationals of other African countries, from attaining permanent positions in the country. Visa restrictions on family members visiting the country and major delays or difficulties in obtaining a visa were identified as major challenges. Recommendations were made for the relevant government departments or universities to liaise with the DHA and/or assist postdocs in this regard, for example by extending postdocs' contracts to at least two years or redefining their tax-exempt status.

# Annexure A: Recruitment email to potential survey respondents

Dear Dr […]

You are hereby kindly invited to take part in South Africa's first national survey of Postdoctoral Research Fellows ("postdocs") in the country, including their work situation, job satisfaction, challenges, as well as career development and career intentions. If you did not hold any postdoc position in South Africa in 2022, or if you already received this invitation, please ignore this email.

The survey is of strategic importance given the need to understand the changing demands of the labour market and whether postdocs in South Africa develop the required attributes to succeed in their future careers. The survey is underwritten and funded by the National Research Foundation and executed by a project team from the DSI-NRF Centre of Excellence in Scientometrics and STI Policy (SciSTIP), hosted by Stellenbosch University.

Your participation will involve the completion of an online questionnaire by […]. **Most of the questions refer to the postdoc position you held in 2022.** We are quite aware of the demands made on individuals to complete questionnaires of this nature. For this reason, we have attempted to keep the time to complete the questionnaire as short as possible (less than 20 minutes).

The survey has been approved by the Research Ethics Committee: Social, Behavioural and Education Research at Stellenbosch University (Project ID#: 26097). Your participation in this survey is voluntary, and there are no foreseeable risks involved in your participation. You have the right to skip questions or to exit the questionnaire (by closing your browser) at any time, even if you have agreed to take part initially. The survey data you provide will be strictly anonymous and presented in the aggregate in reports or publications produced from the survey.

Entering into a prize draw for a R2 500 Takealot voucher is available to respondents. If you are interested in entering into the prize draw, you will be redirected (upon completion of the questionnaire) to a separate site where you may choose to enter your name and email address into a drawing for a chance to win. However, this information will not be linked to your responses, and will be treated as confidential.

Please refer to an information leaflet for more detail on ethics-related aspects of the survey.

In order to complete the survey, please click on the following link: […]. By doing so you are confirming that you have read and understood the explanation above and the information leaflet, and voluntarily agree to participate in this survey.

Note: this invitation is personal and non-transferable. Please DO NOT forward this email to other postdocs, as it may lead to duplication of data collection.

Yours sincerely,

Profs Heidi Prozesky and Johann Mouton

## Prof Heidi Prozesky PhD (Sociology)

Associate Professor
**Centre for Research on Evaluation, Science and Technology**
+27 21 8089464 |+27 83 6663166

Room 4032, Krotoa Building, Ryneveld Street
Stellenbosch Campus | South Africa
Research profile link | ORCID ID
www.sun.ac.za | Find us on social media

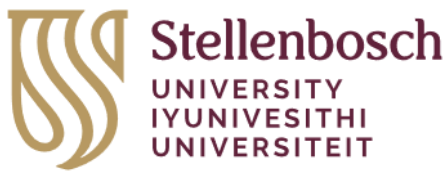

# Annexure B: Respondent information leaflet

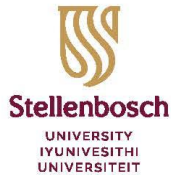 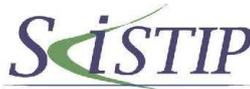

## Respondent information leaflet for a national survey of postdoctoral research fellows in South Africa

- The questionnaire does not collect identifying information such as your name, email address or IP address, nor does it collect personal, biographical information. The data you provide are therefore completely anonymous.

- Although you will gain no personal, direct benefits from participating in this survey, it aims to benefit all members of the South African postdoc community, by identifying important issues within that community and informing and equipping those who advocate policy, to respond to these issues.

- You are being asked to participate because in 2022 you were a postdoctoral fellow at one of South Africa's public universities.

- Your email address was obtained either from your host institution, or from the public domain.

- In the questionnaire, you will be asked questions about your postdoctoral career trajectory, current postdoctoral position, as well as your future career preparation and expectations.

- The questionnaire touches on some sensitive subjects, such as your income bracket and experiences of discrimination.

- Once you have submitted your completed questionnaire online, you will no longer be able to withdraw your responses as there will be no way of linking your responses back to you.

- The anonymous data you provide could serve as a baseline if we decide to repeat the study in future to determine changes over time.

- The survey will be conducted according to the ethical guidelines and principles of South Africa's Department of Health Ethics in Health Research: Principles, Processes and Studies (2015).

- If you have any questions about this study or encounter any problems, please contact the principal investigator of this survey, Prof Heidi Prozesky, at hep@sun.ac.za.

- If you have questions, concerns or complaints specifically with regard to your rights as a research participant, please contact Mrs Clarissa Robertson [cgraham@sun.ac.za; 021 808 9183] at Stellenbosch University's Division for Research Development.

- Please save a copy of this information leaflet.

# Annexure C: Survey questionnaire (MS Word version)

**SECTION 1: POSTDOC CAREER TRAJECTORY**

1. Did you go straight from PhD graduation to a postdoc position?
    - Yes
    - No
    - Other (please specify)

2. Did you have another postdoc position right before you started your current position?
    - No
    - Yes, in another country
    - Yes, in South Africa, but in a different institution
    - Yes, in South Africa and at the same institution

3. [If Q2≠No] What is the total number of postdoc positions that you have held since completing your PhD? *Please include your current position in your answer*. [Open form]

4. [If Q2≠No] At the end of 2021, what was the total number of years you had spent in all your postdoc positions taken together? [Open form]

5. [If Q2≠No] To what extent do you agree or disagree with the following statement: "Poor job prospects have led to me holding more than one postdoc position since my PhD"
    - Strongly agree
    - Agree
    - Neither agree nor disagree
    - Disagree
    - Strongly disagree

**SECTION 2: CURRENT POSTDOC POSITION**

6. At which institution is your current postdoctoral position? [Drop-down list, as per universities who were sent the letter from the NRF]

7. Which scientific field or discipline best matches the work you do primarily, in your current postdoctoral position? *Please be as specific as possible, for example, waste water management OR theoretical physics OR educational management*

8. What is your current, individual, gross annual income in South African Rand (ZAR)? *If your income is not in ZAR, please use a currency converter (e.g. http://www.xe.com/currencyconverter/)*
    - Less than R200 000
    - R200 000 to R299 999
    - R300 000 to R399 999
    - R400 000 to R499 999
    - R500 000 to R599 999
    - R600 000 to R699 999
    - R700 000 to R799 999
    - R800 000 or more
    - I'd prefer not to answer

9. To what extent to you agree or disagree with the following statements? [matrix question; 1=strongly agree; 2=agree; 3=neither agree nor disagree; 4=disagree; 5=strongly disagree]
    - The money I receive is adequate to cover my expenses and provide for reasonable leisure activities, considering the cost of living in my community
    - I can currently save the amount of money I want to, from my remuneration
    - I receive the pay I am due on time and without bureaucratic problems

10. How frequently do you receive your remuneration?
    - Monthly
    - Quarterly
    - Bi-annually
    - It varies
    - Other (please specify)

11. [If Q10≠Monthly] Does non-monthly payment of your remuneration pose any challenges to you personally?
    - A lot
    - Somewhat
    - Not at all

12. Currently, in accordance with the Income Tax Act of South Africa, postdocs are exempt from paying tax. This "tax exempt" status of postdocs has some drawbacks, though: postdocs are treated more as students than as employees (due to the non-taxable, bursary status of their remuneration), and their access to employment benefits are limited to the minimum or are non-existent. If given the choice, would you prefer to:
    - Keep your tax-exempt status?
    - Change your status to that of a tax-paying employee meaning less net pay, but access to employment benefits?
    - Unsure
    - Other (please specify)
    - Not applicable (I have the tax-paying status of an employee)

13. Which of the following benefits are available to you at your institution? *Select as many as apply*.
    - Medical scheme
    - Paid vacation ("annual") leave
    - Paid sick leave
    - Paid maternity / parental leave
    - Paid family or "compassionate" leave
    - Transport allowance
    - Relocation costs
    - Retirement fund / Pension plan
    - Disability benefits / Workplace insurance
    - Group life insurance scheme
    - Exemption from tuition fees
    - Subsidised childcare
    - Other (please specify)

14. Do you have a memorandum of agreement (MoA) or other similar contract (e.g. a letter of appointment) with your host institution?
    - Yes
    - No
    - Unsure

15. [If Q14=Yes] Were you able to negotiate the terms and conditions of your MoA or contract to the extent that you would have preferred?
    - Yes, fully
    - Yes, somewhat
    - No, not at all

16. In your current postdoc position, have you contributed, or do you contribute to any teaching (e.g. classroom or small-group teaching)?
    - No
    - Yes, undergraduate
    - Yes, graduate
    - Yes, both

17. Do you want to teach?
    - No
    - Yes, undergraduate
    - Yes, graduate
    - Yes, both
    - Unsure

18. In your current postdoc position, have you contributed, or do you contribute to formal supervision of master's students?
    - No
    - Yes

19. [If Q18=Yes] Is your supervision of master's students formally acknowledged, i.e. you appear on the students' supervision record and thesis?
    - Yes, always
    - Yes, but only sometimes
    - No, never

20. Regarding master students' supervision, which one of the following statements best fits your current situation?
    - I am allowed to supervise and co-supervise
    - I am allowed to co-supervise but not to be the main supervisor
    - I am not allowed to do any type of supervision
    - Other (please specify)

21. Do you want to supervise master's students?
    - No
    - Yes
    - Unsure

22. In your current postdoc position, have you contributed, or do you contribute to formal supervision of doctoral students?
    - No
    - Yes

23. [If Q22=Yes] Is your supervision of doctoral students formally acknowledged, i.e. you appear on the students' supervision record and thesis?
    - Yes, always
    - Yes, but only sometimes
    - No, never

24. Regarding doctoral students' supervision, which one of the following statements best fits your current situation?
    - o I am allowed to supervise and co-supervise
    - o I am allowed to co-supervise but not to be the main supervisor
    - o I am not allowed to do any type of supervision
    - o Other (please specify)

25. Do you want to supervise doctoral students?
    - o No
    - o Yes
    - o Unsure

26. What is the primary reason you chose to take your current postdoc position?
    - o To gain additional research training / experience in my doctoral field
    - o To gain research training / experience in a different field of research
    - o To enhance my future employment prospects within a university or research institute
    - o I feel it is a necessary step to obtain a desired permanent position
    - o To develop my research portfolio through focussed research
    - o I was unable to find a different, suitable position
    - o Other (please specify)

27. Thinking about your current postdoc position, how satisfied are you with the following? [matrix question; 1=completely satisfied, 2=somewhat satisfied, 3=neither satisfied nor dissatisfied, 4=somewhat dissatisfied, 5=completely dissatisfied; 6=NA]
    - o Personal sense of accomplishment
    - o Opportunities to work on interesting projects
    - o Opportunities to interact with high-quality researchers from other departments and institutions
    - o Ability to influence decisions that affect me
    - o Level of integration into the academic/research community at my host institution
    - o My host institution's provision of career-development opportunities, assistance or advice
    - o My host institution's provision of workplace-sponsored training opportunities in skills that postdocs need
    - o Your current postdoc position overall

28. To what extent to you agree or disagree with the following statements? (matrix question; 1=strongly agree; 2=agree; 3=neither agree nor disagree; 4=disagree; strongly disagree)
    - o There are no differences between the treatment of male and female postdocs by their supervisors and colleagues
    - o There is no discrimination against postdocs based on their race or ethnic background

29. Do you feel that you yourself have experienced discrimination or harassment in your current postdoc position?
    - o Yes
    - o No
    - o I'd prefer not to say

30. Have you observed discrimination or harassment during your current postdoc position? *This relates to observations of others, e.g. colleagues, not your personal experience*.
    - o Yes
    - o No
    - o I'd prefer not to say

31. To what extent do you agree or disagree with the following statements? [matrix question; 1=strongly agree; 2=agree; 3=neither agree nor disagree; 4=disagree; 5=strongly disagree]
    - My supervisor is meeting my initial expectations
    - I have learned much from my supervisor about how to succeed as a scientist
    - I consider my supervisor to be a mentor
    - My supervisor demands to be first or senior author on all the journal articles we publish
    - My supervisor understands that I have family and/or personal obligations
    - My supervisor is supportive of my career plan
    - My research work and scientific contributions are valued by my supervisor

32. How satisfied are you with [matrix question; 1=very satisfied; 2=satisfied; 3=somewhat satisfied; 4=not very satisfied; 5=not at all satisfied]
    - The amount or level of supervision you receive from your supervisor
    - The amount of communication with your supervisor
    - Your supervisor's overall performance as your supervisor

33. How much control do you have over decisions about the following aspects of your research? [matrix question; 1=completely under my control; 2= mostly under my control; 3=shared control with supervisor; 4=mostly under supervisor's control; 5= completely under supervisor's control]
    - Planning new research projects
    - Choosing collaborators
    - Writing papers
    - Writing grant proposals
    - Determining authorship of papers

34. Are you allowed to apply for funding as a principal (lead) investigator of a research project?
    - Yes
    - No
    - I don't know

35. Are you required, either formally or informally, to produce a certain number of peer-reviewed journal articles per annum?
    - Yes
    - No
    - Unsure

36. [If Q35=Yes] How many peer-reviewed journal articles are you required to produce per annum? [Open form]

37. How important do you deem the following factors/attributes to be in contributing to a successful postdoc experience? [matrix question; 1=extremely important; 2=very important; 3=moderately important; 4=slightly important; 5=not at all important] [12]
    - The security of knowing that my training will push my career forward
    - Having a supervisor with adequate funding/grants
    - University facilities (e.g. comprehensive collection of journals and books; high-quality research tools; well-maintained buildings; lab technical support; university services)
    - Opportunities for networking (e.g. attending scientific meetings or meeting influential researchers)
    - Mentoring
    - Direction and vision
    - Accommodations made by spouse, partner, and/or family
    - Training
    - Communication
    - Compensation and benefits that are aligned with my specialised experience and expertise

- Quality research
- Supportive colleagues

**SECTION 3: FUTURE CAREER PREPARATION AND EXPECTATIONS**

38. In the past year, have you applied for any permanent positions?
    - Yes
    - No

39. [If Q38=Yes] To what extent do you agree or disagree with the following statement: "The position/s I applied for are related to my current research"?
    - Strongly agree
    - Agree
    - Neither agree nor disagree
    - Disagree
    - Strongly disagree

40. What is your perception of the job market in your field?
    - Excellent
    - Good
    - Fair
    - Poor
    - Not sure

41. How well prepared for non-academic career opportunities do you consider yourself?
    - A lot
    - Somewhat
    - Not at all

42. What do you think is the greatest challenge for your personal career progression? *Please select only one answer*.
    - The economic impact of COVID-19
    - Competition for funding
    - Unwillingness / inability to sacrifice personal time / time with family
    - Unwillingness / inability to relocate
    - Discrimination / bias
    - My desire to stay in academia
    - My desire to leave academia
    - Lack of appropriate networks / connections
    - Language skills
    - Lack of relevant skills
    - Other (please specify)
    - None

43. [If Q42=Lack of relevant skills] You suggested that a lack of relevant skills is the greatest challenge for your personal career progression. Which skills do you feel you're lacking in?
    - Interpersonal / communication skills
    - Writing skills
    - People management / leadership skills
    - Computational skills
    - Statistical skills
    - Specific experimental techniques
    - Other (please specify)

44. Do you hope to pursue an academic career?
    - Yes
    - No
    - Unsure

45. [If Q44=Yes] When thinking of your future, long-term career plans, in which role would you most prefer to work?
    - Research and teaching combined (i.e. a "tenure track" or professor position)
    - Primarily or only research
    - Primarily or only teaching
    - Not yet decided
    - Other (please specify)

46. [If Q44=No] When thinking of your future, long-term career plans, in which role would you most prefer to work?
    - Research
    - Research management / administration / policy
    - Science communication / Scientific publishing or writing
    - Patent law / Technology transfer
    - Consulting
    - Professional practice
    - Not yet decided
    - Other (please specify)

47. Up to this point in time, how well do you feel your postdoc training has: [matrix question; 1=very well, 2=well, 3=indifferent, 4=not very well, 5=not well at all, 6=not sure, 7=I have not been in a postdoc position long enough to judge)
    - Prepared you for achieving your career goals?
    - Given you skills that would be marketable in a variety of careers?
    - Exposed you to career opportunities outside of academia?

48. Do you plan to leave South Africa upon completion of your current postdoc position?
    - Yes
    - No
    - Unsure

49. [If Q48=Yes] Why do you plan to leave South Africa? *Please select all that apply*.
    - Better/more job opportunities elsewhere
    - Variety of experience / exposure to new people and ideas
    - Prefer to stay but cannot because of immigration rules
    - Lack of support (e.g. funding) for research and development in South Africa
    - Better facilities/technology/researchers elsewhere
    - Financial reasons
    - Personal reasons
    - Other (please specify)

50. In which country do you plan to pursue your career? [dropdown list]

**SECTION 5: ONE LAST QUESTION**

51. Please feel free to share any additional comments or remarks below:

**SCREEN-OUT MESSAGE**

Almost finished!

Many thanks for the time you took to participate in South Africa's first national survey of postdoctoral fellows! If you would like us to share with you a report detailing our results of this survey, please email [hep@sun.ac.za](mailto:hep@sun.ac.za) with "Request for report" in the subject line.

Click the button below to submit your responses. You will be given a link to the PRIZE ENTRY website where you will be prompted for your email address. If you would like to be included in a drawing with a chance to win a R2 500 Takealot voucher, please provide your email address. This information is not linked in any way to your questionnaire responses.

Again, thank you and best wishes for your career!

# Annexure D: Respondents' comments on the survey

In this annexure, we summarise the comments we received from respondents on the survey and/or the questionnaire. We start with the positive comments, followed by the respondents' hopes of impact that the survey results may have. We conclude with constructive criticism that was offered by respondents, which we believe would be useful for future surveys of this nature.

A number of respondents were "happy" or glad" that the survey – considered "long overdue" – was "finally" being conducted. It was considered a very important survey, because of the "precarious situation many postdocs face", and because, "in South Africa, postdoctoral researchers are not treated very well". Many more respondents thanked us for conducting the survey. One reason was that "it will help shed light on the current status of postdoctoral fellows in South Africa", but more frequently, we were thanked for giving the respondents the opportunity to air or share their thoughts, and to communicate their experiences as postdocs. For one respondent, doing so was "somewhat therapeutic, and helps for self-re-evaluation". Another even apologised "for the long reply" to the request for comments at the end of the questionnaire.

As to the questionnaire, positive feedback from two respondents was that it covered "all" or at least "most of" their "concerns". Then there were general statements, such as finding the questionnaire "interesting", or even "remarkable". Further describing the questionnaire as "informative", one respondent believed that it "will help to develop an understanding about the current problems of research-oriented work, and their future solutions". One respondent looked "forward to reading the findings".

Two respondents expressed the hope that the survey would impact decision making. Specifically, one felt that those decisions should "create a pathway for persons that have undergone postdoc training in South African universities to be integrated into academia so that they can also train and motivate individuals interested in academia or research-only". More generally, it was also hoped that the survey would "help propel postdoc experiences and expectations in South Africa"; improve "the postdoc ecosystem"; and "prove helpful in bettering the lives of future postdocs".

Contrary to two respondents' view that the questionnaire covered "all" or at least "most of" their "concerns", a few respondents felt that this was not the case. It "does not take some very important points into consideration", one stated. Another was more specific, claiming that "the disadvantages of being a postdoc in South Africa" were "not captured well" in the questionnaire. However, a third completely disagreed, as they found the questionnaire to be "a little biased towards revealing negative postdoctoral experiences, without giving much room to share positive experiences". That respondent recommended a question such as, "What do you enjoy most about postdoc life?", with several options. Other aspects we did not take into consideration, according to one respondent, were "the complexities of multidisciplinary projects and multiple supervisors" when "phrasing questions about supervision".

Another bias that was highlighted is the questionnaire's "focus on the hard sciences". The respondent further explained that, for postdocs in some other faculties there are fewer opportunities outside academia, and we are generally paid much less as well, making some of your questions irrelevant and laughable". For another, most of the questions did not apply to their situation, as they were "not at the onset of an academic career". On the other hand, a respondent felt that, because they were "still at an early stage of [their] postdoc", "not much can be shared".

We were criticised on technical points as well, including "grammatical errors", some of the options not being well set up or described; and some items that request multiple options to be selected, while the software had been programmed to allow only one option to be selected. We again checked all those questions that ask for more than one response and they were all programmed to allow for multiple responses. It may be an issue with the UP administration of the survey. One mistake that we are aware of, and which a respondent identified, is that we omitted an "other" option at Question 26 ("What is the primary reason you chose to take your current postdoc position?"). Thus, respondents could not specify other reasons beyond the six that we listed. Finally, one respondent mentioned that the questionnaire is "relatively long" and therefore "the first time I started it, I did not finish".